\documentclass[twocolumn]{aa}
\usepackage[varg]{txfonts}
\usepackage{graphicx}
\usepackage{natbib}

\bibpunct{(}{)}{;}{a}{}{,}

\newcommand{\ca}{\ion{C}{ii}}
\newcommand{\mg}{\ion{Mg}{ii}}
\newcommand{\si}{\ion{Si}{iv}}
\newcommand{\cgs}{erg~s$^{-1}$~cm$^{-2}$~sr$^{-1}$}
\newcommand{\sn}{$\mathrm{S}/\mathrm{N}$}

\usepackage{color}

\begin{document}

\title{Statistical analysis of UV spectra of a quiescent prominence observed by IRIS}
\author{
S. Jej\v ci\v c\inst{1,2}
\and
P. Schwartz\inst{3}
\and
P. Heinzel\inst{1}
\and
M. Zapi\'or\inst{1}
\and
S. Gun\'ar\inst{1}
}
\institute{
Astronomical Institute, The Czech Academy of Sciences, 25165 Ond\v{r}ejov, Czech Republic
\and
Faculty of Mathematics and Physics, University of Ljubljana, 1000 Ljubljana, Slovenia
\and
Astronomical Institute of Slovak Academy of Sciences, 05960 Tatransk\'a Lomnica, Slovak Republic     
}
\date{}
\abstract
{The paper analyzes the  structure and dynamics of a quiescent prominence that occurred on October 22, 2013 and was observed by several
instruments including the Interface Region Imaging Spectrograph (IRIS).}
{We aim to determine the physical characteristics of the observed prominence using \mg\ k and h (2796 and 2803~\AA) , 
\ca\ (1334 and 1336~\AA), and \si\ (1394~\AA) lines observed by IRIS. In addition we study the dynamical behavior of the prominence.}
{We employed the one-dimensional non-LTE  (departures from the local thermodynamic equilibrium – LTE) modeling of \mg\ lines assuming static 
isothermal-isobaric slabs. 
We selected a large grid of models with realistic input 
parameters expected for quiescent prominences (temperature, gas pressure, effective thickness, microturbulent velocity, height above the solar surface) 
and computed synthetic \mg\ lines. The method of Scargle periodograms was used to detect possible prominence oscillations.}
{We analyzed 2160 points of the observed prominence in five different sections along the slit averaged over ten pixels due to low signal to noise ratio in 
the \ca\ and \si\ lines. We computed the integrated intensity for all studied lines, while the central intensity and reversal ratio was determined only for both \mg\ and 
\ca\ 1334 lines. We plotted several correlations: time evolution of the integrated intensities and central intensities, 
scatter plots between all combinations of line integrated intensities, and reversal ratio as a function of integrated intensity. 
We also compared \mg\ observations with the models. Results show that more than two-thirds of \mg\ profiles and about one-half of \ca\ 1334 profiles are reversed. 
Profiles of \si\ are generally unreversed. The \mg\ and \ca\ lines are optically thick, while the \si\ line is optically thin.}
{The studied prominence shows no global oscillations in the \mg\ and \ca\ lines. Therefore, the observed time variations are caused by 
random motions of fine structures with velocities up to 10 km~s$^{-1}$. The observed average ratio of \mg\ k to \mg\ h line intensities can be used to determine 
the prominence's characteristic temperature. 
Certain disagreements between observed and synthetic line intensities of \mg\ lines point to the necessity of using more complex two-dimensional multi-thread modeling
in the future.}

\authorrunning{S. Jej\v ci\v c et al.}
\titlerunning{Behavior of IRIS UV lines}


\maketitle
\section{Introduction}
         \label{s-int}

Prominences are relatively dense and cool plasma structures embedded in the low-density, hot corona. Prominence plasma is supported against gravity by the coronal 
magnetic fields. Prominences can be seen in emission as bright prominences above the limb or as dark filaments visible in absorption against the
solar disk. Depending on their location, nature, and lifetime we classify them as quiescent or active region prominences. 
Quiescent prominences are long-lasting structures 
with lifetimes of several days or weeks. On the other hand, active region prominences are short-lived structures lasting for several hours or days. However,
quiescent prominences may also sometimes become unstable and suddenly erupt. Reviews of the physics of solar prominences can be found in \citet{lab10}, \citet{mac10},
in the book "Solar Prominences" \citep[][]{vial15}, or in the proceedings of the International Astronomical Union (IAU) 300 Symposium \citep[][]{iau14}. Observations of quiescent prominences were 
reviewed by \citet{hei07}, their observational characteristics by \citet{par14}, and the modeling of prominences and their fine structures was reviewed
by \citet{gun14}. A distinct class of eruptive prominences was found to be embedded in the core of the coronal mass ejections (CMEs). These prominences have relatively 
hot temperatures (around 10$^5$~K), low electron density and low gas pressure \citep[see e.g.][]{hei16cme,jej17}.

Observations made by space-borne and ground-based instruments show that even the quiescent prominences, which have relatively stable
large-scale structures, show highly dynamical small-scale structures with plasma flows, small amplitude periodic oscillations, and apparent lifetimes as 
short as several minutes. Typical prominence fine structures consist of a large number of horizontal and quasi-vertical threads and knots of cool plasma,
with typical widths of several hundred kilometers \citep[or even less, see, e.g.,][]{lin05}. Low amplitude oscillations with periods of  several tenths of minutes are 
observed in line intensities, line widths, and line-of-sight (LOS) velocities, while theoretical models show temporal variations of temperature,
electron density, and  velocity \citep[see, e.g.,][]{oli02,arr12,hei14b}. These periodic oscillations are most
likely related to standing or propagating magnetohydrodynamic (MHD) waves and are under investigation.

The Interface Region Imaging Spectrograph \citep[IRIS,][]{pon14} is capable of taking high resolution spectra and slit-jaw images
(SJIs) in the far ultraviolet (FUV) and near ultraviolet (NUV) regions. Several ultraviolet (UV) spectral lines observed by IRIS can be used to study individual
prominence fine structures in great detail. This allows us to gain a better understanding of their nature, evolution, and behavior. With
a combination of spectroscopic observations and non-LTE (departures from the local thermodynamic equilibrium – LTE) radiative transfer prominence models we can derive several key parameters of
prominence plasma, such as the temperature, electron density (which is proportional to the gas pressure), LOS velocity, integrated line intensity, 
and optical thickness of the prominence plasma in the used spectral line.

The spectroscopic analysis of the IRIS profiles of \mg\ k and h lines made by \citet{sch14} and \citet{lev16,lev17} in combination with
the results of the one-dimensional (1D) non-LTE modeling of \citet{hei14,hei15} show that these two lines are relatively strong, optically thick (up to
10$^3$ or 10$^4$), and sensitive to Doppler brightening or dimming effects for velocities larger than 20 km~s$^{-1}$.
Observations show that \mg\ profiles can be unreversed (single), reversed, or composite profiles formed by the superposition of several
individual Doppler-shifted single profiles emitted along a single LOS. Models show that unreversed profiles occur at lower values of
gas pressure, while at higher pressure values the reversal becomes more pronounced with the temperature. The non-LTE models that
included the prominence-corona transition region (PCTR) can explain the shape of some reversed profiles \citep{hei15}. In
addition, \ca\ and \si\ IRIS spectra were studied by \citet{lev16}. These authors found that \si\ lines are very
faint and have a low signal to noise (\sn) ratio.

In the present paper we analyze IRIS observations of a quiescent prominence and study the correlations between the intensities of
\mg, \ca, and \si\ lines, together with the correlations between different plasma parameters derived by modeling. We also
study the dynamics of this prominence and investigate the presence of possible oscillations. 

The paper is organized in the following way. In Sect.~\ref{s-obs} we present the used UV observations of \mg, \ca, and \si\ lines
obtained by IRIS. Section~\ref{s-ana} deals with the analysis of observed IRIS UV spectra. Section~\ref{s-dyn} describes the
dynamics of the observed prominence structures: time evolution of the LOS velocities, integrated intensities, and
the central intensity of selected UV lines together with the potential prominence oscillations. In Sect.~\ref{s-int} we show the results of 
our statistical analysis, which demonstrate correlations
between \mg, \ca, and \si\ lines together with 1D non-LTE modeling of \mg\ lines and CHIANTI (an atomic database for spectroscopic diagnostics 
of astrophysical plasmas) computations of \si\ line. Discussion and conclusions are presented in Sect.~\ref{s-con}.

\section{IRIS observations}
        \label{s-obs}

On October 22 and 23, 2013 a coordinated observing campaign was carried out with two space-borne spectrographs: SUMER (Solar
Ultraviolet Measurements of Emitted Radiation; \citeauthor{wil95} \citeyear{wil95}) on board the Solar and Heliospheric Observatory (SoHO) and 
IRIS (Interface Region Imaging Spectrograph; \citeauthor{pon14} \citeyear{pon14}). Several ground\discretionary{-}{-}{-}based instruments were also involved,
among them the Multichannel Flare Spectrograph (MFS; \url{http://radegast.asu.cas.cz/MFS/prominence_archiv/sls.html}) and the Horizontal Sonnen Forschung Anlage 2 
(HSFA2; \citeauthor{kot09} \citeyear{kot09}) at the Astronomical Institute of the Czech Academy of Sciences, Ond\v{r}ejov, and the COronal
Multichannel Polarimeter for Slovakia (COMP-S; \citeauthor{kuc10} \citeyear{kuc10}) at the Astronomical Institute of the Slovak Academy of Sciences. A quiescent, although rather dynamic, prominence was observed at the NW solar limb on October 22, 2013 by IRIS in both the FUV and NUV parts of the spectra between 8:41 and 10:49\,UT and by SUMER in the hydrogen Lyman lines between 07:00\,and\,08:33\,UT. The spectrographs at the Astronomical Institute of the Czech Academy of Sciences observed in the
H$\alpha$, H$\beta$, D$_{3}$, and \ion{Ca}{ii} H (3968\,\AA) lines between 10:38\, and\,11:39\,UT.

\begin{figure}  
\resizebox{\hsize}{!}{
\includegraphics{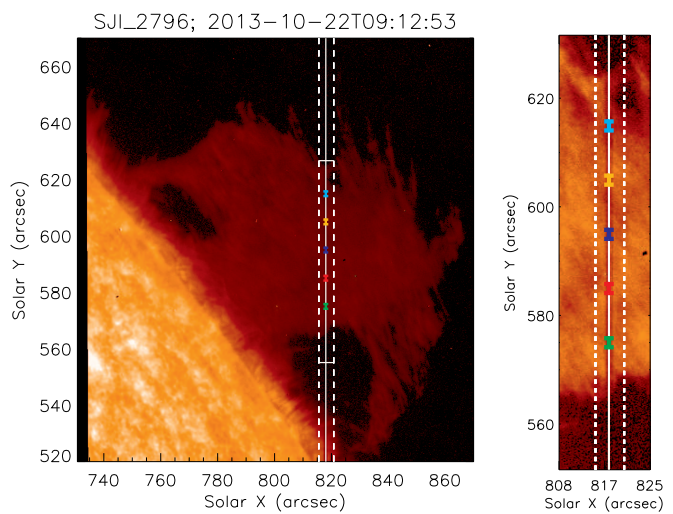}}
\caption{SJI in \ion{Mg}{ii} 2796\,\AA\ channel obtained at 09:12:53\,UT corresponding
 to eighth slit position of eighth raster of prominence observation.
 An area around the slit position is zoomed for better visibility. Five sections selected for analysis are marked
by color bars in the right panel. The movie corresponding to this figure is available as the online material of the journal where the white box indicates 
the range of data we used for analysis.} 
\label{f-prom}
\end{figure}

\begin{figure*}[h]    
\centerline{\includegraphics[width=0.33\textwidth,clip=]{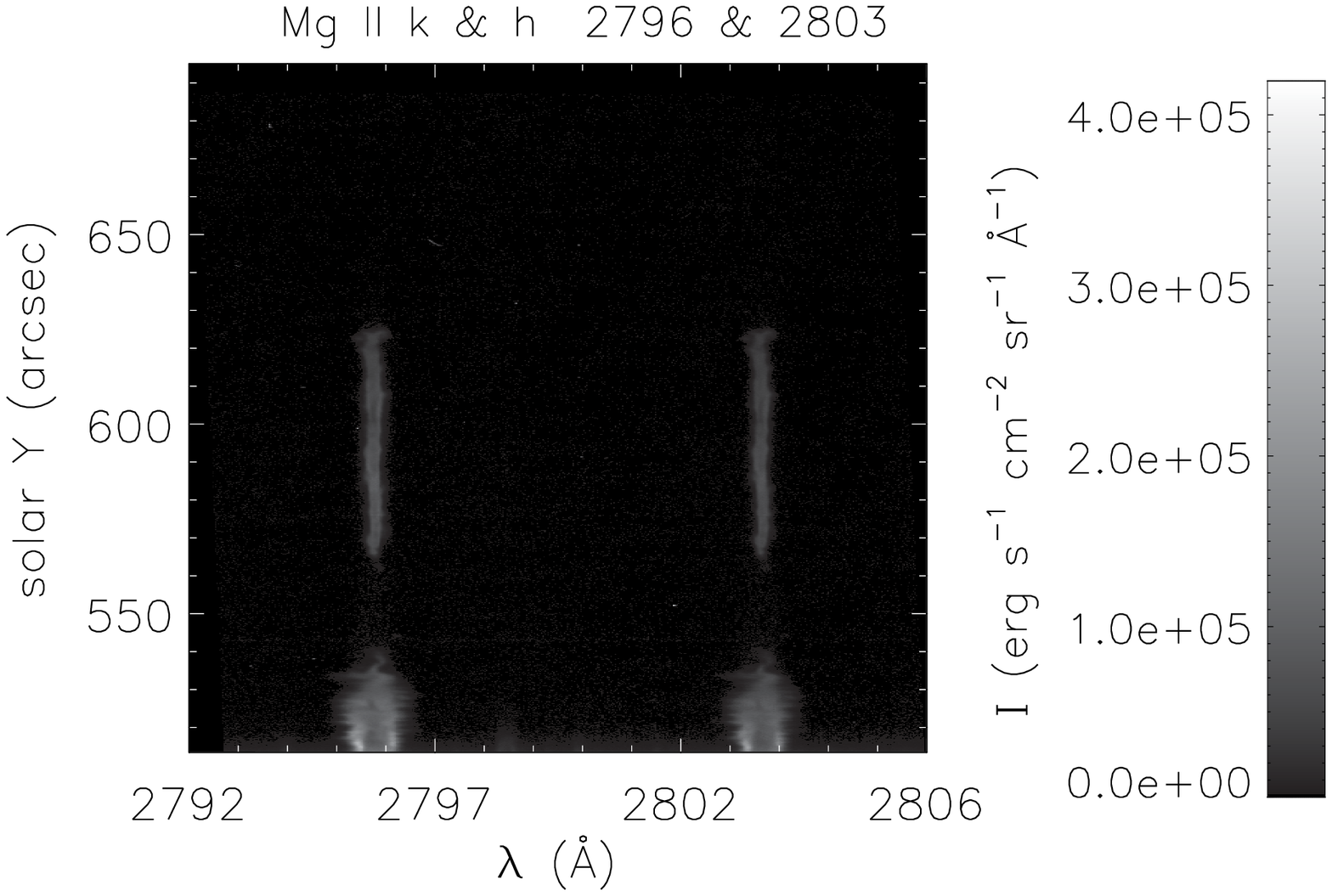}
            \hspace*{-0.0\textwidth}
            \includegraphics[width=0.33\textwidth,clip=]{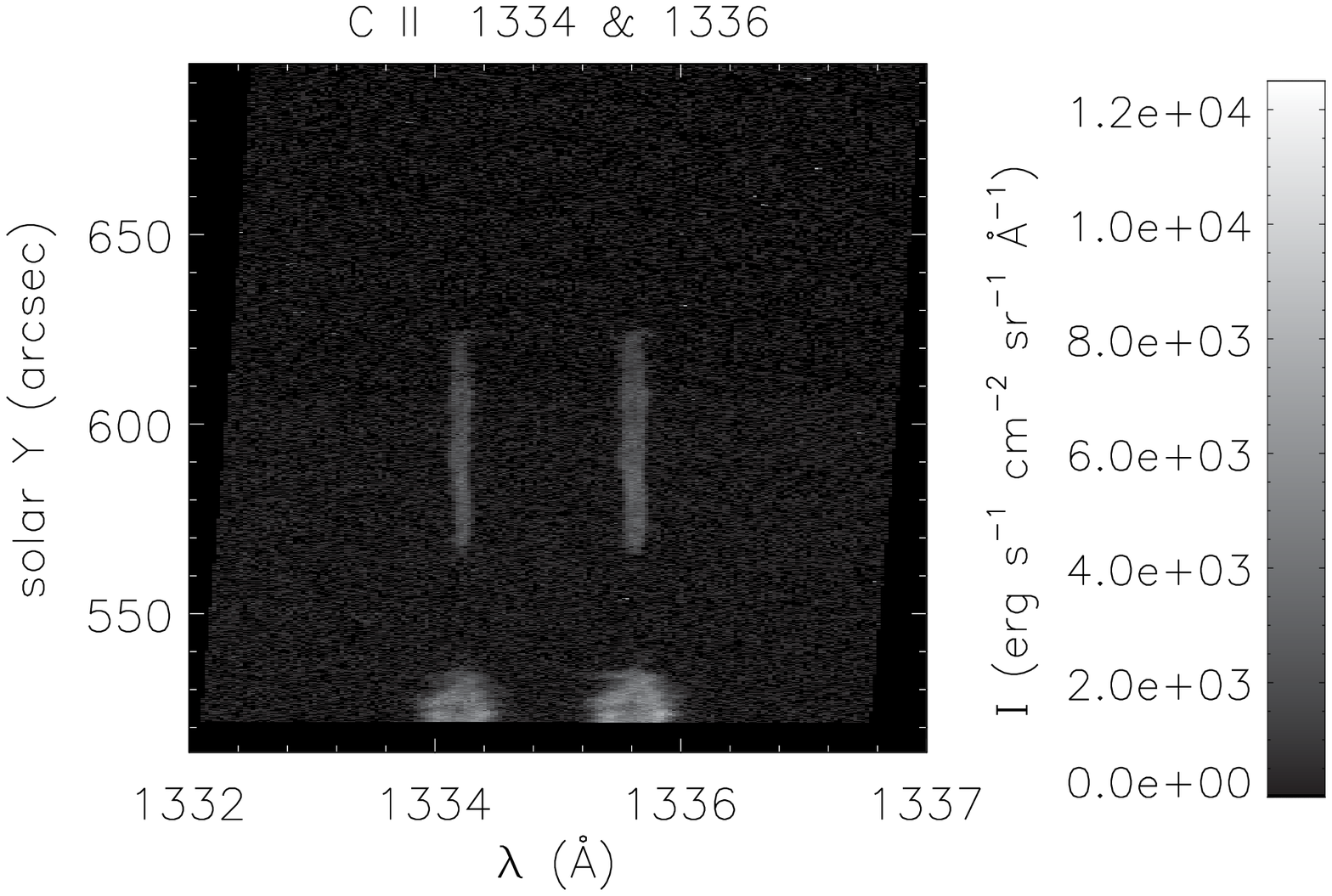}
            \hspace*{-0.0\textwidth}
            \includegraphics[width=0.33\textwidth,clip=]{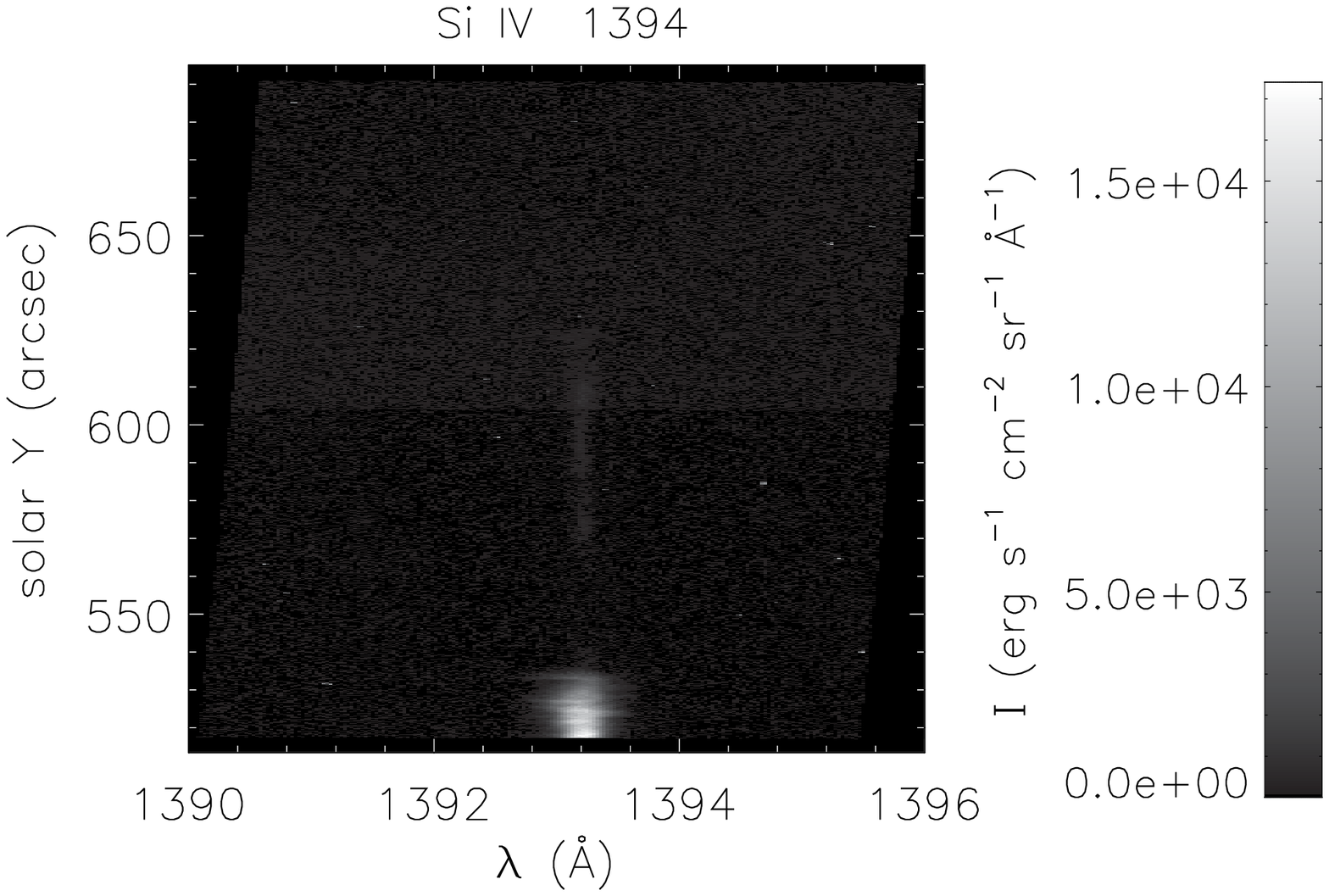}
            }
\caption{ Example of \mg\  k and h spectra at 2796 and 2803~\AA~(left panel), \ca\ spectra at 1334 and 1336~\AA~(middle panel),
and \si\ spectrum at 1394~\AA~(right panel) at 9:12:53 UT as indicated in Fig.~\ref{f-prom}.}
\label{f-sp}
\end{figure*}

\begin{figure*} [h]   
\centerline{\includegraphics[width=0.34\textwidth,clip=]{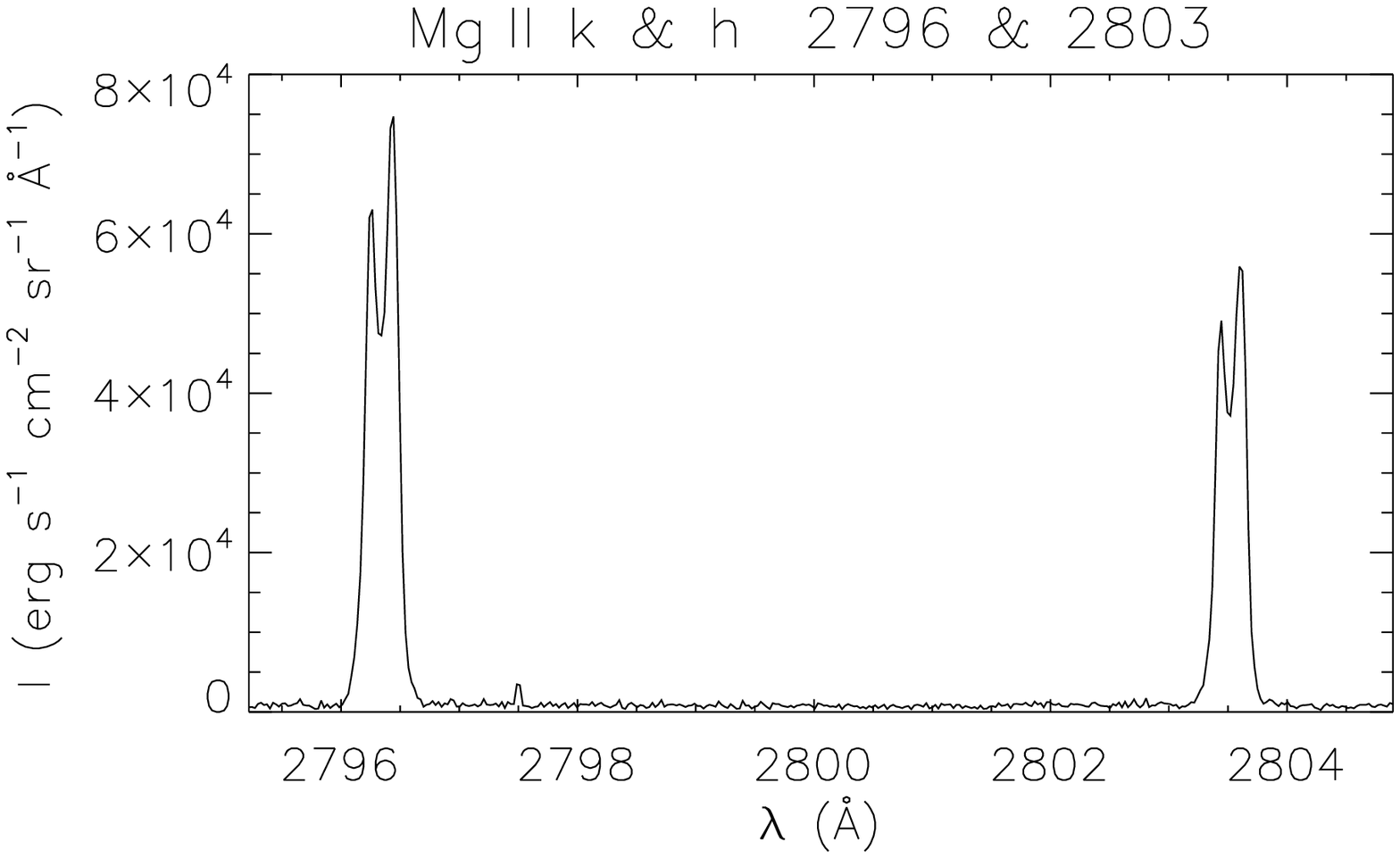}
            \hspace*{-0.01\textwidth}
            \includegraphics[width=0.34\textwidth,clip=]{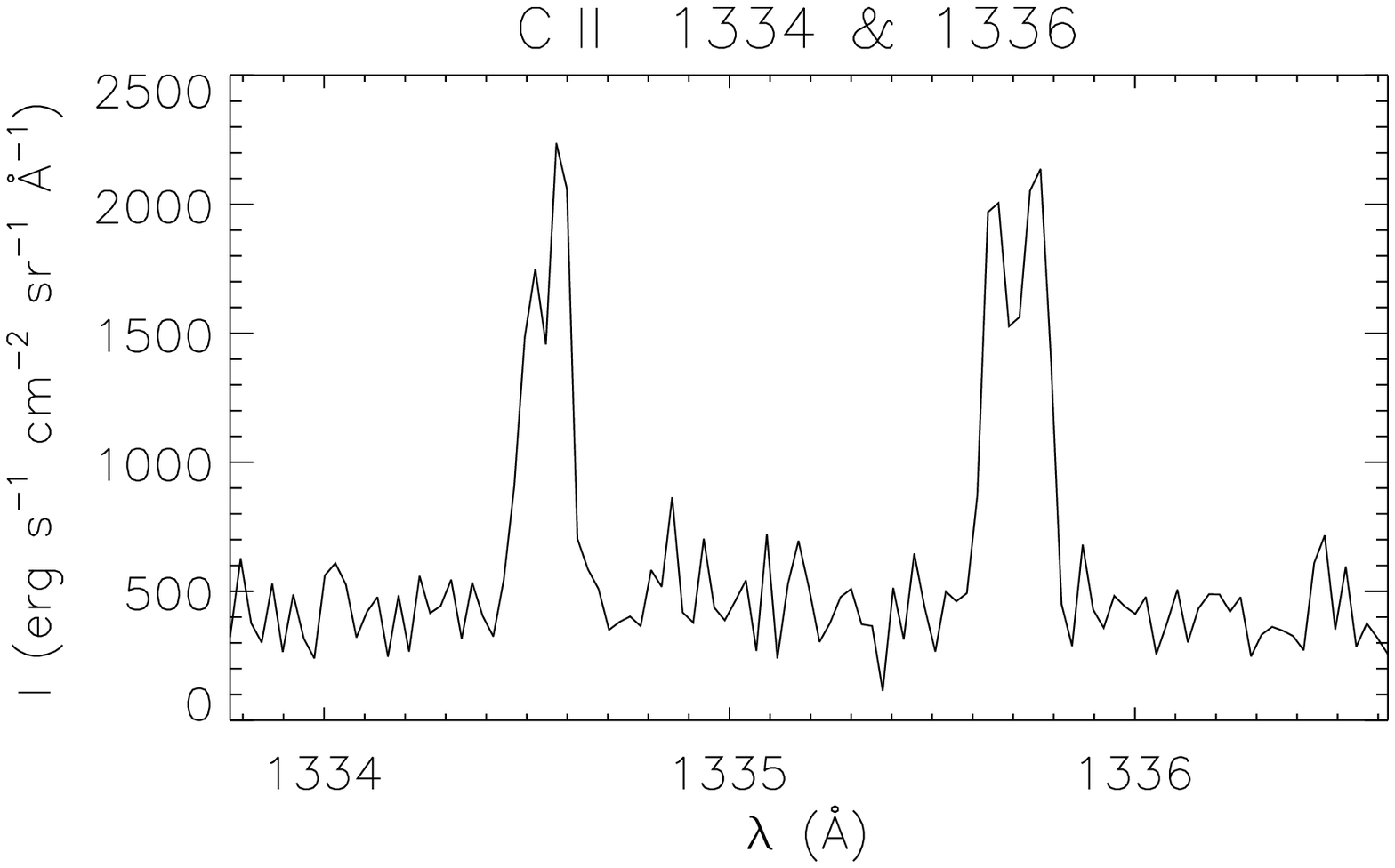}
            \hspace*{-0.015\textwidth}
            \includegraphics[width=0.34\textwidth,clip=]{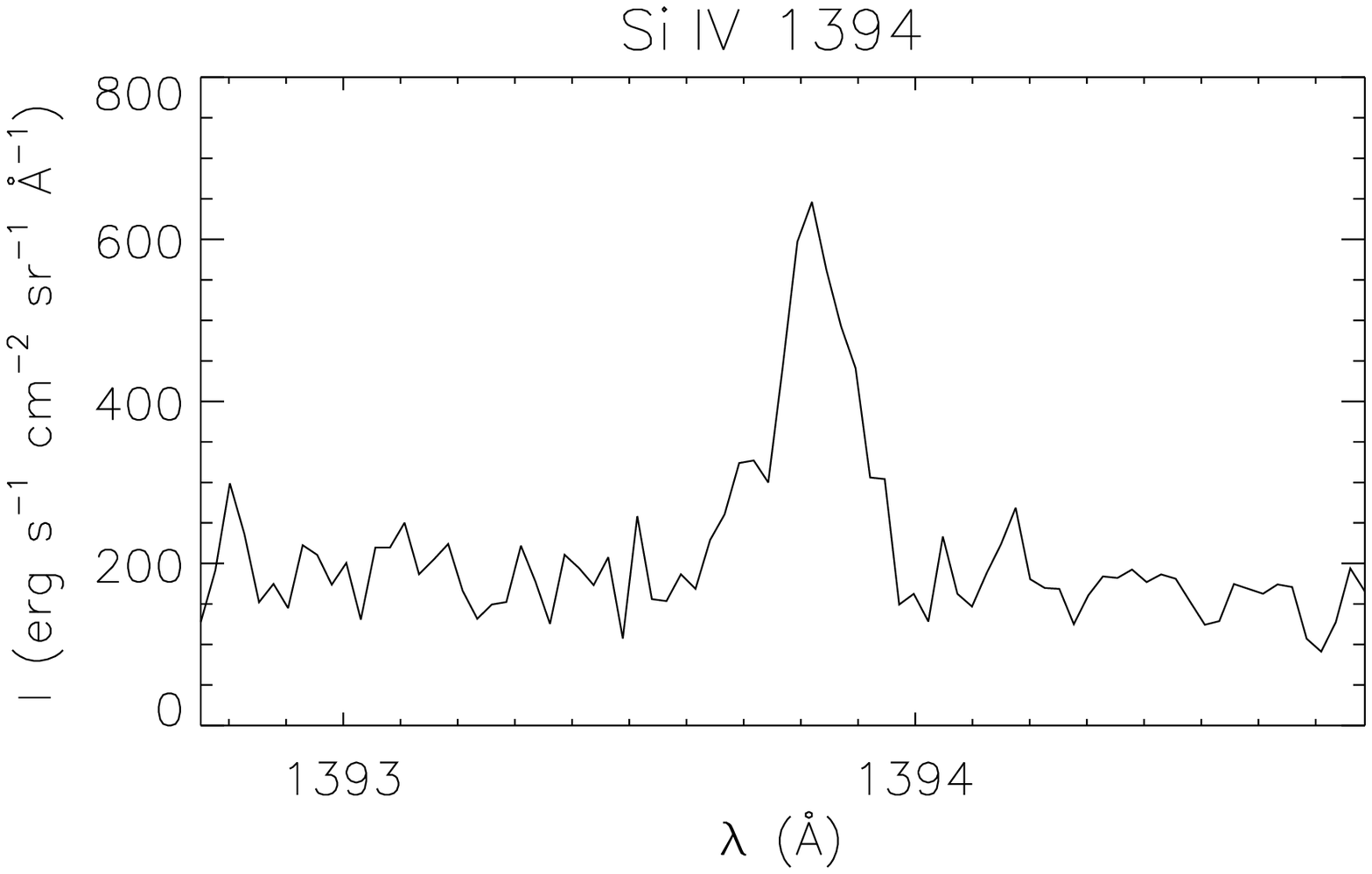}
            }
\caption{Example of \mg\  k and h profiles at 2796 and 2803~\AA~(left panel), \ca\ profiles at 1334 and 1336\,\AA~(middle panel),
and \si\ profile at 1394\,\AA~(right panel) for the middle section along the slit at 9:12:53 UT (see Fig.~\ref{f-prom}).}
\label{f-prof}
\end{figure*}

IRIS observed spectra in three different wavelength bands: two in FUV with wavelength ranges 1332\,\AA\,--\,1358\,\AA\ and 1389\,\AA\,--\,1407\,\AA\ and one in NUV 
within the range 2783\,\AA\,--\,2835\,\AA. These observations include lines of \ca\ 1334\,\AA\ and 1336\,\AA, \ion{Fe}{xii} 1349\,\AA, \ion{O}{i} 1356\,\AA, 
\si\ 1394\,\AA\ and 1403\,\AA, and \mg\ k and h (2796 and 2803~\AA), together with SJIs in \si\ 1403\,\AA\ and \mg\ 2796~\AA~channels. The field of view
(FOV) was 5\,arcsec\,$\times$\,174\,arcsec for each spectrum and 167\,arcsec\,$\times$\,174\,arcsec for each SJI. The IRIS observations consist of 30 repetitions 
(rasters) each with 16 different slit positions within one raster, with steps of 0.35~arcsec, exposure time 15\,s, and step cadence 16.6\,s,
which gives in total 480 spectra. Exposure time of all SJIs was 33\,s, which gives in total 240 images. The spectral resolution of the \ca\ spectra was 51.9~m\AA, while 
for \mg\ and \si\ it was 50.9~m\AA. Spatial resolution along the slit was 0.332~arcsec. The line intensities of level\discretionary{-}{-}{-}2 data expressed
in counts per second, which have been already corrected for dark current, flat\discretionary{-}{-}{-}field, and geometrical distortion, were calibrated into absolute 
radiometric units using the procedure published, for example, by \citet{kle16}. An example of one SJI obtained in the \mg\ 2796\,\AA\ channel at 09:12:53\,UT is shown
in Fig.~\ref{f-prom}. The corresponding slit position is marked by a vertical full white line while two white vertical dashed lines show boundaries of rastering.

\section{Characteristics and analysis of IRIS UV spectra}
        \label{s-ana}

Here we focus on the analysis of both FUV and NUV spectra observed by the IRIS spectrograph. Two FUV spectral bands of IRIS   
contain two weak \ca\ lines at 1334\,\AA~and 1336\,\AA~as well as one even weaker \si\ line at 1394\,\AA. The \ca\ line at 1336~\AA~actually consists of 
two components that are mutually blended. In the NUV region we analyze the two strong \mg\ k and h lines at 2796\,\AA~ and 2803\,\AA, respectively. 
Hereafter we will skip the wavelength units in the names of all selected lines. 

Examples of \mg, \ca,\ and \si\ spectra observed at 9:12:53\,UT are shown in Fig.~\ref{f-sp} for the slit position indicated in Fig.~\ref{f-prom}. 
In the lower part of all spectra a bright solar limb is visible together with dynamical spicules while the middle part corresponds to 
the prominence location. These prominence spectra are typical for a quiescent prominence because there are no significant Doppler motions.
We chose five sections along the prominence slit to have information about the structure and dynamics of different parts of the prominence (see Fig.~\ref{f-prom}).  
Because the prominence is rather weak with low \sn\ ratios  in \ca\ and in \si\ line intensities, we 
averaged the signal over ten pixels along the slit for all five selected sections. Due to this averaging we reduced the spatial 
resolution along the slit. The right panel of Fig.~\ref{f-prom} shows the zoomed region around the given slit position 
with the five sections marked by different colors. All the analysis is made for these five averaged sections and all plots will be presented using this color coding.   
 
Examples of the \mg, \ca,\ and \si\ line profiles for observations obtained at 9:12:53\,UT from the middle section (see Fig.~\ref{f-prom}) 
are shown in Fig.~\ref{f-prof}. In this example the \mg\ and \ca\ line profiles are reversed, while the faint \si\ line exhibits only one emission peak 
without any central reversal. 
Both \mg\ lines show a similar shape of the asymmetrical and reversed profiles where the blue peak is lower than the red one. 
Although profiles of the \ca\ 1334 and 1336 lines are also reversed and asymmetrical, they differ in shape from each other because  
the profile of the latter is a composition of two blended components. However, the asymmetry of the \ca\ 1334 line is qualitatively the same as that of the \mg\ lines.
Asymmetrical reversed profiles generally give us information about the dynamics along the LOS of the prominence.
A comparison between all profiles shows that noise is negligible only for the \mg\ lines where  \sn\ $\geq$ 500. 
The level of \sn\ $\sim$~15 in spectral intensities of the two \ca\ lines is still acceptable,   
while for the \si\ line the ratio \sn\ $\sim$~4 is rather low for any detailed profile analysis.  
All other lines observed by IRIS at the prominence are even fainter and noisier and therefore they are not used in our analysis.  

\begin{figure}[t]
\centering
\includegraphics[width=\columnwidth]{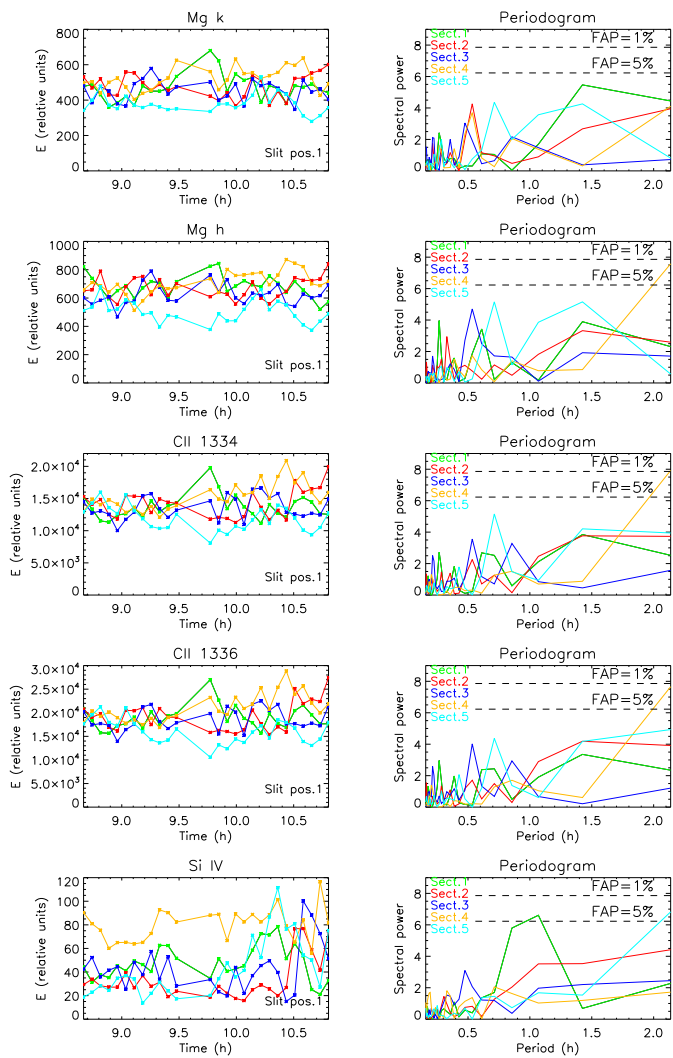}
\caption{Examples of temporal evolution of integrated intensity of \mg, \ca, and \si\ lines for fourth slit position and five selected sections (left panel) 
and corresponding Scargle periodograms (right panel). Color coding is the same as in Fig.~\ref{f-prom}. Dashed lines show 1\% and 5\% FAP levels.}
\label{f-oint}
\end{figure}

\begin{figure*} 
\centerline{\includegraphics[width=0.34\textwidth,clip=]{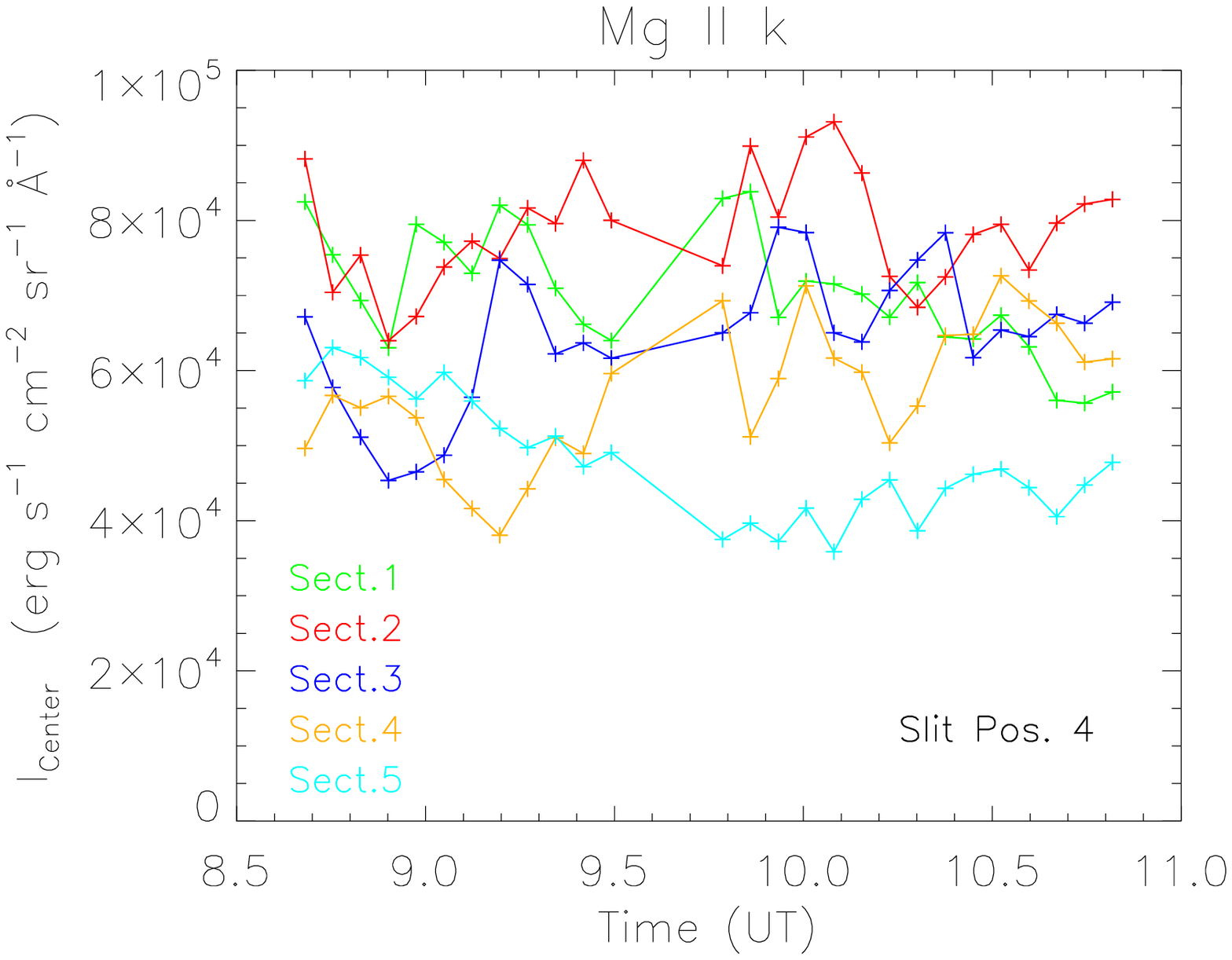}
            \hspace*{-0.01\textwidth}
            \includegraphics[width=0.34\textwidth,clip=]{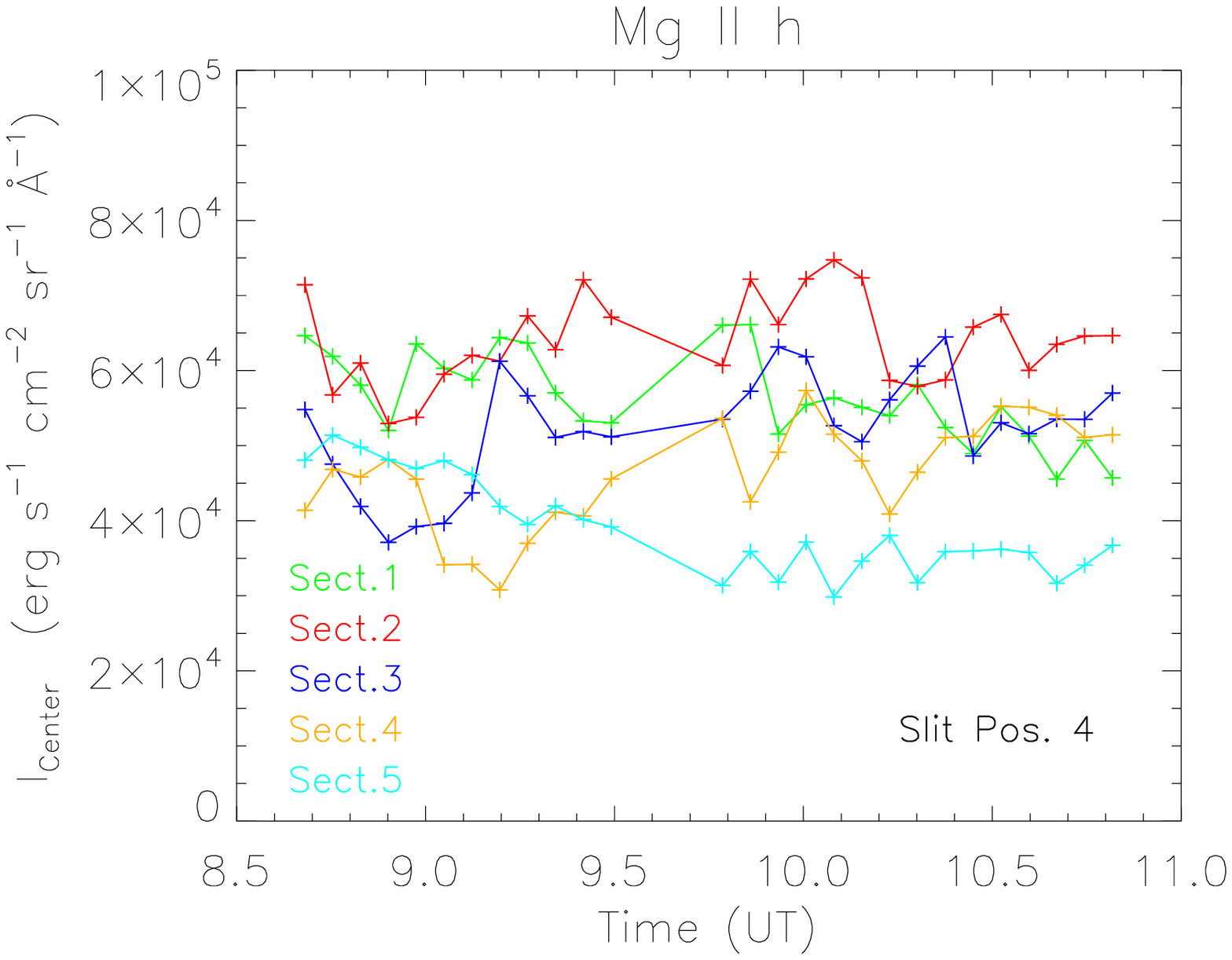}
            \hspace*{-0.01\textwidth}
            \includegraphics[width=0.34\textwidth,clip=]{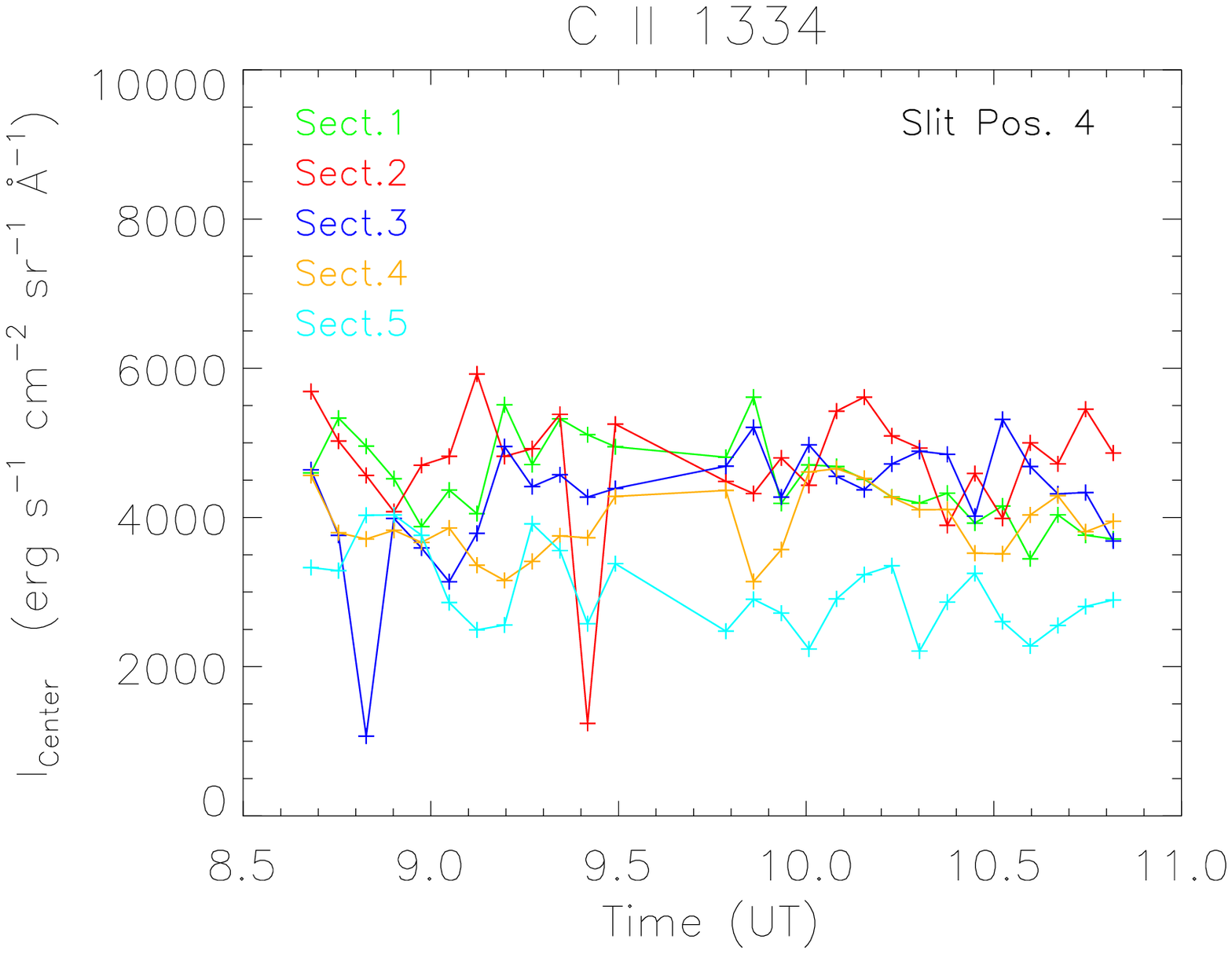}
            }
\caption{Temporal evolution of central intensity  of \mg\,k (left panel), \mg\,h (middle panel), and 
\ca\,1334 (right panel) lines for fourth slit position and five selected sections.}
\label{f-it}
\end{figure*} 

For a comparison of the observed \mg\ spectral line profiles with the synthetic ones obtained by a non-LTE model, in many cases it is more 
useful to compare characteristics of the profiles -- such as integrated intensity $E$, asymmetry, central reversal, width -- instead of 
a direct comparison of the profiles. For spectral lines with large optical thickness in their cores, one of the most important profile characteristics is 
the depth of the central reversal. 
Moreover, there is also a good connection between this 
characteristics and the physical parameters of the prominence plasma, such as temperature, density, and ionization degree. 
In the works of \citet{gun10}, \citet{ber11}, and \citet{sch15} where prominence 
observations in the hydrogen Lyman series were statistically compared with the two-dimensional (2D)\ non\discretionary{-}{-}{-}LTE models of the prominence fine structures, the 
depth of the central reversal together with the statistical population of reversed profiles showed a clear connection with the prominence plasma density. 
The depth of the central reversal was defined in those works as a ratio of minimal intensity at the reversal to the average of the maximal intensities at two peaks 
(by definition reversed profiles have this ratio lower than one). 
Here we follow the paper of \citet{hei14} and compute the reversal ratio as the ratio of the averaged line-peak to line-center intensity, thus reversed profiles have a reversal 
ratio larger than one. Integrated intensities of the \mg, \ca,\ and \si\ lines are computed in absolute radiometric units for five selected 
sections along the slit after subtracting the continuous background. Central intensities $I_{\rm Center}$ and reversal ratios are computed only for the \mg\ and 
\ca\ 1334 lines because the \ca\ 1336 line is blended and the   \si\ 1394 line is unreversed and noisy. Doppler velocities of the \mg\ and \ca\ lines are 
computed as well but without the \si\ 1394 line because it is too faint and noisy.

\begin{table*}[h]  
\centering
\caption{Correlation coefficients for a given section between both integrated intensities of 
\mg\ lines and both \ca\ lines together with combinations of \mg, \ca,\ and \si\ lines.}
\label{t-cor}
\begin{tabular}{ccccccccccc}
\hline
\hline
{\it Section}& \mg\ k & \ca\ 1334 & \mg\ k & \mg\ k & \mg\ h & \mg\ h & \mg\ k & \mg\ h & \ca\ 1334 & \ca\ 1336 \\
             & \mg\ h & \ca\ 1336 & \ca\ 1334 & \ca\ 1336 & \ca\ 1334 & \ca\ 1336 & \si\ 1394 & \si\ 1394 & \si\ 1394 & \si\ 1394 \\
\hline
1  & 0.994 & 0.786 & 0.665 & 0.559 & 0.677 & 0.568 & 0.400 & 0.396 & 0.396 & 0.410 \\
2  & 0.994 & 0.802 & 0.667 & 0.625 & 0.676 & 0.625 & 0.578 & 0.553 & 0.438 & 0.372 \\
3  & 0.998 & 0.736 & 0.585 & 0.545 & 0.581 & 0.527 & 0.268 & 0.206 & 0.324 & 0.365 \\
4  & 0.991 & 0.811 & 0.689 & 0.690 & 0.693 & 0.696 & 0.183 & 0.166 & 0.274 & 0.220 \\
5  & 0.992 & 0.822 & 0.636 & 0.698 & 0.642 & 0.700 & 0.457 & 0.428 & 0.510 & 0.485 \\
\hline
\end{tabular}
\end{table*}

\begin{table}[h]  
\centering
\caption{Correlation coefficients for a given section between integrated intensities and central 
intensities for \mg\ k line, \mg\ h line,  and \ca\ 1334 line.}
\label{t-cori}
\begin{tabular}{cccc}
\hline
\hline
{\it Section}&  $E$~~\mg~k &  $E$~~\mg~h &  $E$~~\ca~1334  \\
              & $I_{\rm central}$ & $I_{\rm central}$& $I_{\rm central}$ \\
\hline
1   & 0.440  & 0.458 & 0.493  \\
2   & 0.303  & 0.292 & 0.307 \\
3   & 0.599  & 0.546 & 0.376 \\
4   & 0.669  & 0.670 & 0.516 \\
5   & 0.607  & 0.606 & 0.472 \\
\hline
\end{tabular}
\end{table}

\begin{figure}
\centering
\includegraphics[width=\columnwidth]{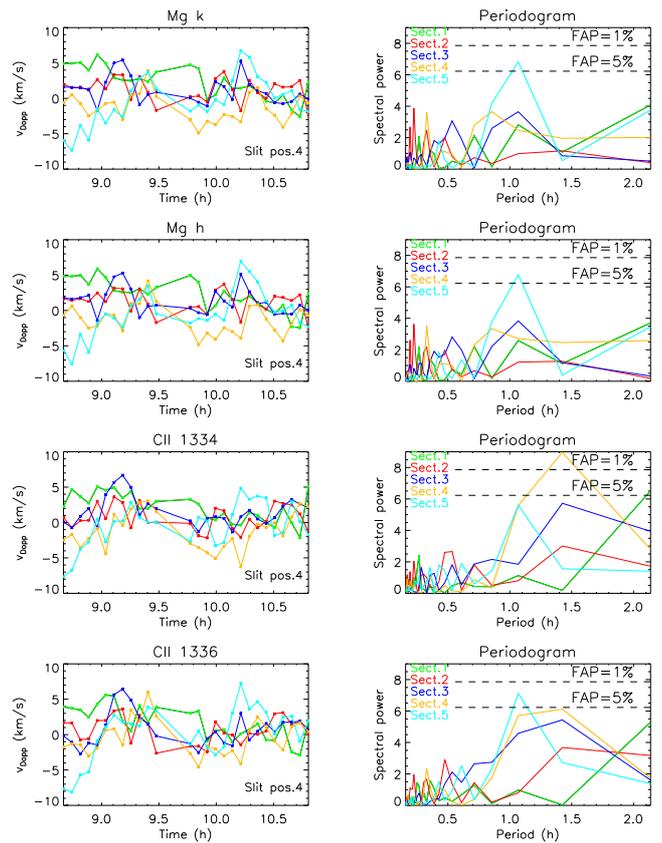}
\caption{Examples of Doppler velocity of \mg\ and \ca\ lines for fourth slit position  and selected sections (left panel) and corresponding Scargle periodograms 
(right panel). Dashed lines show 1\% and 5\% FAP levels. }
\label{f-odop}
\end{figure}

\section{Dynamics of the observed prominence}
        \label{s-dyn}

The width of one raster scan in a horizontal direction is about 5.6 arcsec, which gives a roughly 4\,000~km-wide strip 
(see the width between two white dashed vertical lines in Fig.~\ref{f-prom}). Five selected sections along the slit (marked with different colors in Fig.~\ref{f-prom}) 
cover only parts of the prominence structure where some dynamics are expected. 
For each selected section we get 27 points (rasters) marked with different colors. 
We excluded three rasters due to high contamination. 
Time variations of integrated intensities of the \mg, \ca,\ and \si\ lines for the fourth slit position and five selected sections along the slit
are shown in the left panel of Fig.~\ref{f-oint}. 
We clearly see that the \ca\ and \si\ lines have a few orders of magnitude lower integrated intensities than \mg\ lines, because \ca\ and \si\ lines are 
generally weak and noisy lines. The blended \ca\ 1336 line is brighter than the single \ca\ 1334 line, the \si\ 1394 line is the weakest, 
and the \mg\ k line is brighter than the \mg\ h line as reported, for example, by \citet{hei14}, \citet{sch14}, \citet{lev16} and \citet{liu15}. 

To check whether these time variations of integrated intensities are real or caused by noise, one has to compute correlation coefficients between the integrated intensities 
of all line combinations for five selected sections, which gives in total 432 points at a given section. 
The results are presented in Table~\ref{t-cor}. Correlation coefficients between both \mg\ lines in selected sections are close to 1 and for both \ca\ lines are around 
0.8. These values indicate that time variations of integrated intensities in the \mg\ and \ca\ lines are real and 
not caused by the noise. Our impression is that these time variations of integrated intensities in the \ca\ and \mg\ lines could be consistent with long period 
linear oscillations as reported by \citet{hei14b} and \citet{zap16}. A detailed analysis of the linear oscillations of the studied prominence is described in Sect.~\ref{s-osc}.

Correlation coefficients between the \mg\ and \ca\ lines are up to 0.7, between \si\ and \mg\ lines up to 0.4, and between \si\ and \ca\ lines up to 0.5. 
These values show that there is a weak correlation between different combinations of all these lines. 
These differences seem to be related to the fact that these lines are formed at different depths along the LOS in the prominence and that the \si\ line exhibits low \sn\ ratio.  
The \mg\ and \ca\ lines are formed in cool parts at a formation temperature of around 10\,000 - 20\,000~K. The \si\ 1394 line is formed in PCTR 
around 80\,000~K. 

Figure~\ref{f-it} shows the time variations of the central intensities for both \mg\ lines and the \ca\ 1334 line for the fourth slit position and five selected sections. 
A detailed analysis for all 432 points at a given section shows no correlation between integrated intensities and central intensities for a given line as presented in 
Table~\ref{t-cori}. These correlation coefficients with values between 0.3 and 0.6 indicate that the  \ca\ and \mg\ lines are optically thick. Therefore the central line
intensity is not proportional to integrated intensity.

We have computed also Doppler velocities of the \mg\ and \ca\ lines for all profiles 
(we omitted the \si\ line because it is too faint and noisy). Single unreversed 
profiles were fitted with a Gaussian while the reversed ones were fitted with the Gaussian using only their wings. The left panel of Fig.~\ref{f-odop} shows 
time variations of the LOS velocities of the \mg\ and \ca\ lines with values up to 10 km~s$^{-1}$.

\begin{figure}
\centering
\includegraphics[width=7cm]{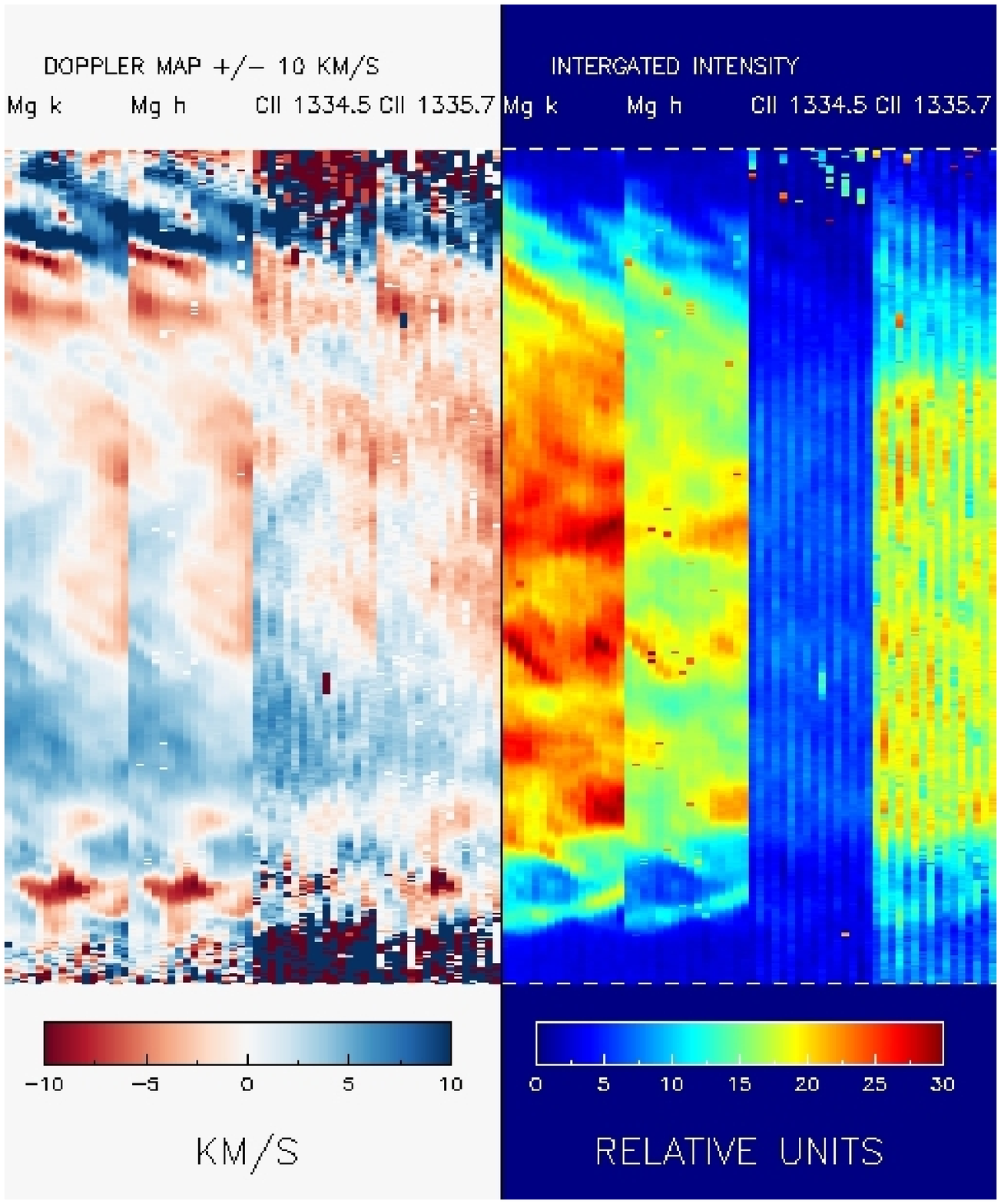}
\caption{Dopplergrams for \mg\ and \ca\ lines (left panel) together with 2D maps of integrated intensity (right panel). The movie corresponding to this 
figure is available as the online material of the journal.}
\label{f-map}
\end{figure}

\begin{figure}
\centering
\includegraphics[width=0.45\columnwidth]{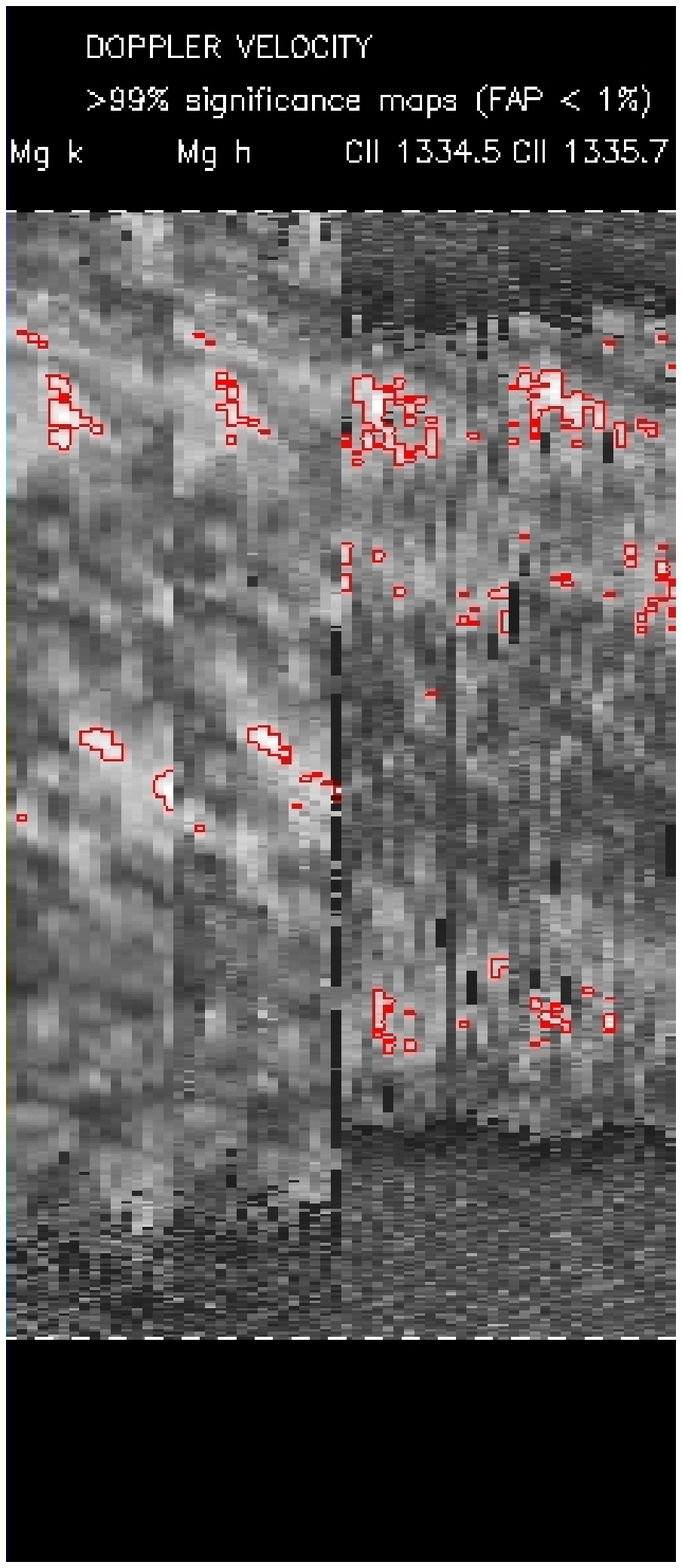}
\includegraphics[width=0.45\columnwidth]{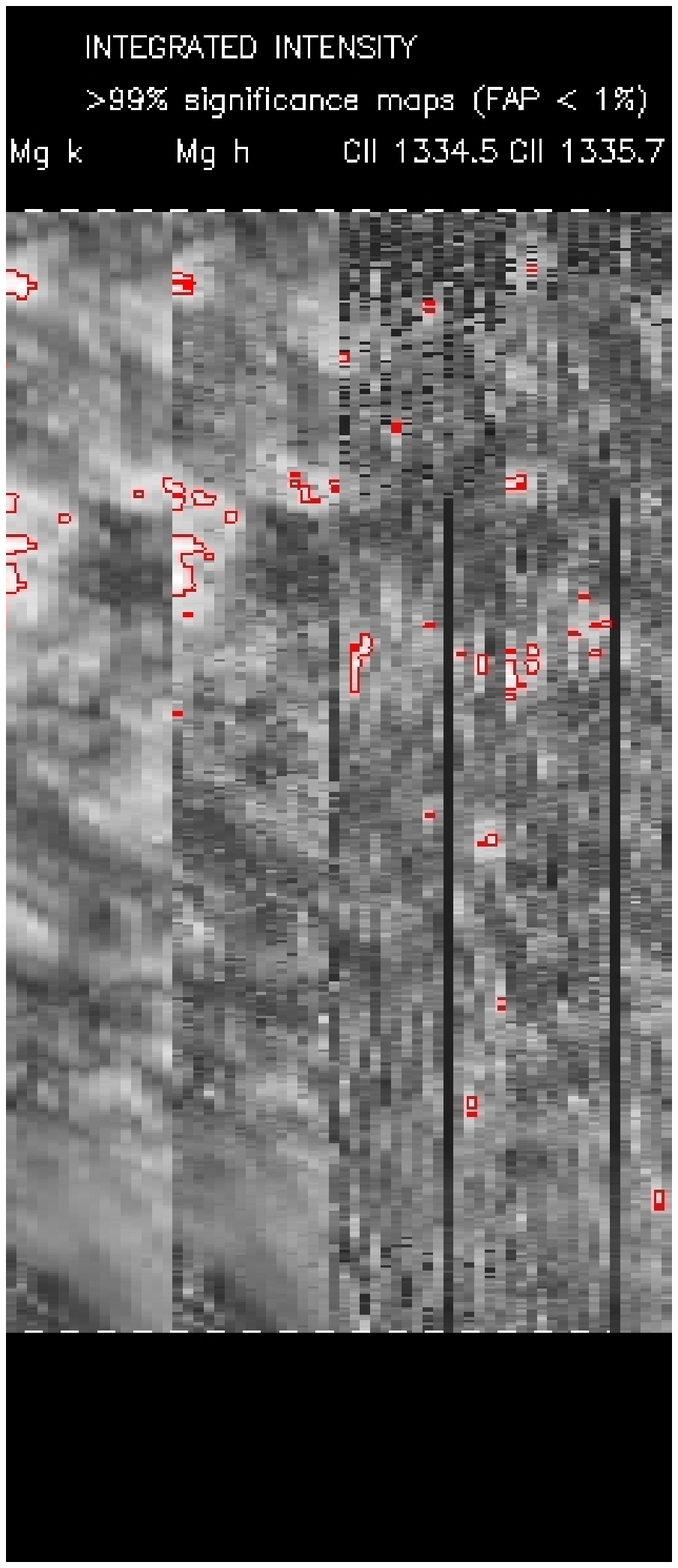}
\caption{Spectral power of the Doppler signal (left panel) and maximum intensity signal (right panel). Each rectangle corresponds to the spectral line indicated 
above. Red curves limit areas with FAP $<1$\%. }
\label{f-pow}
\end{figure}

\subsection{Prominence oscillations}
\label{s-osc}

From the spectra collected in the raster mode, we calculated the Doppler line shift for all slit positions averaged over ten pixels (see Sect.~\ref{s-ana}) 
for both \mg\ and \ca\ lines. We skip the weak and noisy \si\ line. 
The collected data series in raster mode allowed us to construct 2D Dopplergrams for the \mg \ and \ca\ lines as well as 2D maps of the integrated 
intensity (see Fig.~\ref{f-map}). They covered an area of 5.6 $\times$ 83 arcsec, which corresponds to 30 $\times$ 430 pixels (see in Fig.~\ref{f-prom} 
the box defined by two short, white, solid  horizontal lines and two white, vertical, dashed lines). 
From a series of individual snapshots we created an online movie corresponding to Fig.~\ref{f-map}. 
With a time evolution of Doppler velocity and integrated intensity we performed a Scargle periodogram analysis \citep{sca82} for each pixel 
independently. 
We used the Scargle periodogram because it allows us to analyze a dataset with gaps, which was our case.
An example of the time evolution of Doppler velocity and integrated intensity together with a Scargle periodogram for selected pixels are presented in the 
Figs.~\ref{f-oint} and~\ref{f-odop}.

For each pixel we took the value of the maximum peak in the periodogram and we constructed 2D maps of spectral power for the \mg\ and \ca\ lines 
(see Fig.~\ref{f-pow}). 
In the map we selected areas where the false alarm probability (FAP) level is lower than 1\%. A FAP equal to 1\% may be interpreted as a 1\% probability 
that a periodogram peak at this level is caused by noise. This value may be interpreted also as a 99\% confidence level. 

Looking  at plots in Figs.~\ref{f-oint} and~\ref{f-odop} one can have an impression of the existence of oscillations in the dataset, but a periodogram 
analysis reveals no oscillations. Areas with a FAP $<$ 1\% are very small. A FAP level below 1\% located in isolated pixels is not sufficient to expect 
any oscillations in the dataset \citep{zap15}.
Looking at the online movie corresponding to Fig.~\ref{f-prom}, we detected plasma elements with different Doppler velocity and intensity crossing 
the FOV where the FAP $<$ 1\%. This causes a change of the observed spectral parameter, which mimics one period of oscillations 
(see for example the plot in Fig.~\ref{f-odop} for section 5 of the \mg\ k line) and 
gives a moderate rise of spectral power.

A passage of plasma elements across the observed FOV and its influence on the dataset needs to be noted. One can claim that 
using our technique it is impossible to detect any periodic signal in a prominence that is dynamic with plasma elements in constant motions. 
For example, if a thread oscillates in the plane of sky (POS), then when analyzing a fixed pixel in the FOV we take the signal from different plasma elements. 
However, even for this situation the periodic signal will be present at least in the intensity domain as an effect of the consecutive presence and 
absence of a particular thread in the analyzed pixel. For a very dense field, any detection of a periodic signal in individual threads may be of 
course impossible, but the global oscillatory mode in the Doppler domain may be still detected as a coherent signal present in the whole FOV. 
We follow the methodology described in \citet{ter02}. Performing an independent periodogram analysis for each pixel for as long as possible 
favors the detection of oscillations.
\citet{ter02} noted that if the detected oscillatory area is large (more than several pixels) this is a hint of 
the solar origin of the oscillations. If the observed prominence were a subject of global oscillations, we would detect an oscillatory pattern over 
the whole prominence body.
Taking this into account we conclude that we did not detect any global oscillations in the observed prominence either in the Doppler shift 
or in the integrated intensity. We did not detect any oscillations of a local character either. However, having in mind previous 
considerations, we cannot exclude the presence of undetectable local oscillations. 

\begin{figure}    
\centerline{\hspace*{0.015\textwidth}
            \includegraphics[width=7cm]{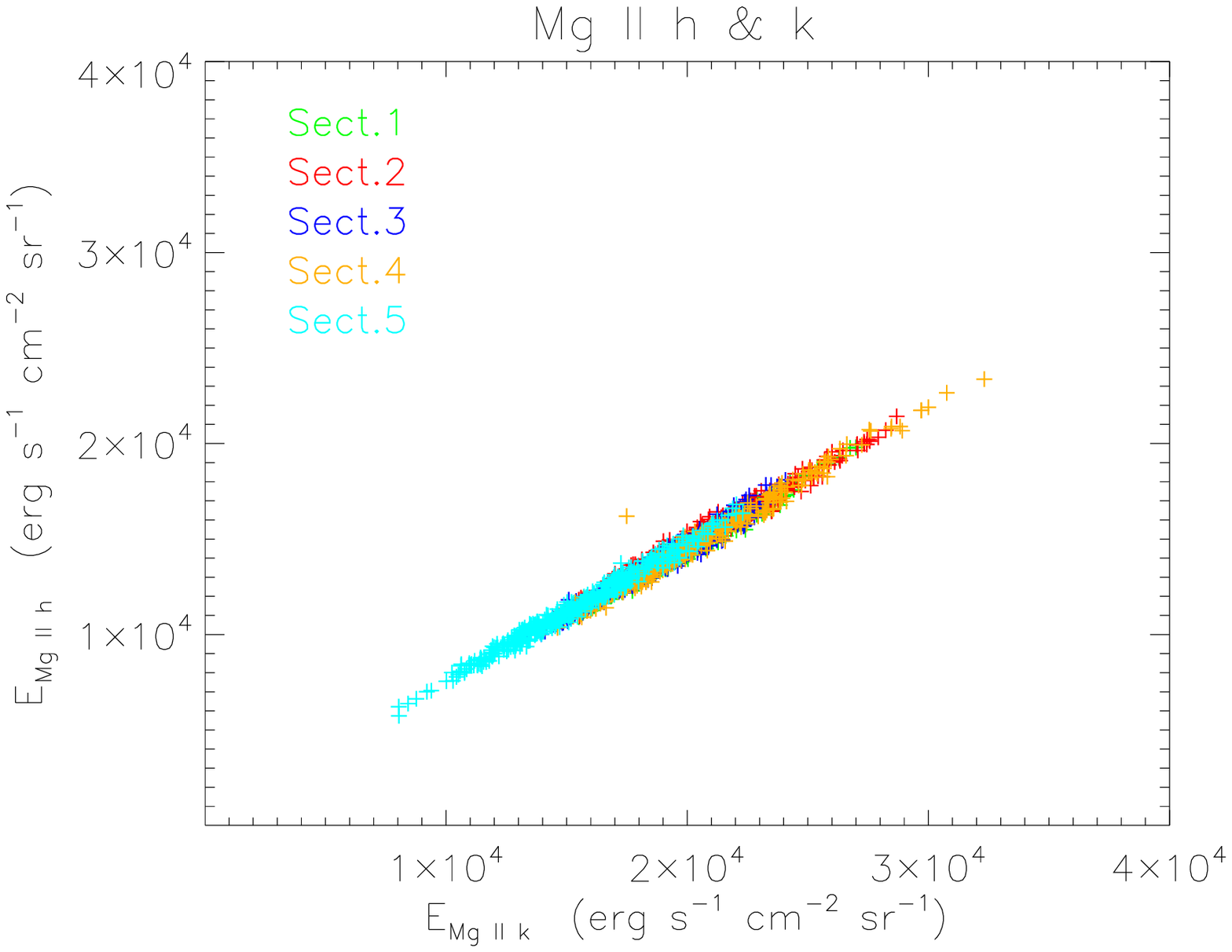}
            \hspace*{-0.02\textwidth}
            }
\vspace{0.01\textwidth}
\centerline{\hspace*{0.015\textwidth}
            \includegraphics[width=7cm]{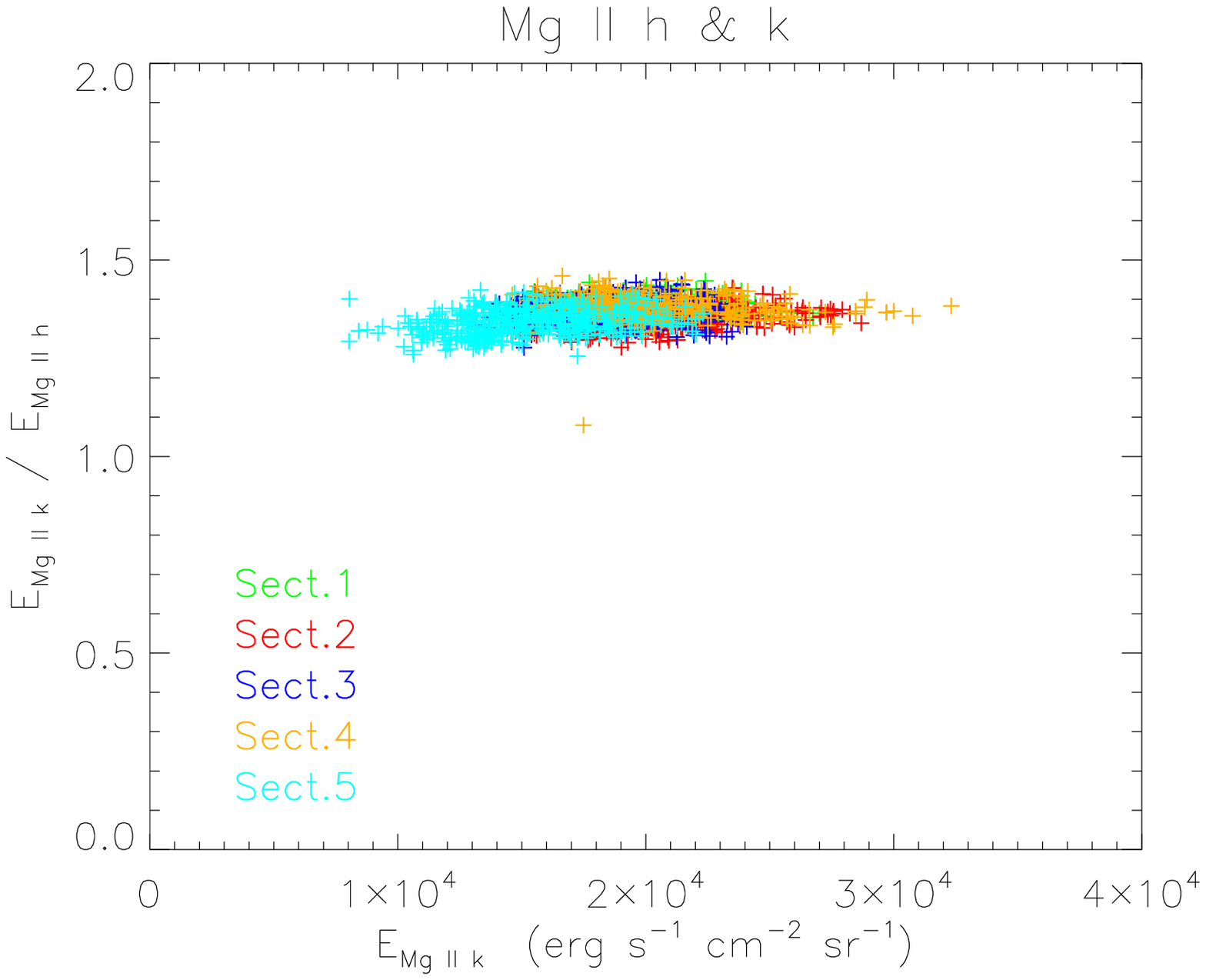}
            \hspace*{-0.03\textwidth}
            }
\caption{Comparison between  integrated intensity emitted in the \mg\ h line~and  integrated intensity emitted in the \mg\ k line 
for five selected sections (upper panel)  and ratio of integrated intensities between \mg\ k and \mg\ h lines as function of integrated intensity 
emitted in \mg\ k line (lower panel).}
\label{f-spmg}
\end{figure}

\begin{figure} 
\centering
\includegraphics[width=7cm]{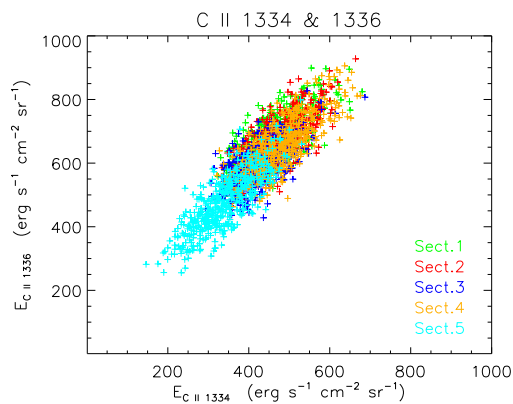}
\caption{Comparison between  integrated intensity emitted in \ca\ 1336 line and integrated intensity 
emitted in \ca\ 1334 line for five selected sections.}
\label{f-spc2}
\end{figure}

\begin{figure*} [h]   
\centerline{\includegraphics[width=0.35\textwidth,clip=]{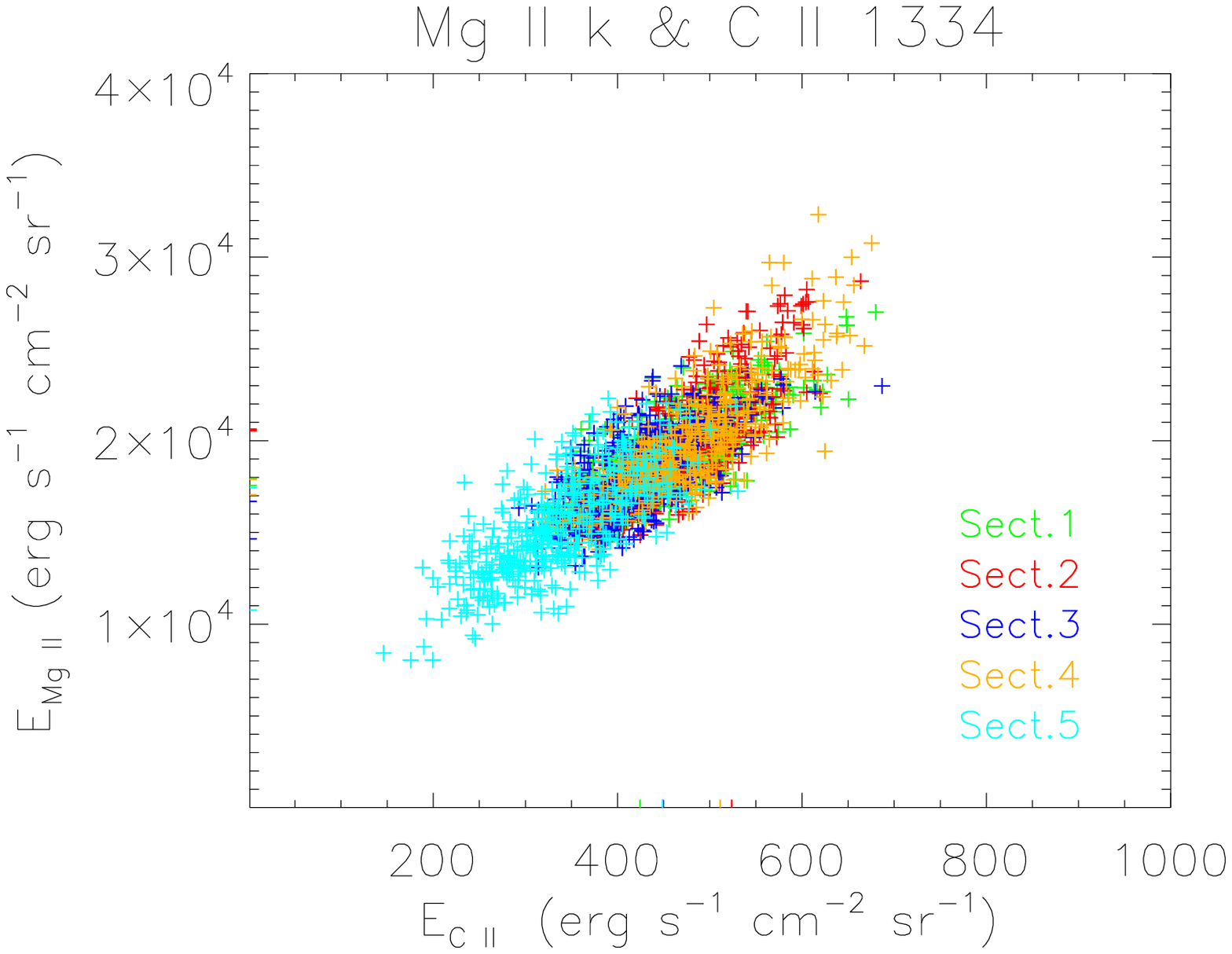}
            \hspace*{-0.02\textwidth}
            \includegraphics[width=0.35\textwidth,clip=]{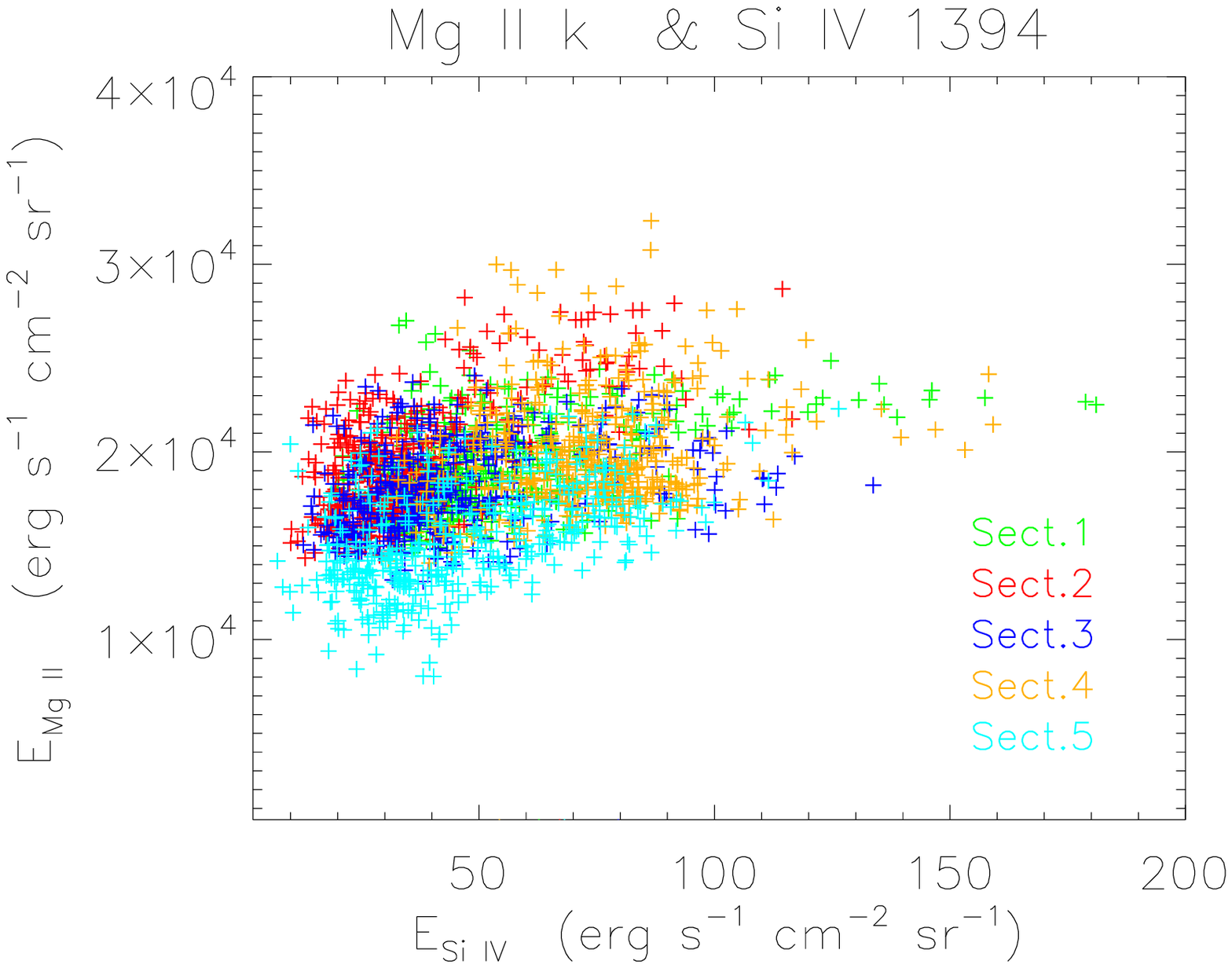}
            \hspace*{-0.02\textwidth}
            \includegraphics[width=0.35\textwidth,clip=]{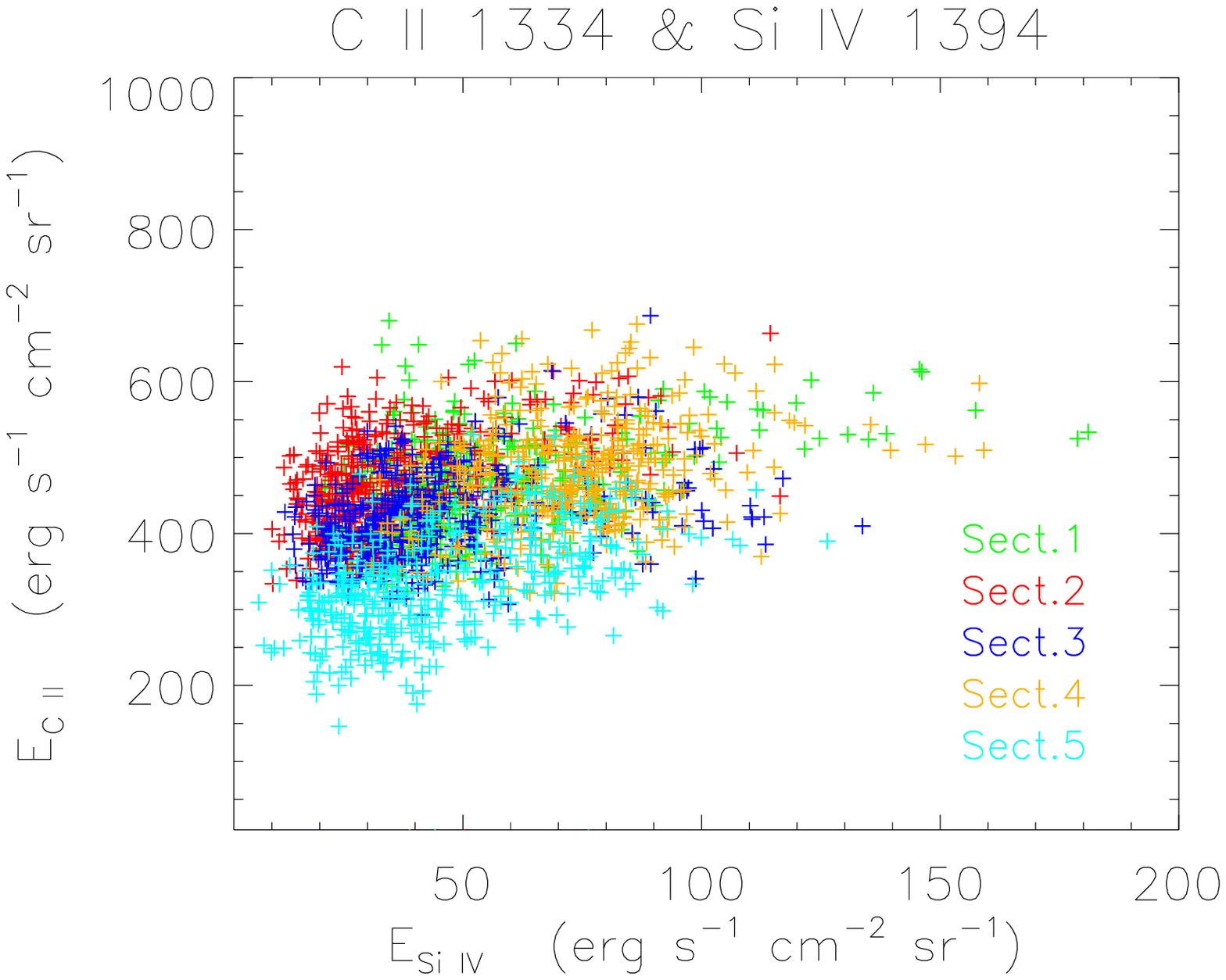}
            }
 \caption{Comparison between integrated intensities of \mg\ k and \ca\ 1334 lines (left panel), \mg\ k line and \si\ line (middle 
panel), and \ca\ 1334 line and \si\ line (right panel) for five selected sections. }
\label{f-spmix}
\end{figure*}

\begin{figure*} [h]   
\centerline{\includegraphics[width=0.35\textwidth,clip=]{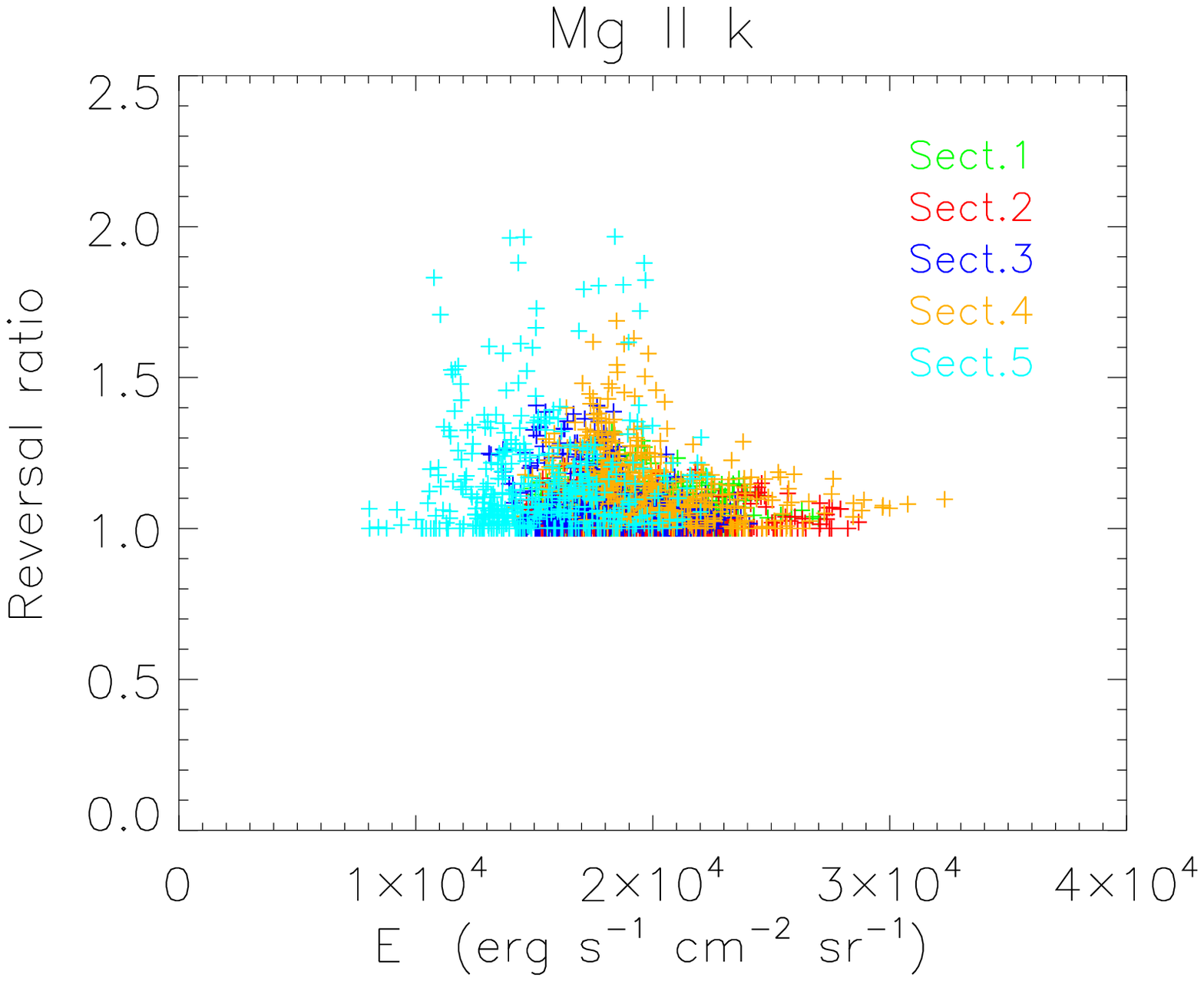}
            \hspace*{-0.025\textwidth}
            \includegraphics[width=0.35\textwidth,clip=]{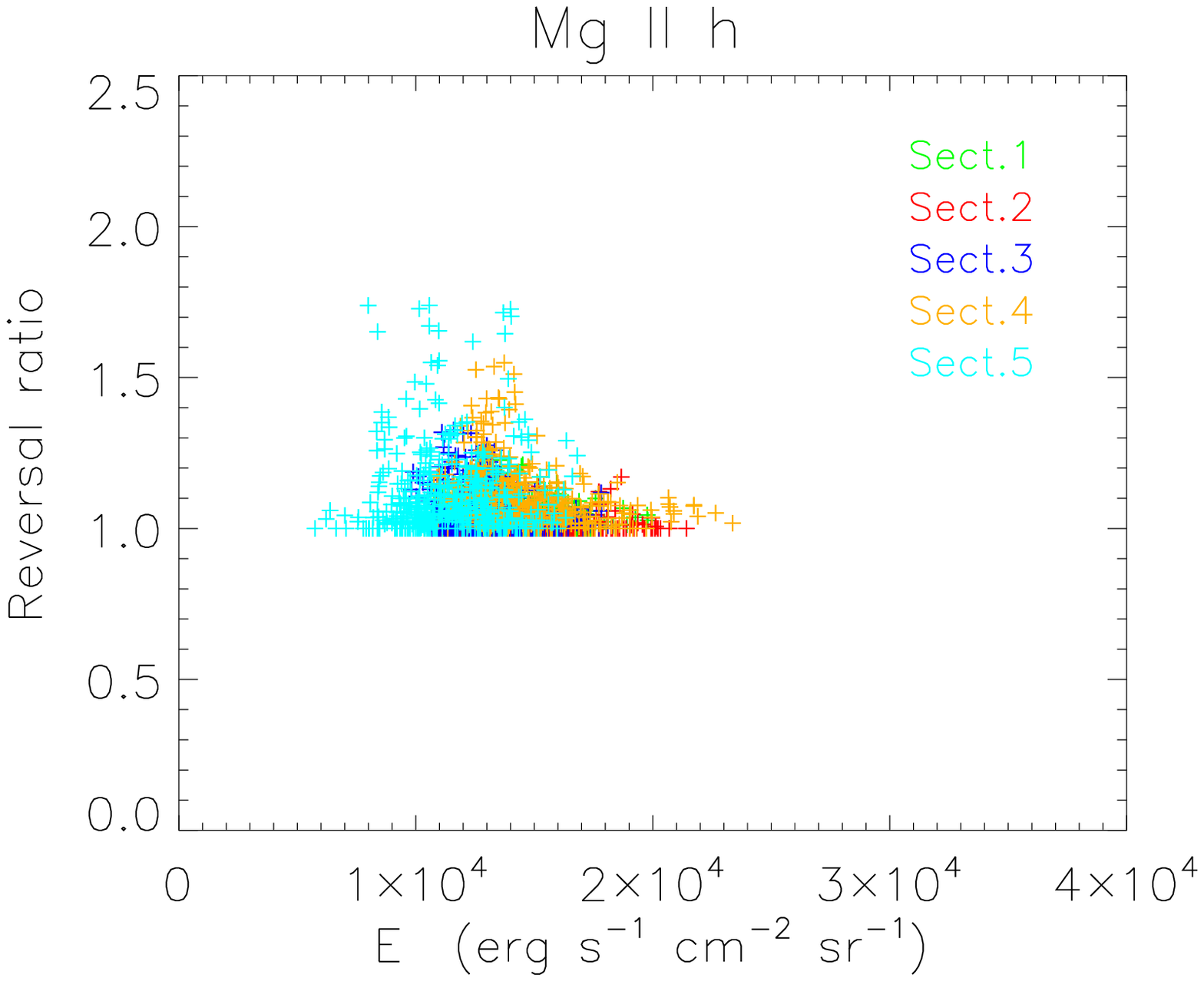}
            \hspace*{-0.025\textwidth}
            \includegraphics[width=0.35\textwidth,clip=]{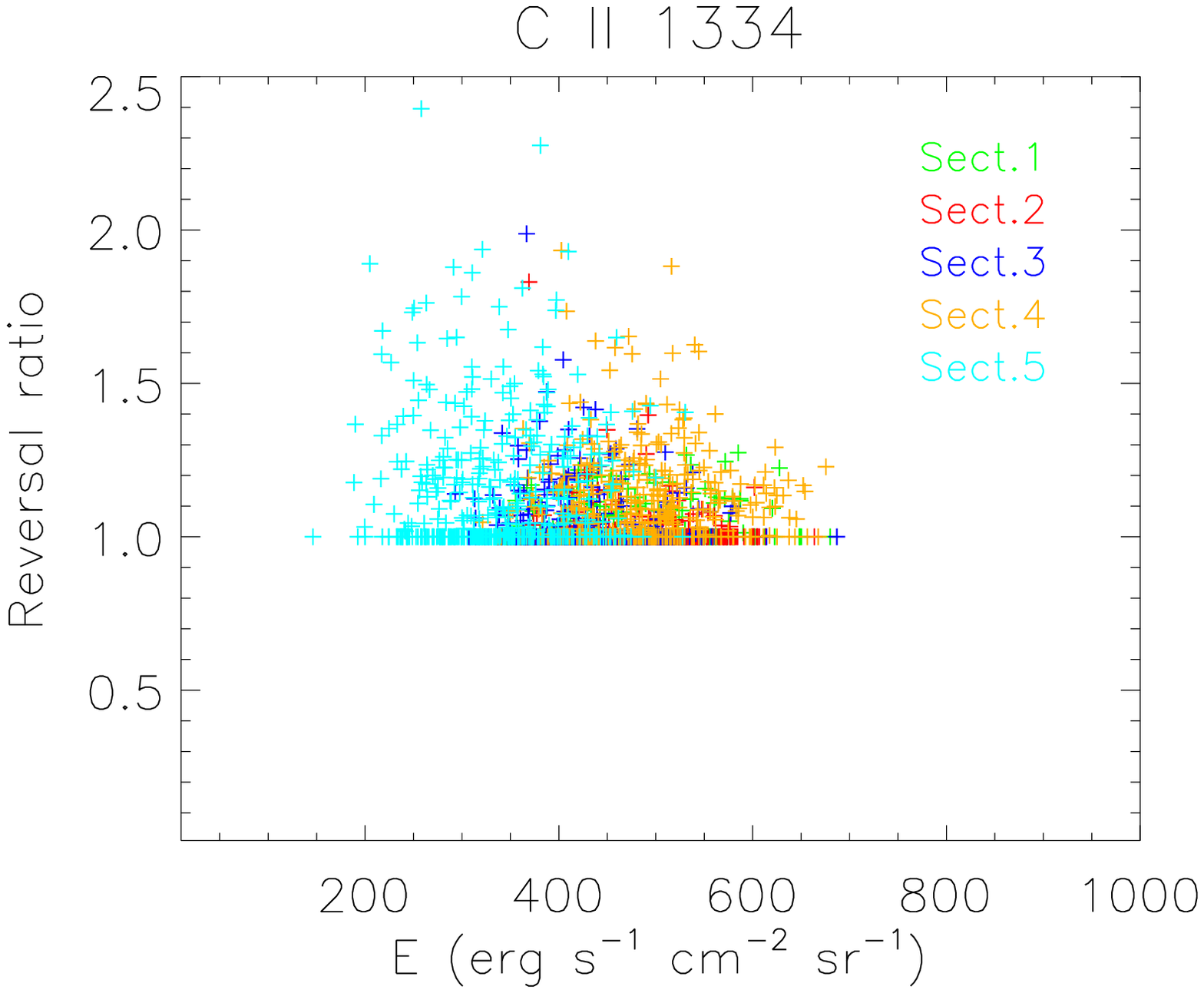}
            }
\caption{Comparison between reversal ratio and integrated intensity of \mg\ k line (left panel), \mg\ h line (middle panel), and 
\ca\ 1334 line (right panel) for five selected sections.}
\label{f-rre}
\end{figure*}

\begin{figure} 
\centering
\includegraphics[width=7cm]{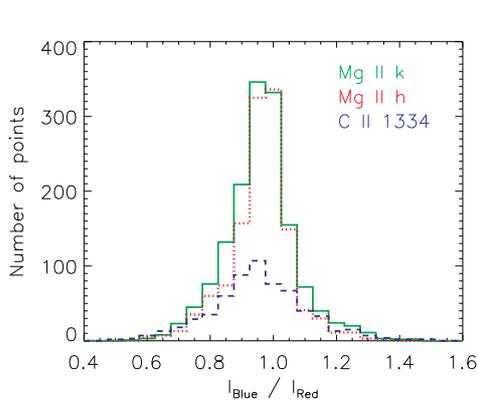}
\caption{Blue to red peak intensity distribution of reversed profiles for \mg\ and \ca\ 1334 lines.}
\label{f-br}
\end{figure}

\begin{figure} 
\centerline{\includegraphics[width=0.26\textwidth,clip=]{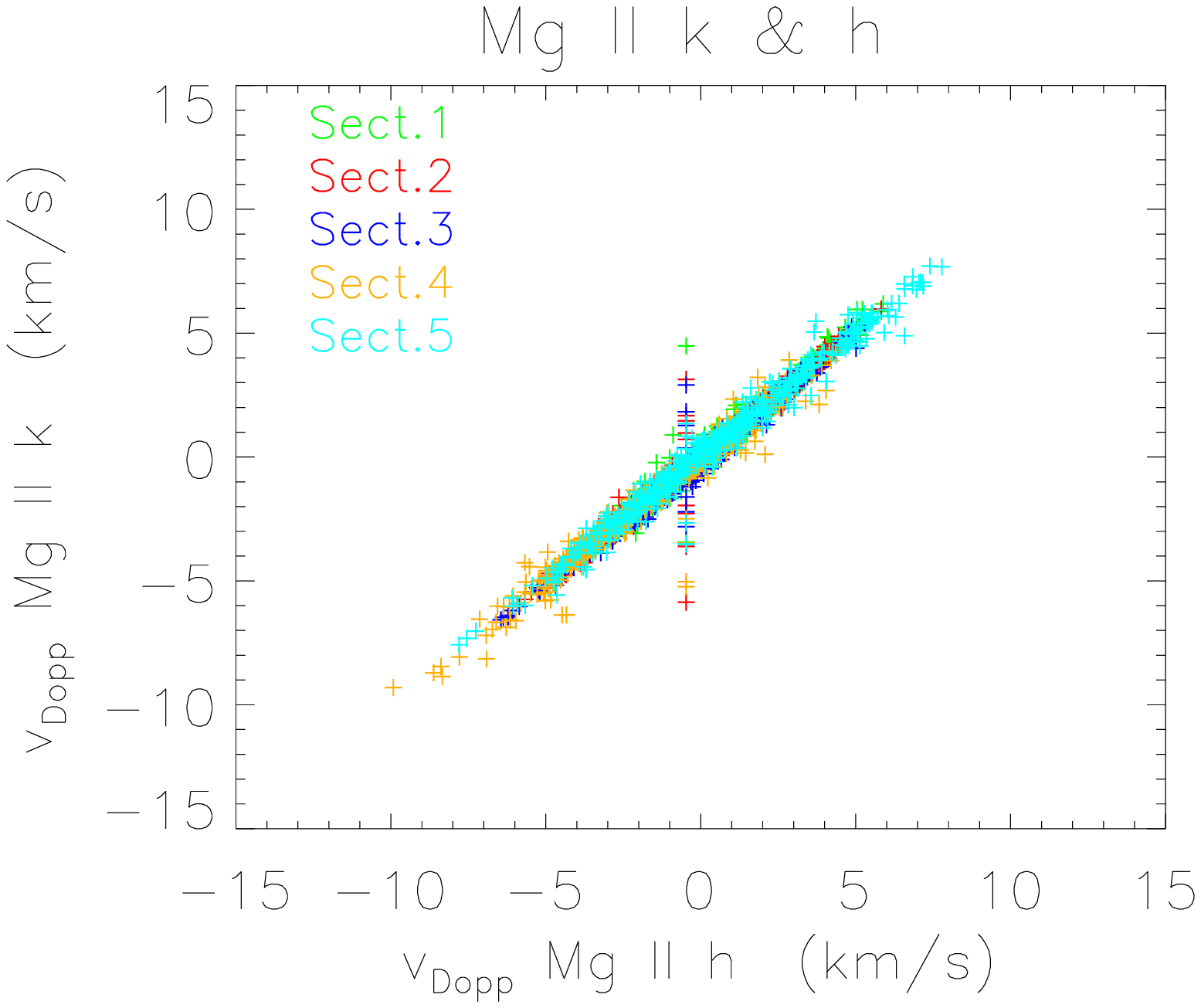}
            \hspace*{-0.02\textwidth}
            \includegraphics[width=0.26\textwidth,clip=]{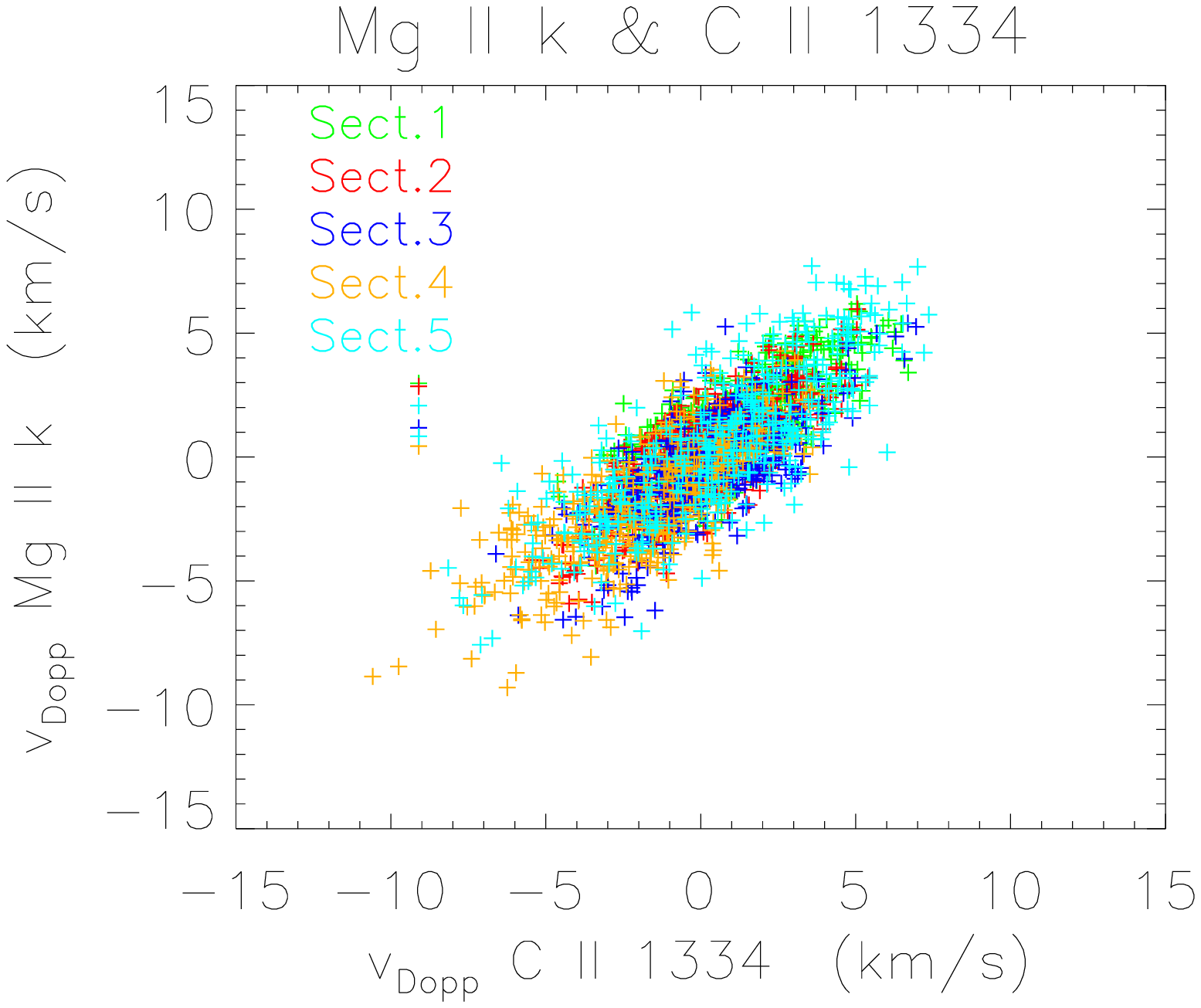}
            }
\caption{Comparison between Doppler velocities of \mg\ lines (left panel) and \mg\ k and \ca\ 1334 lines (right panel) for 
five selected sections.}
\label{f-dv}
\end{figure}

\begin{figure}    
\centerline{\includegraphics[width=0.26\textwidth,clip=]{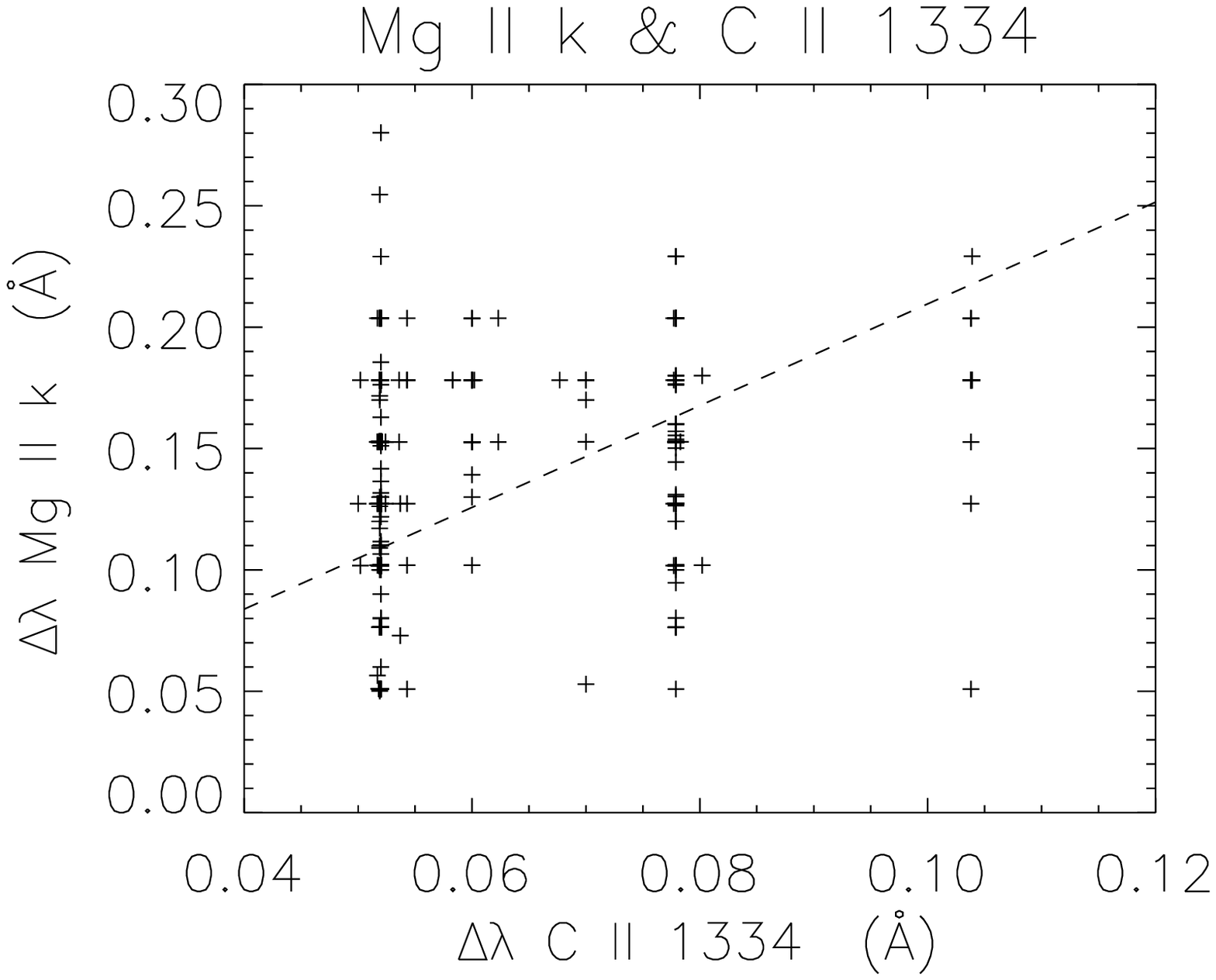}
            \hspace*{-0.02\textwidth}
            \includegraphics[width=0.26\textwidth,clip=]{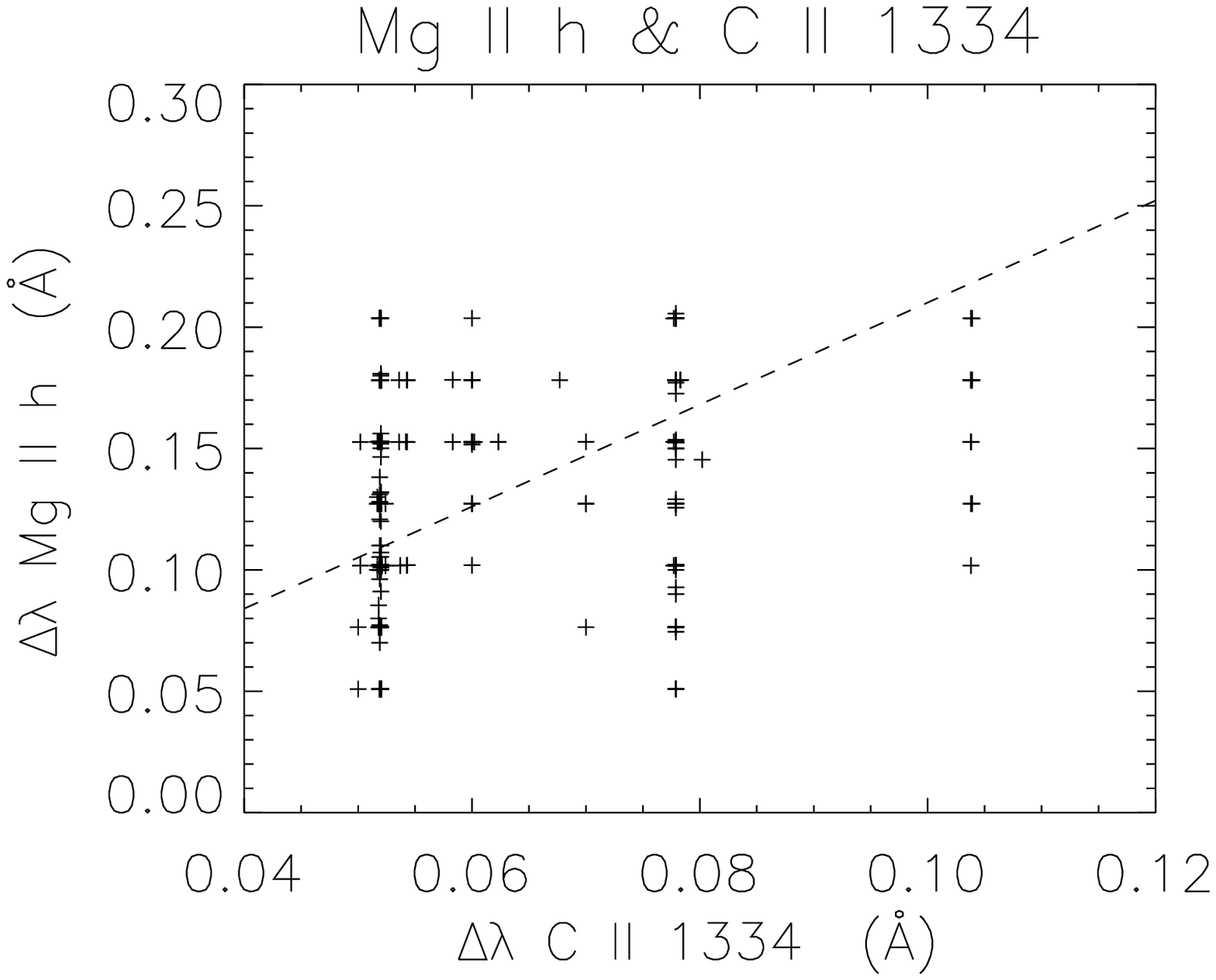}
            }
\caption{Peaks separation of \mg\ k line versus \ca\ 1334 line (left panel) and  \mg\ h line versus \ca\ 1334 line
(right panel). The dashed straight line shows the theoretical relation between the lines.}
\label{f-dop}
\end{figure} 

\section{Statistical analysis of IRIS UV line intensities}
          \label{s-int}

Here we present the behavior of observed integrated intensities of the \mg, \ca, and \si\ lines for five sections, 16 slit positions, and 27 rasters, 
which give in total 2160 points. A scatter plot between both integrated intensities of the \mg\ lines in the upper panel of Fig.~\ref{f-spmg} shows that observed points 
practically lie on a straight line. This result is sufficiently  robust due to high \sn\ ratio. Integrated intensities of the \mg\ k line 
vary between 8\,000 and 32\,500~\cgs, while in the \mg\ h line they vary between 5\,500 and 23\,500~\cgs. The lower panel in Fig.~\ref{f-spmg} shows the ratio between the \mg\ k and h 
integrated intensities as a function of the integrated intensity of the \mg\ k line. 
All observed points show a uniform distribution with the average value of 1.365. This value is consistent with the theoretical values presented in
\citet{hei14}. 
Figure~\ref{f-spc2} shows the scatter plot between integrated intensities of the \ca\ 1336 line and \ca\ 1334 line. The scattering of the values in this plot is mainly 
due to the noise because \ca\ lines are weak. Integrated intensities of the \ca\ 1334 line vary between 150 and 700~\cgs, while blended \ca\ 1336 line vary between 250 
and 950~\cgs.
Figure~\ref{f-spmix} shows scatter plots between the \mg\ k and \ca\ 1334 line, the \mg\ k and \si\ line, as well as the \ca\ 1334 and \si\ line. In the 
left panel the scattering is less pronounced than in the middle and right panels indicating that the \si\ line is formed under different conditions; we will
discuss this in Section 5.2. Integrated intensities of the \si\ line vary between 10 and 180~\cgs.
Scatter plots between the reversal ratio and  integrated intensity of both \mg\ lines and the \ca\ 1334 line are presented in Fig.~\ref{f-rre}. By 
our definition, the reversal ratio is equal to one for single-peak and flat-core profiles, while the reversed profiles have a reversal ratio higher than one.
Statistics show that about 80\% of \mg\ k profiles, 65\% of \mg\ h profiles, and 40\% of \ca\ 1334 profiles are reversed. 
The rest show single-peak or flat characteristics. 
Here we clearly see that the scatter points form a cloud above above that of the limit one. We believe that this cloud is formed due to different 
plasma conditions in optically thick structures along the LOS, such as the temperature, velocity, and gas pressure. In the following subsection we compare synthetic profiles
obtained from 1D non-LTE isothermal-isobaric models with the observations of \mg\ lines to assess our assumption.
Figure~\ref{f-br} shows the histograms of the ratio of blue to red peak intensities of all reversed profiles for both \mg\ lines and the \ca\ 1334 line. The red peak  
is slightly dominant for all lines but the histograms look quite symmetrical. This might confirm our conclusions from the previous section that the observed prominence
does not exhibit any large-scale systematic motions like oscillations or waves, which could produce significant line asymmetry. However, random fine-structure
motions could be responsible for the shape of these histograms as we discuss later. 
In Fig.~\ref{f-dv} we present the correlation between LOS velocities of both the \mg\ lines and  \mg\ k and \ca\ 1334 lines. The plots are similar to 
those in the upper panel of Fig.~\ref{f-spmg} and the left panel of Fig.~\ref{f-spmix}, again showing that correlation between \mg\ lines is better than 
between \mg\ and \ca\ lines because \mg\ lines are less noisy than \ca\ lines. However, the relatively larger scatter between \mg\ and \ca\ lines could be 
due to the fact that the wings of both lines are not formed at the same depth. 
Finally, Fig.~\ref{f-dop} shows the scatter plot between the peaks separation of the \mg\ k or h line, against those for the \ca\ 1334 line. 
This was aimed as a test of the idea that the two peaks might be due to Doppler shifts of the single-peaked profiles. In that case, the peaks separation
would correlate in the \mg\ and \ca\ 1334 lines and the correlation coefficient (dashed line) would be equal to a ratio of 2.1 of \mg\ and \ca\ 1334 line wavelengths.
However, we clearly see that this is not the case and thus we believe that the majority of double-peak profiles are due to true reversal caused by the opacity
effects, which do not obey the above correlation rule.


\subsection{One-dimensional non-LTE isothermal-isobaric models}
           \label{s-mod1}

\cite{hei14} considered 27 isothermal-isobaric models and synthesized the \mg\ lines using the non-LTE code for 1D vertical slabs illuminated by 
the solar disk radiation. Here we compute a larger grid of such models in order to compare the models with the observations. The input parameters of each model 
are the temperature $T$, gas pressure $p$, effective thickness of the slab $D$, microturbulent velocity $\xi,$ and height above the solar surface $H$. 
We neglect the radial flow velocity $v_{\rm rad}$ because we assume static models of a quiescent prominence. The synthetic \mg\ profiles are convolved 
with the IRIS instrumental profile having the Gaussian full width at half maximum (FWHM) equal to 52 m\AA~\citep{hei15}. 
The microturbulence is treated here in a  standard way, 
representing the velocity distribution of uresolved plasma elements. 
Additional line broadening can be introduced by randomly distributed LOS velocities of fine structures integrated along a given line of sight. 
The effects of such random LOS velocities were considered in the frame of 2D prominence models by \citet{gun08}. These authors showed, for example, that 
the LOS motions of fine structures can produce line asymmetries that are qualitatively comparable with observations.

\begin{table} \small  
\centering
\caption{Large grid of input parameters used for 1D-slab isothermal-isobaric models. All combinations gives in total 343 models for each value of 
microturbulent velocity.}
\label{t-md}
\begin{tabular}{cccccc}
\hline
\hline
{\it T} &{\it p} & {\it D} & $\xi$ & $v_{\rm rad}$ & {\it H} \\
(K) & (dyn~cm$^{-2}$) & (km) & (km~s$^{-1}$) & (km~s$^{-1}$) & (km) \\
\hline
5\,000 & 0.01 & 200 & & &  \\ 
6\,000 & 0.02 & 500 & & & \\
8\,000 & 0.05 & 1\,000 & 0& & \\
10\,000 & 0.1 & 2\,000 & 5 & 0 & 10\,000 \\
12\,000 & 0.2 & 5\,000 & 8& & \\
15\,000 & 0.5 & 10\,000 & & & \\
20\,000 & 1 & 20\,000 & & & \\
\hline
\end{tabular}
\end{table}

We fixed the height of the prominence slab to 10\,000~km to be consistent with \cite{hei14}.
The set of input parameters of the grid of models includes seven characteristic values of temperatures, gas pressures, and effective thicknesses, which gives altogether 
343 models for a given microturbulent velocity as is shown in Table~\ref{t-md}.
From the 1D non-LTE model we obtain the electron density $n_{\rm e}$, the integrated intensity, reversal ratio, and optical 
thickness at the center $\tau$ of each \mg\ line.
All model plots are made in the range of observed integrated intensities of the \mg\ lines up to 5~$\times$~10$^4$~\cgs\ with 201 models for a given microturbulent velocity, 
for seven different temperatures marked with different colors. In the following plots, the left panels show the \mg\ k line at microturbulent velocities 0, 5, 
and 8~km~s$^{-1}$ and the right panels show the \mg\ h line at the same microturbulent velocities.

\begin{figure}    
\centerline{\includegraphics[width=0.27\textwidth,clip=]{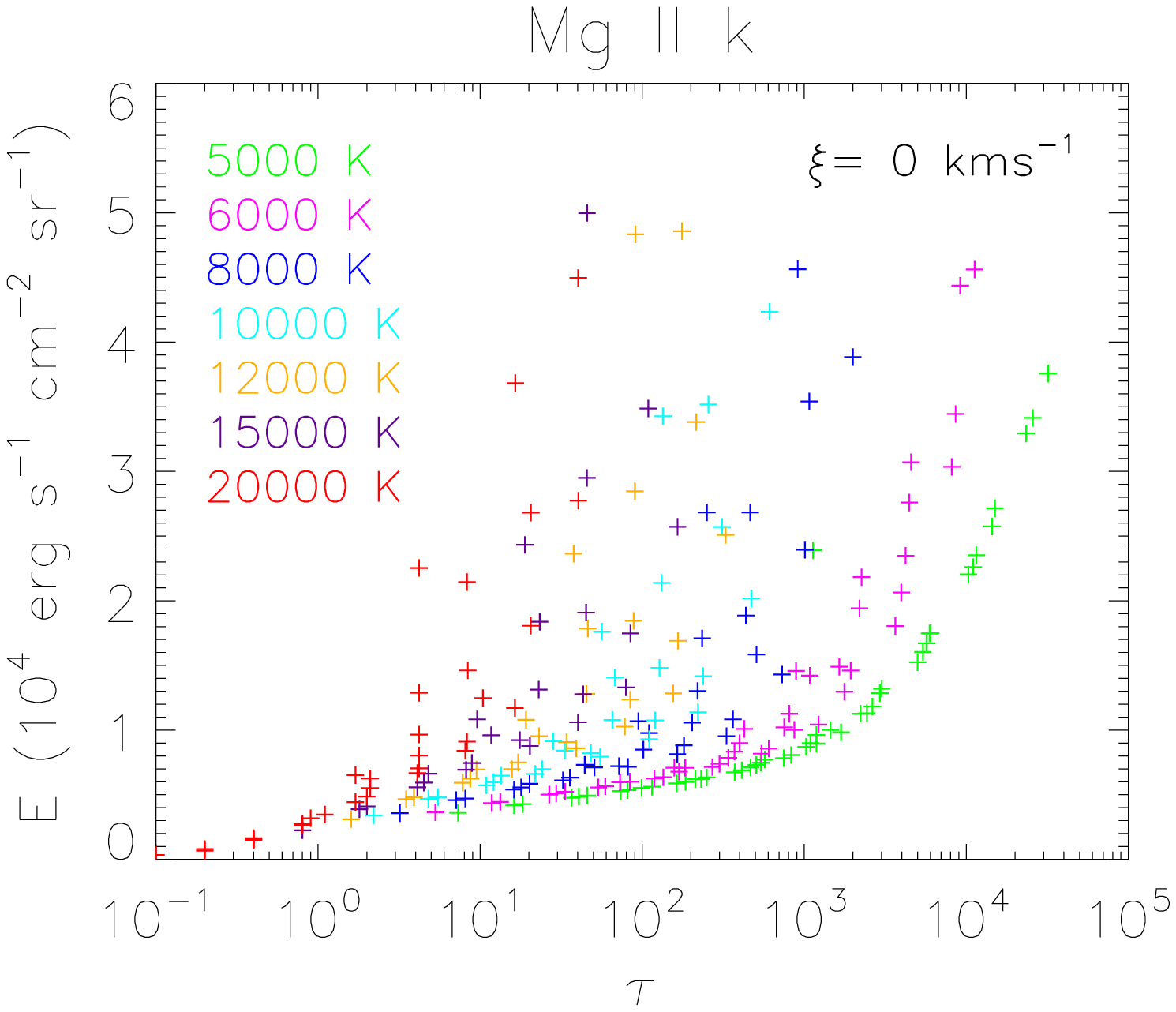}
            \hspace*{-0.03\textwidth}
            \includegraphics[width=0.27\textwidth,clip=]{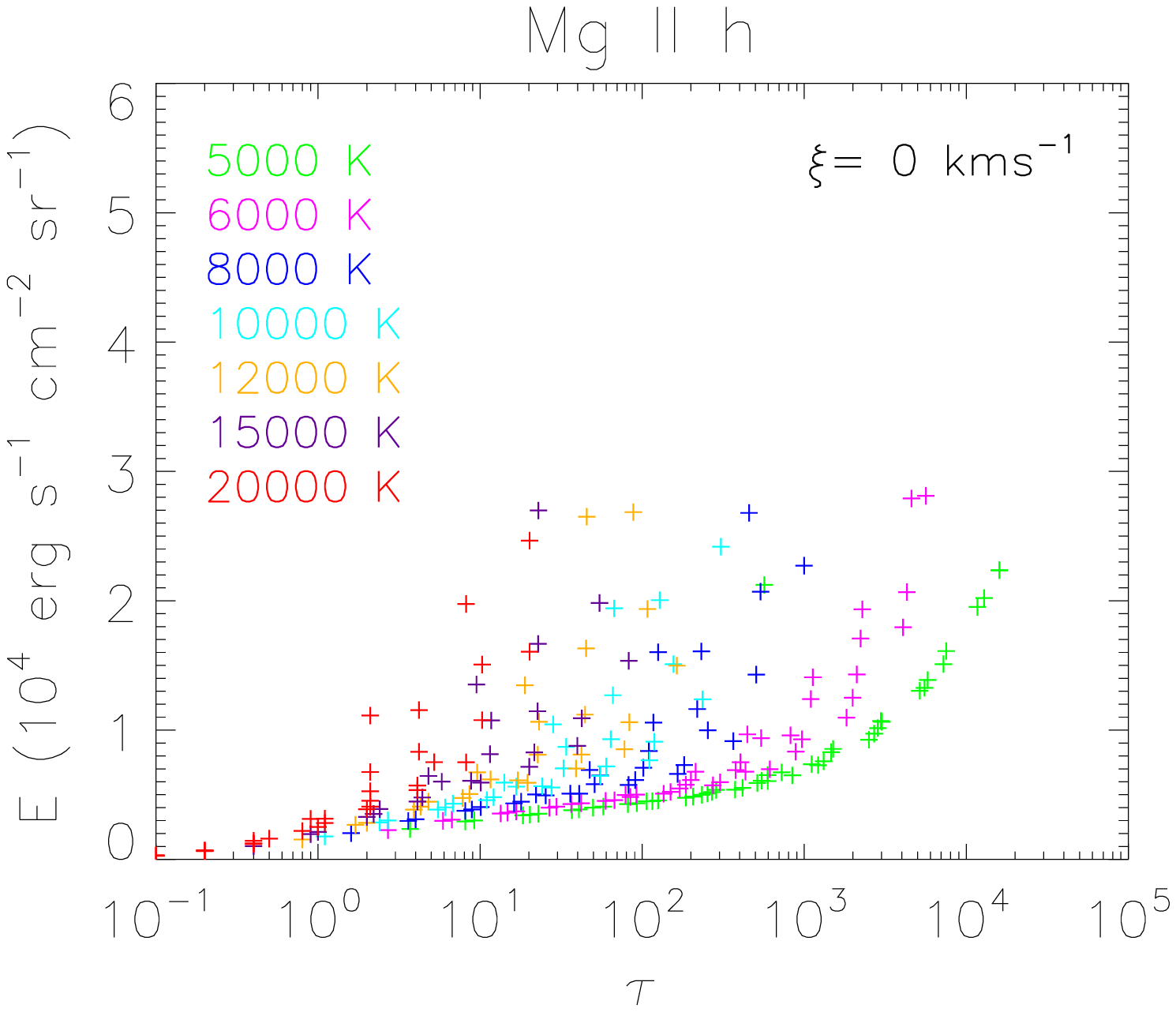}
            }
\vspace{0.01\textwidth}
\centerline{\includegraphics[width=0.27\textwidth,clip=]{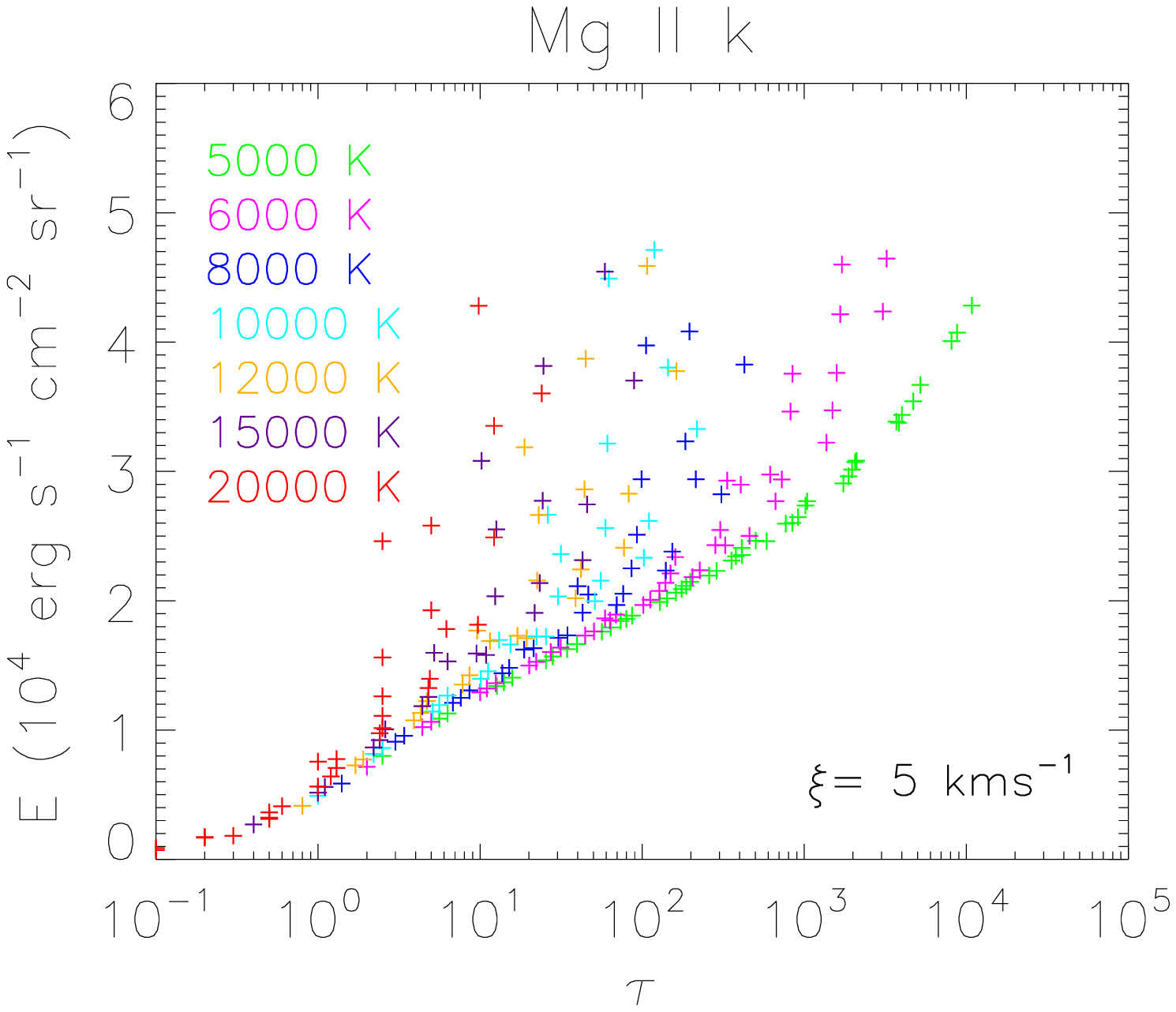}
            \hspace*{-0.03\textwidth}
            \includegraphics[width=0.27\textwidth,clip=]{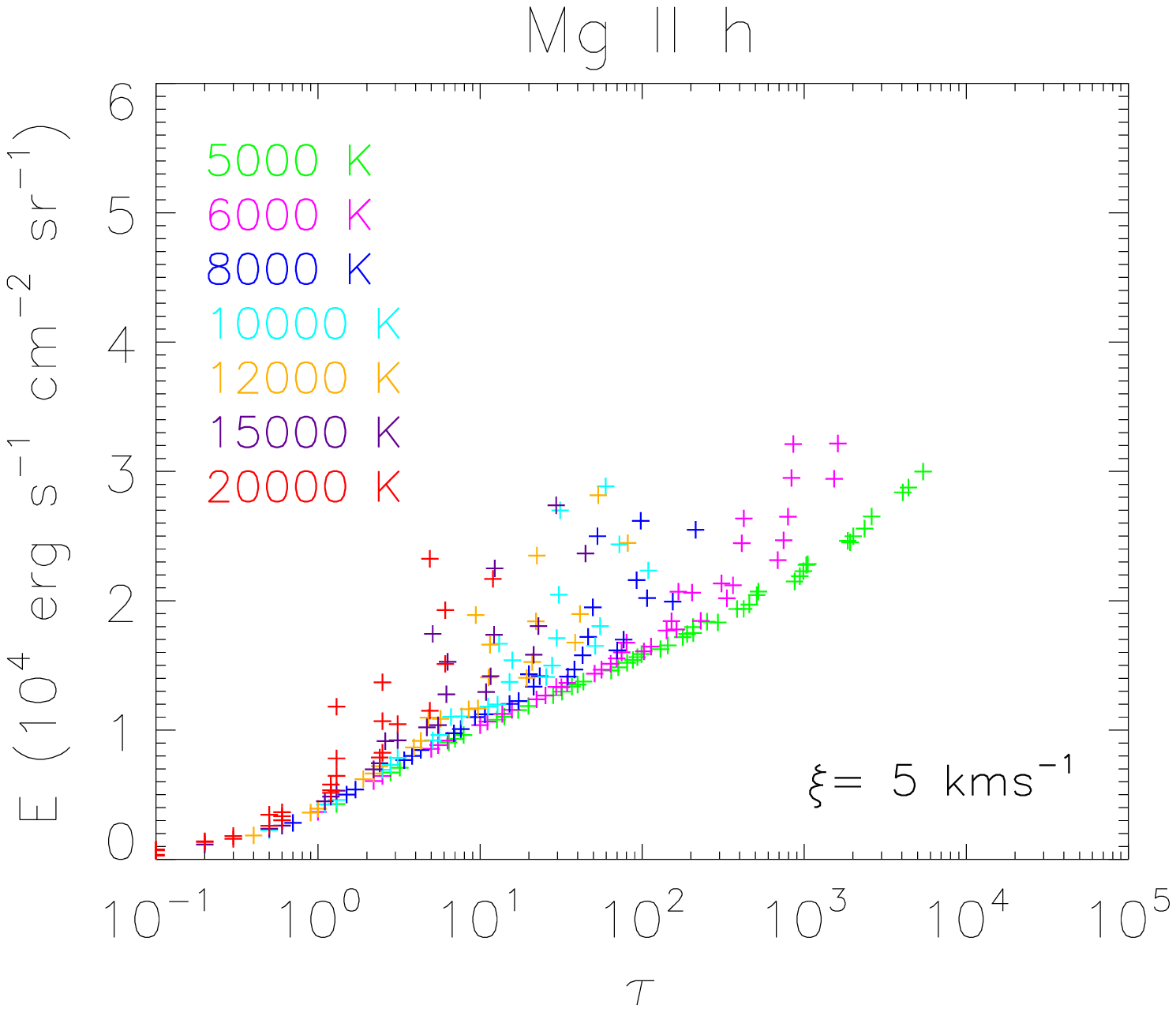}
            }
\vspace{0.01\textwidth}
\centerline{\includegraphics[width=0.27\textwidth,clip=]{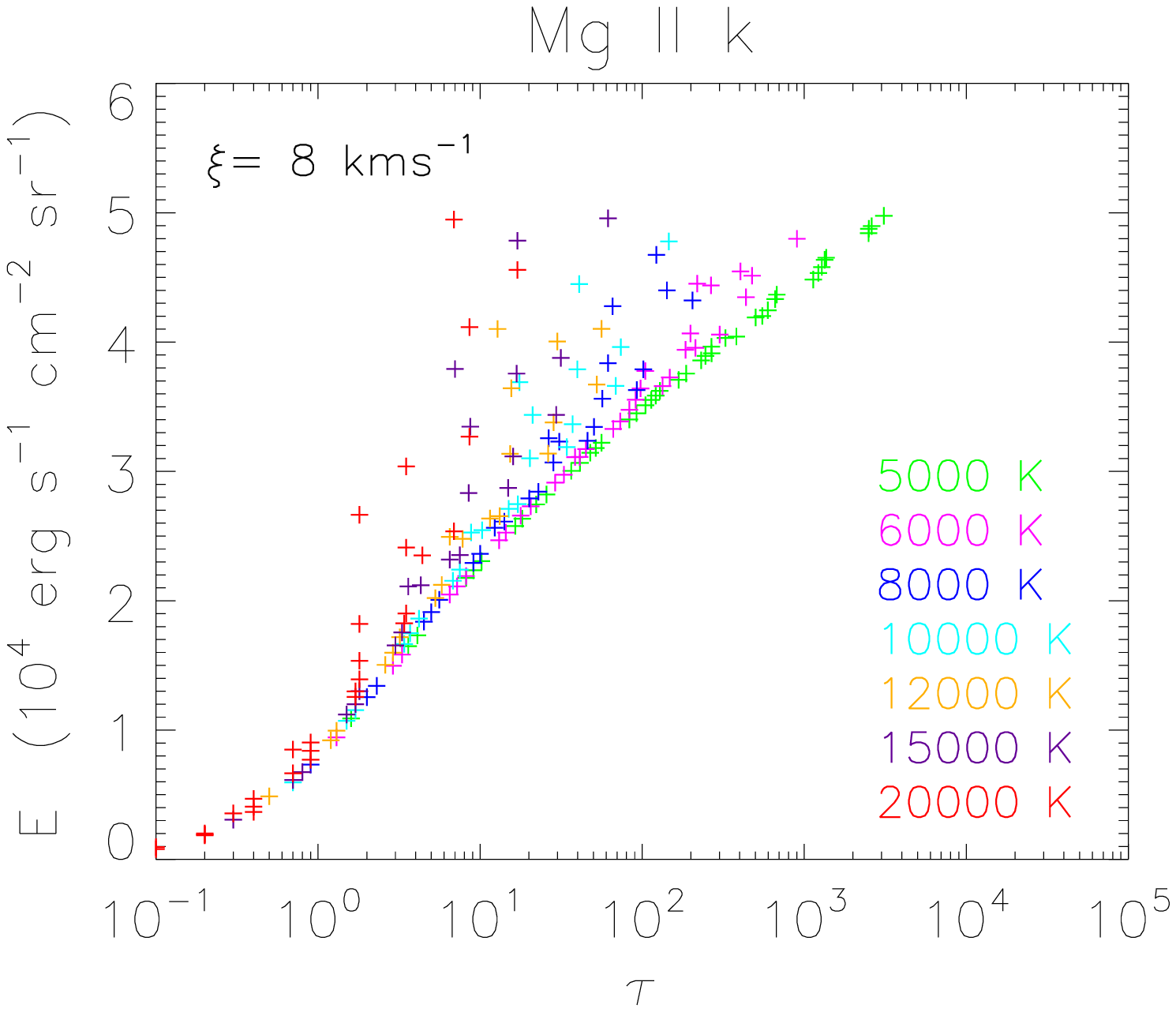}
            \hspace*{-0.03\textwidth}
            \includegraphics[width=0.27\textwidth,clip=]{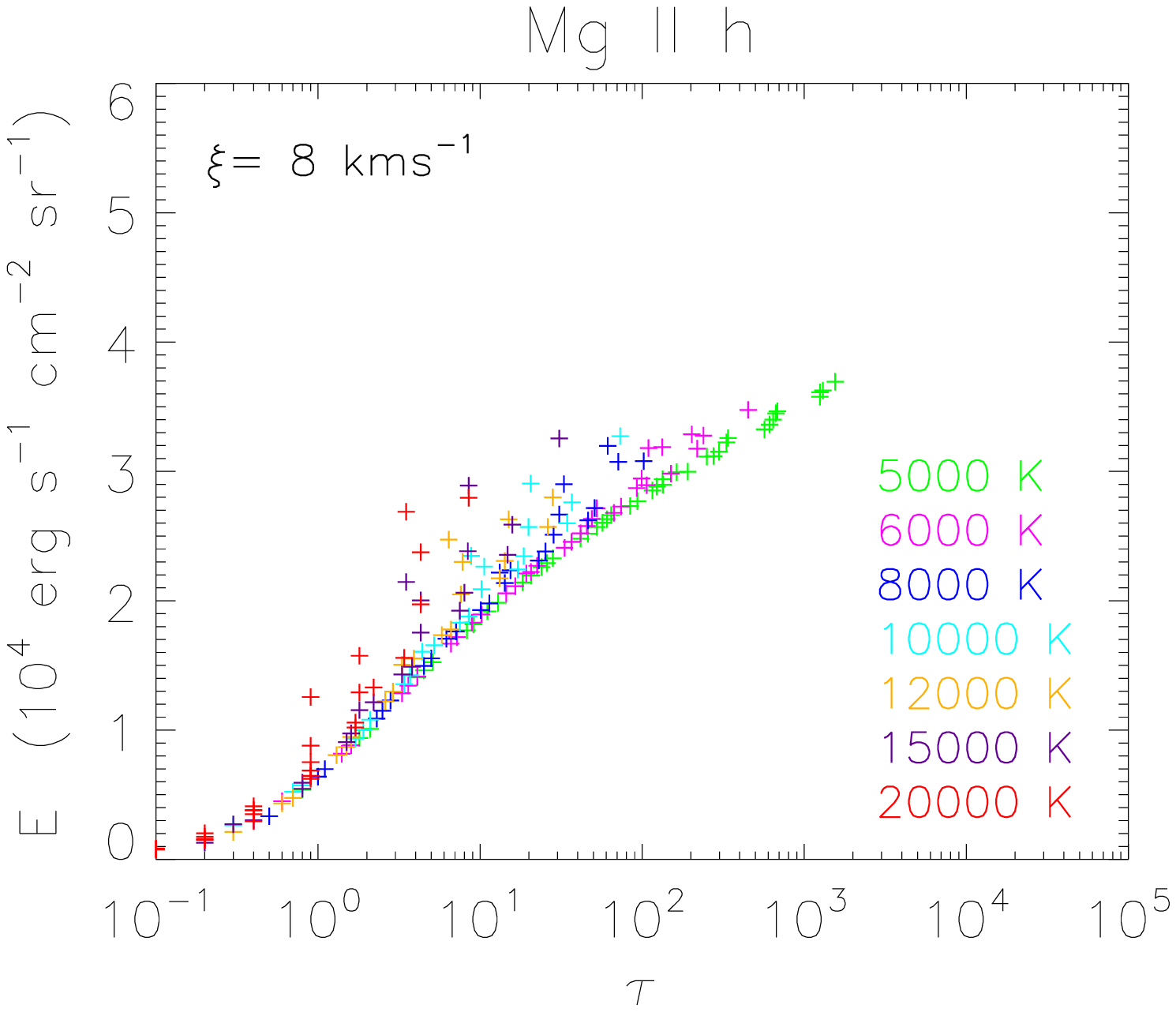}
            }
\caption{Integrated intensity of \mg\ lines as a function of optical thickness for different temperatures at three representative microturbulent velocities.}
\label{f-taue}
\end{figure}

\begin{figure}    
\centerline{\includegraphics[width=0.27\textwidth,clip=]{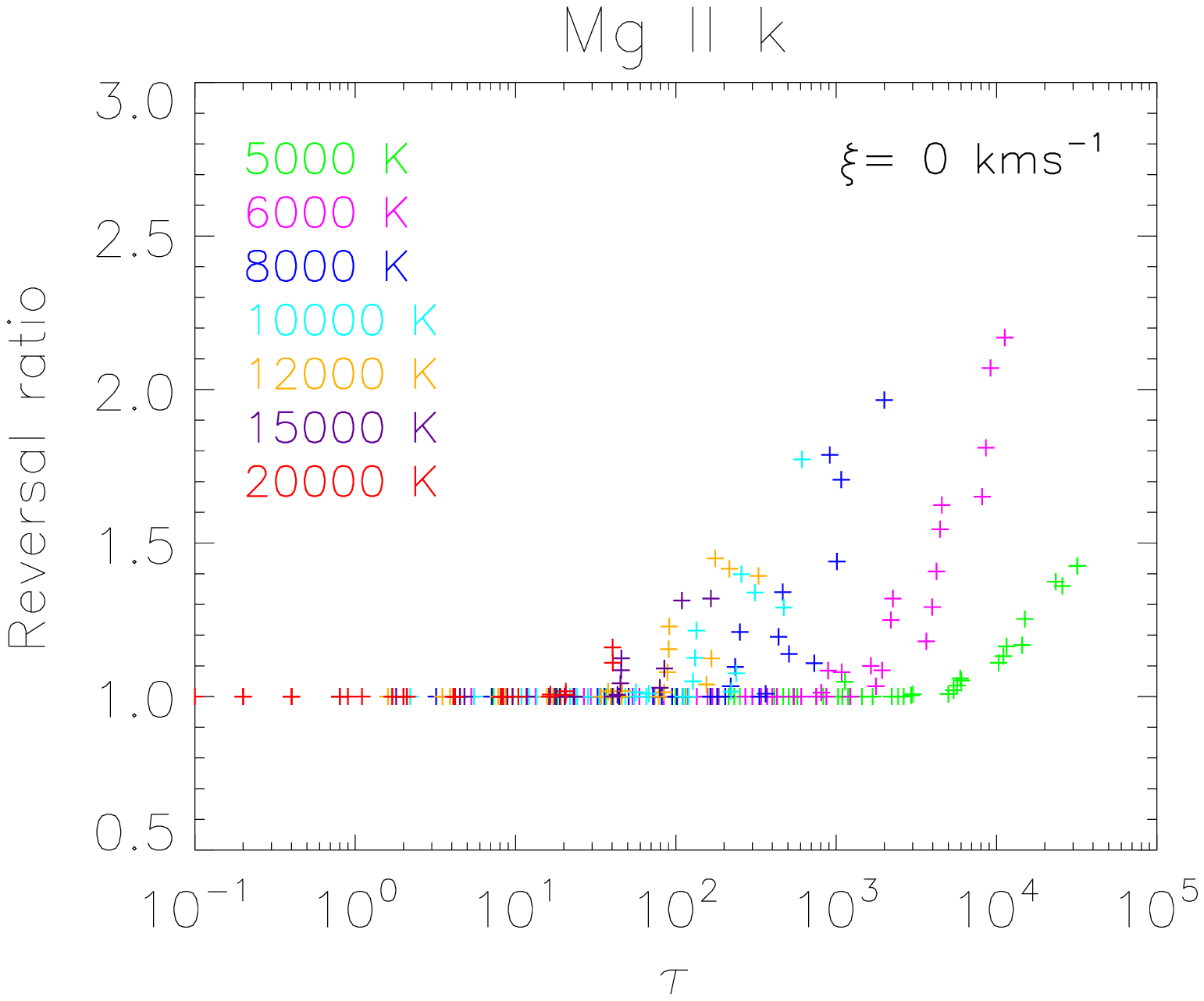}
            \hspace*{-0.03\textwidth}
            \includegraphics[width=0.27\textwidth,clip=]{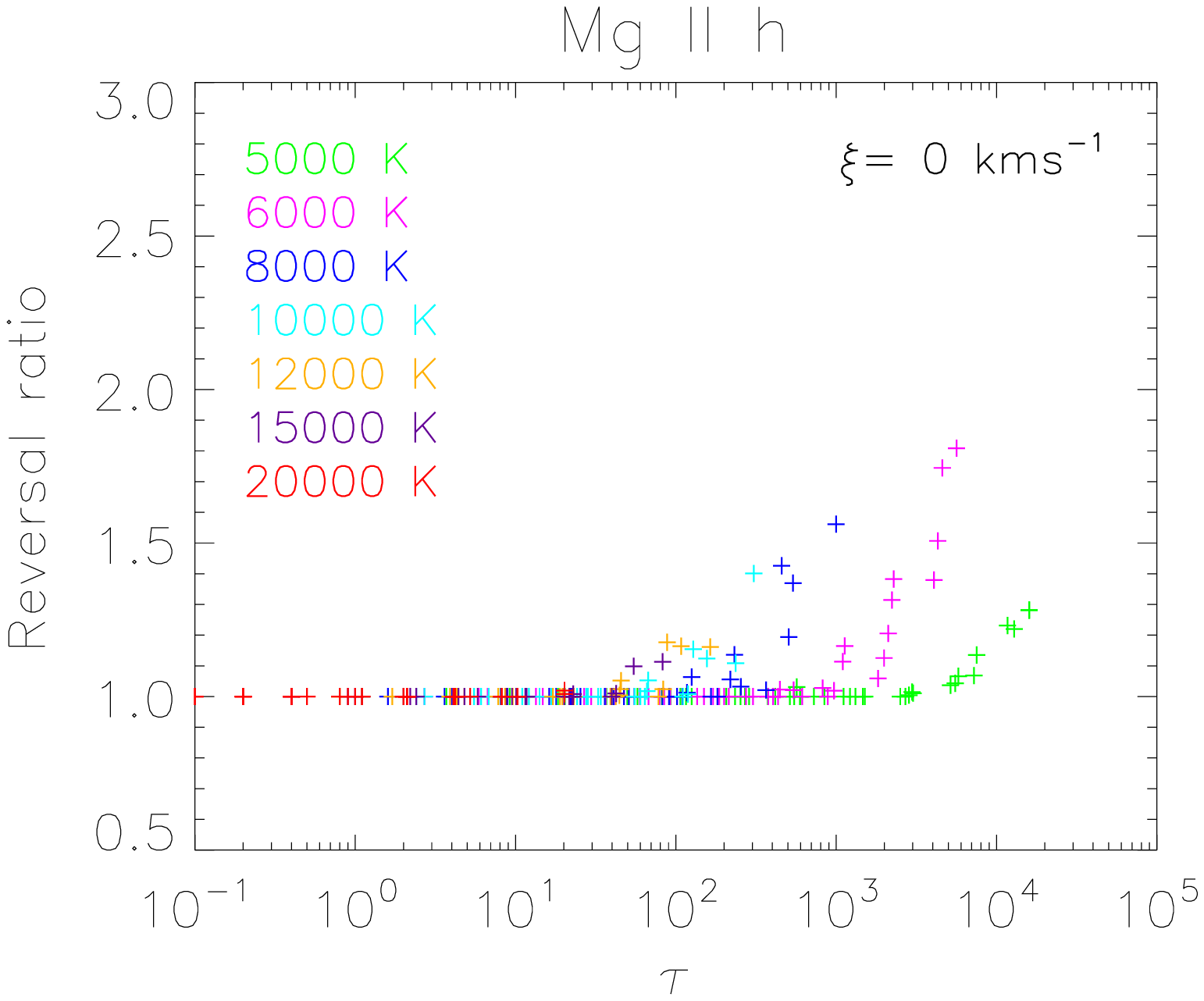}
            }
\vspace{0.01\textwidth}
\centerline{\includegraphics[width=0.27\textwidth,clip=]{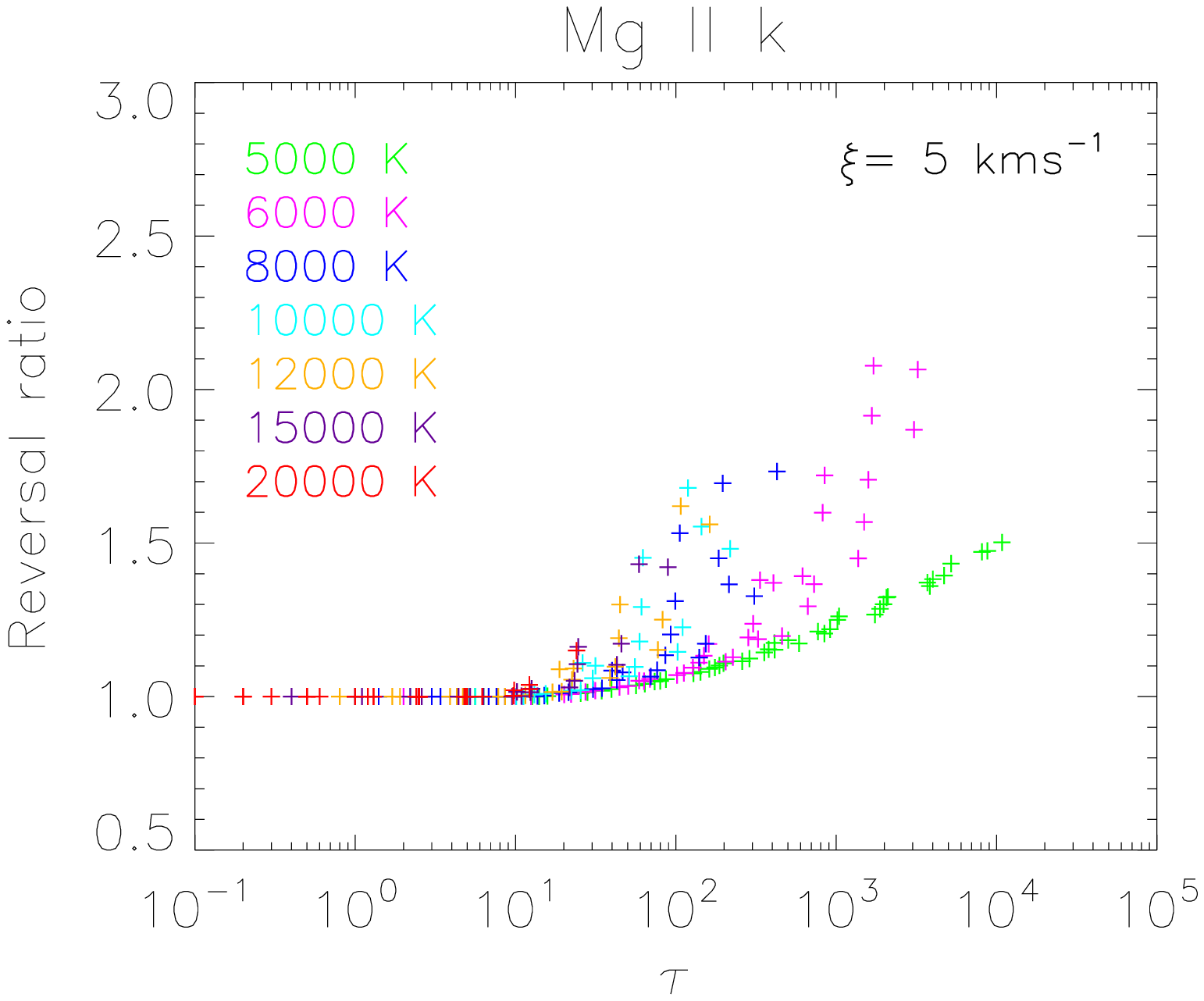}
            \hspace*{-0.03\textwidth}
            \includegraphics[width=0.27\textwidth,clip=]{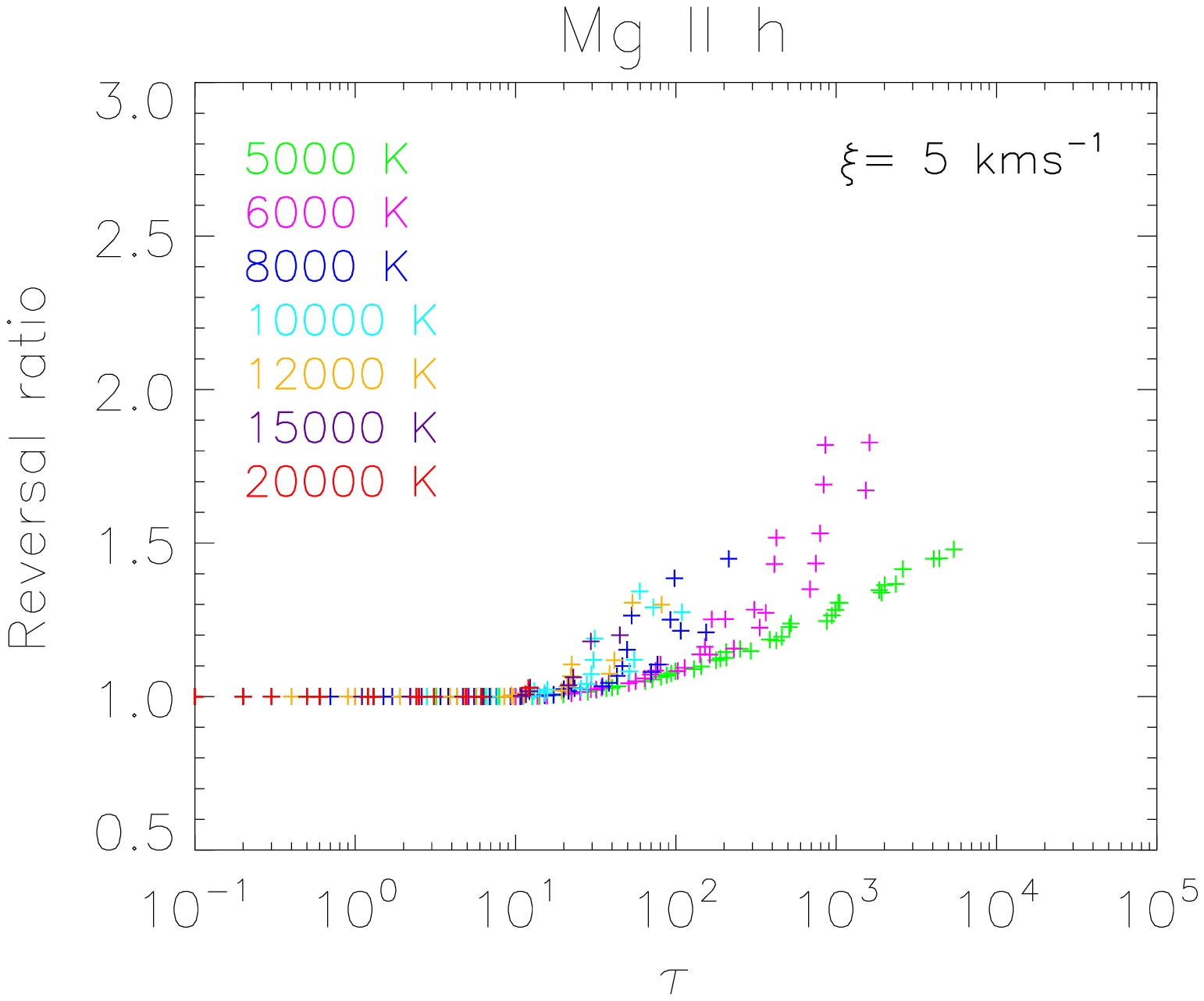}
            }
\vspace{0.01\textwidth}
\centerline{\includegraphics[width=0.27\textwidth,clip=]{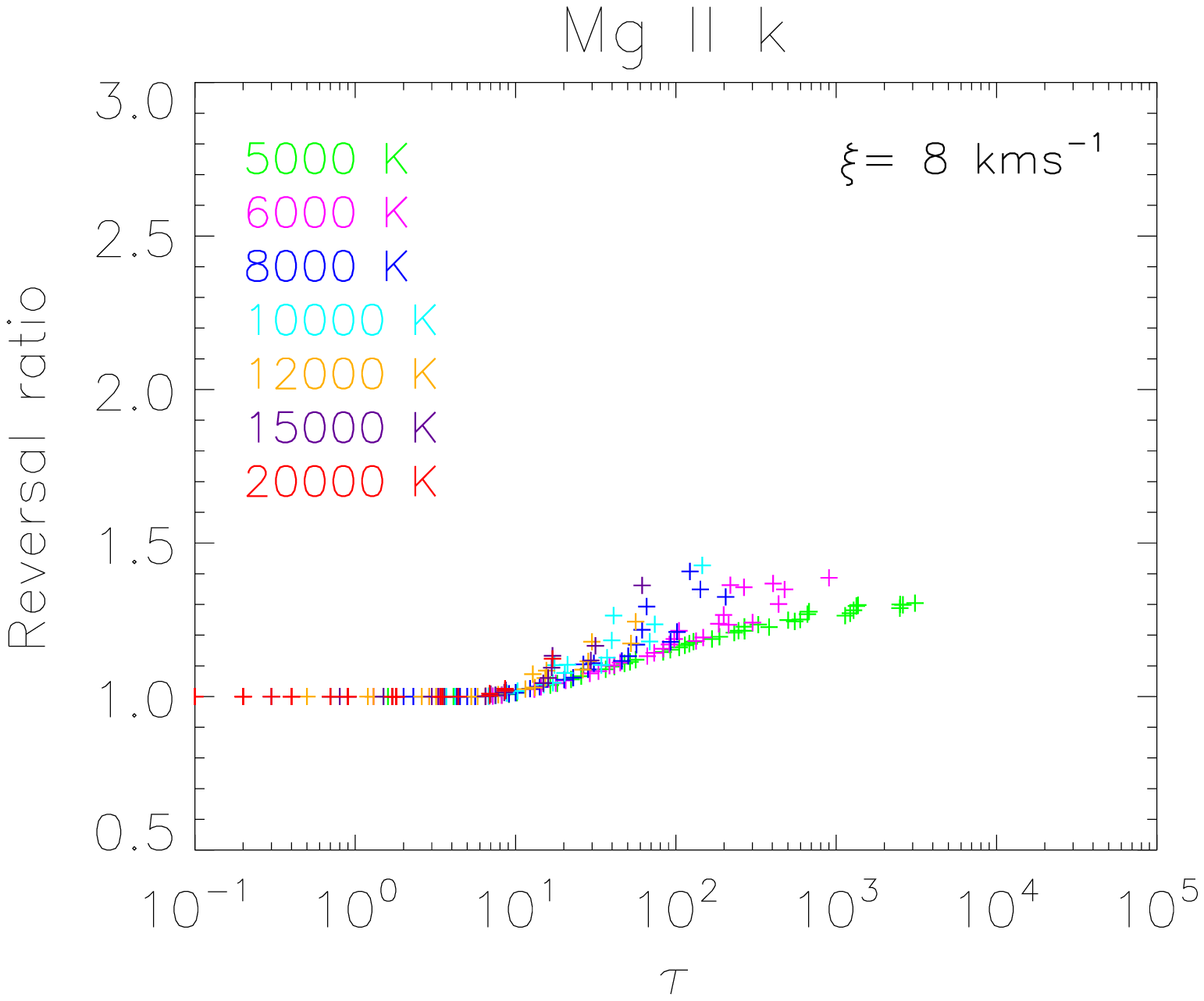}
            \hspace*{-0.03\textwidth}
            \includegraphics[width=0.27\textwidth,clip=]{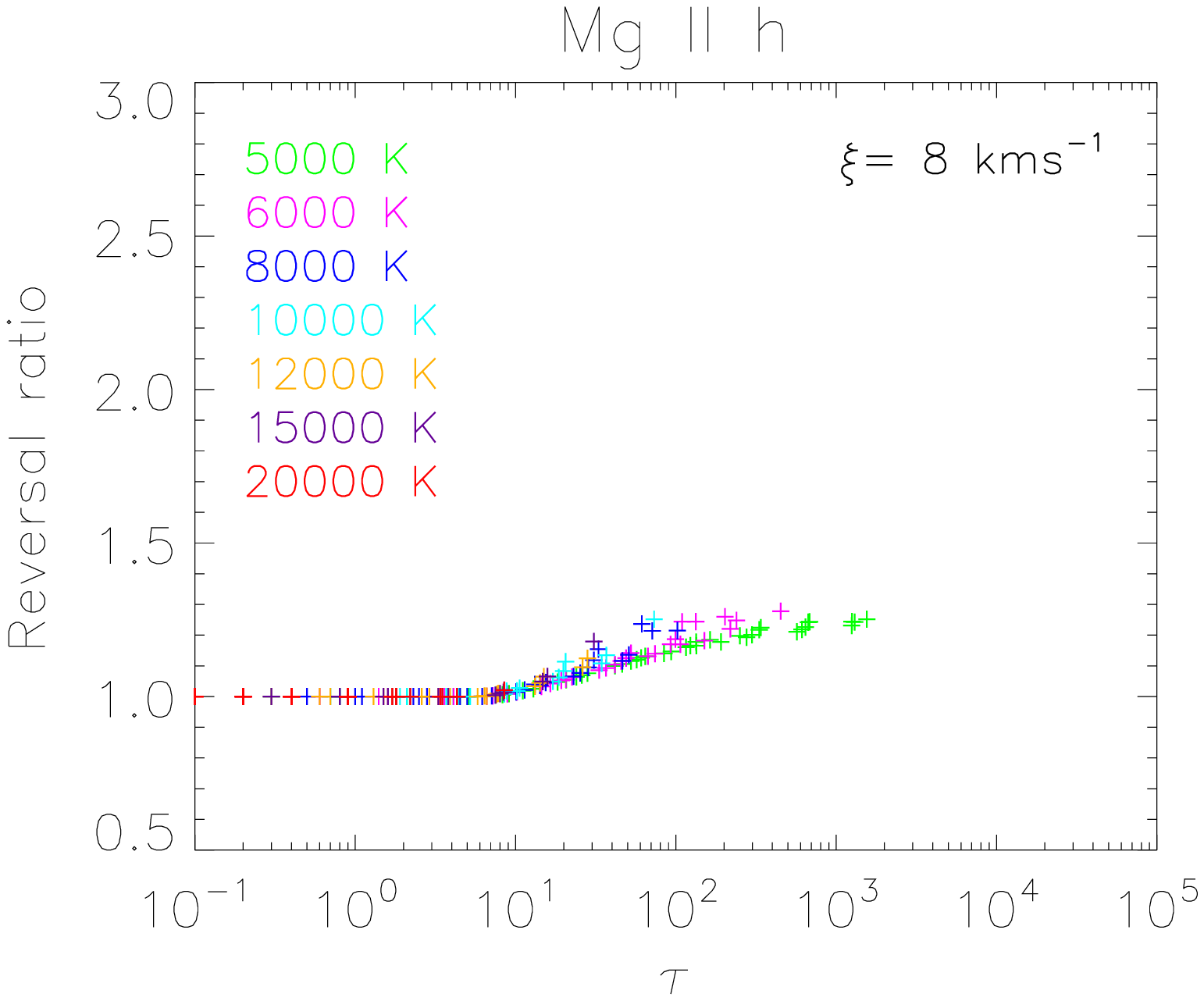}
            }
\caption{Reversal ratio of \mg\  lines as a function of optical thickness for different temperatures at three representative microturbulent velocities.}
\label{f-taur}
\end{figure}

\begin{figure}    
\centerline{\includegraphics[width=0.27\textwidth,clip=]{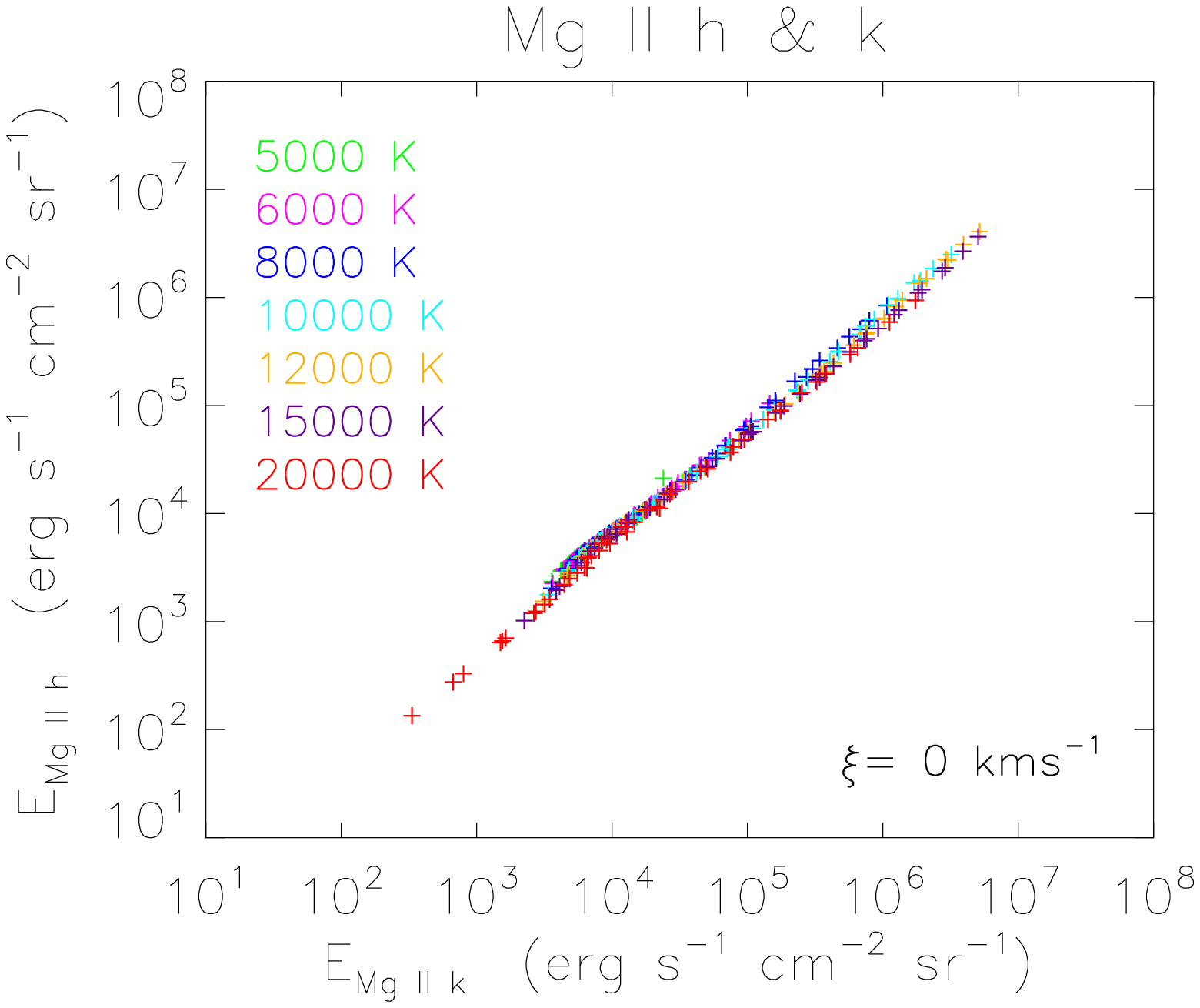}
            \hspace*{-0.03\textwidth}
            \includegraphics[width=0.27\textwidth,clip=]{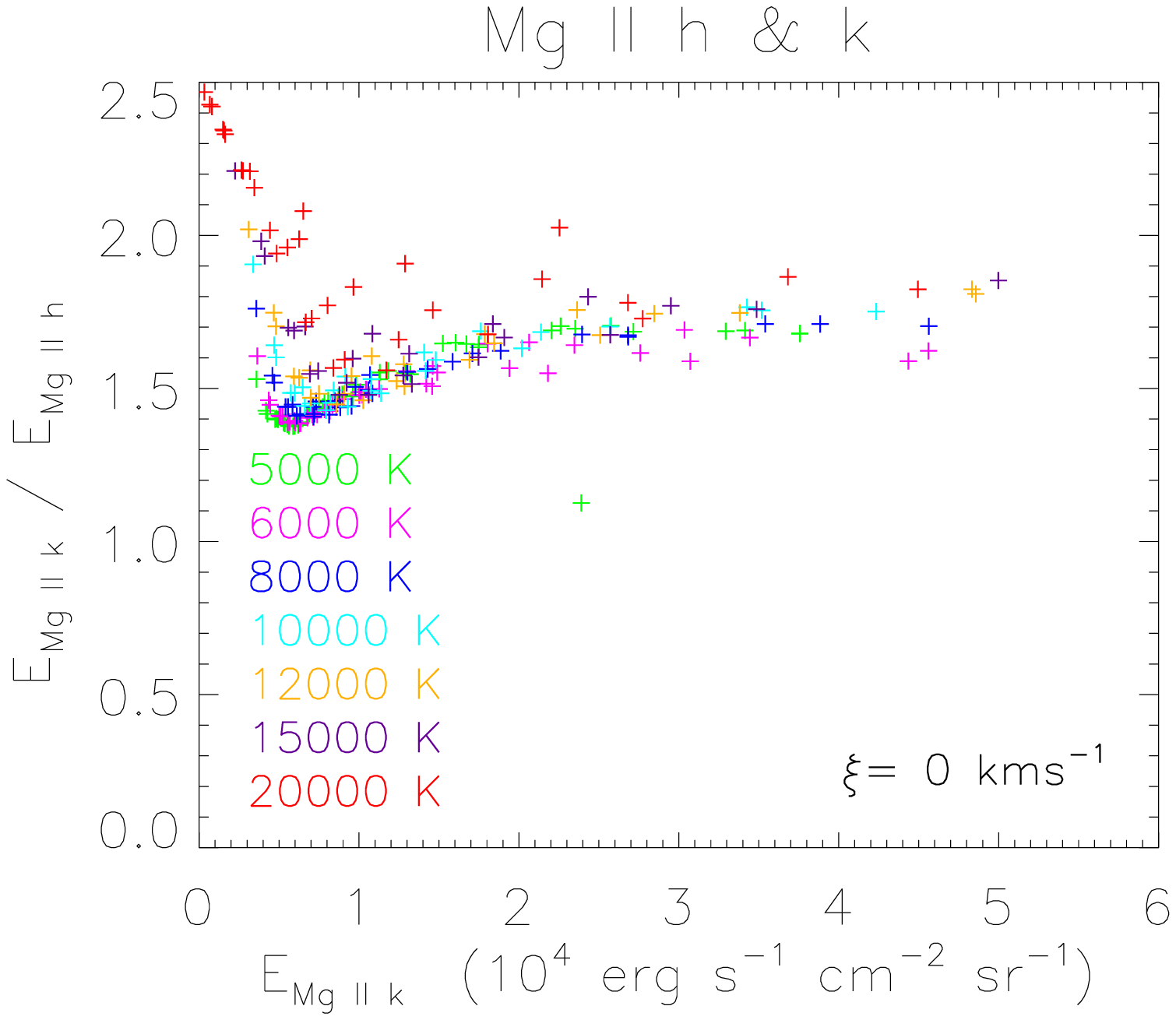}
            }
\vspace{0.01\textwidth}
\centerline{\includegraphics[width=0.27\textwidth,clip=]{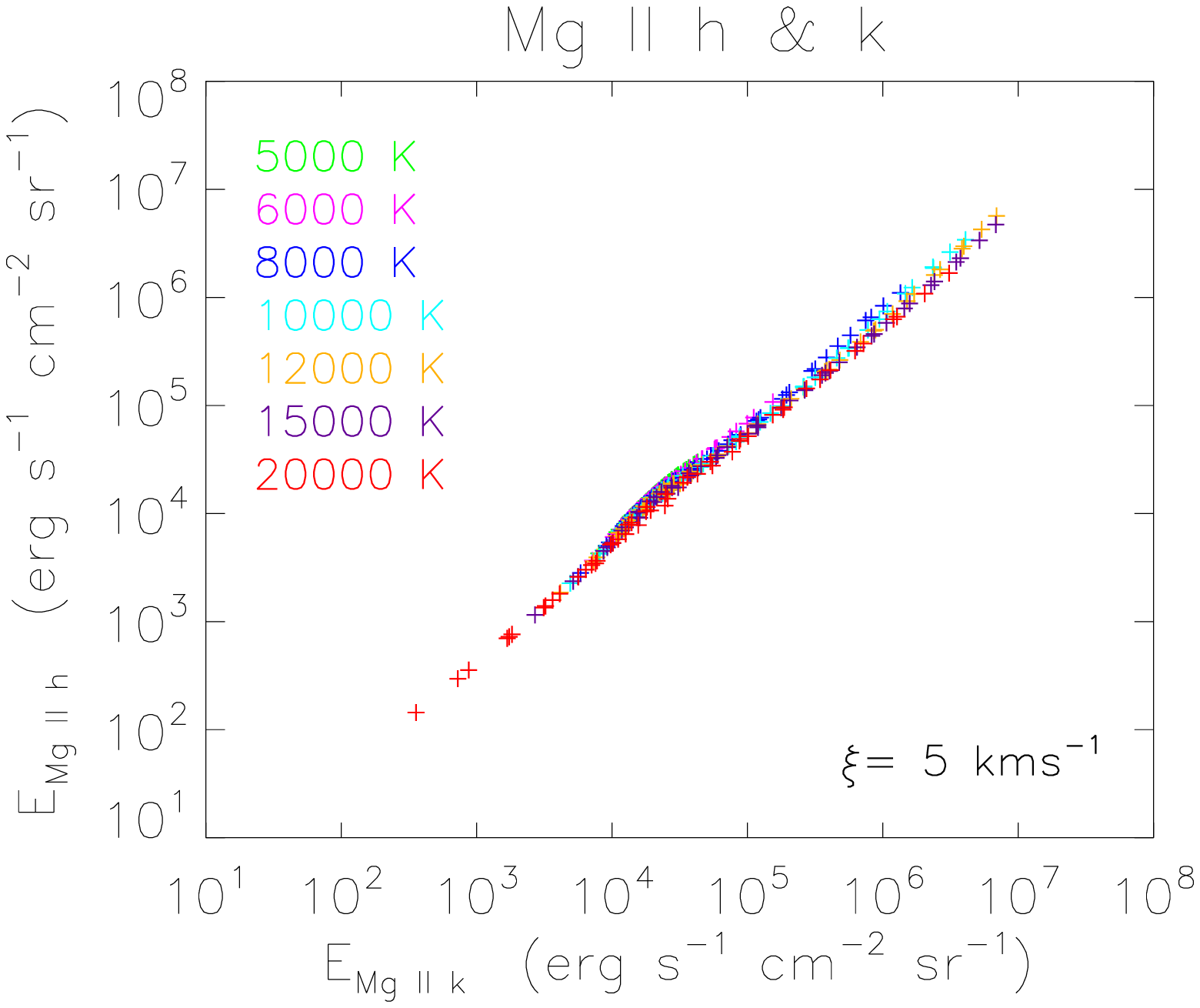}
            \hspace*{-0.03\textwidth}
            \includegraphics[width=0.27\textwidth,clip=]{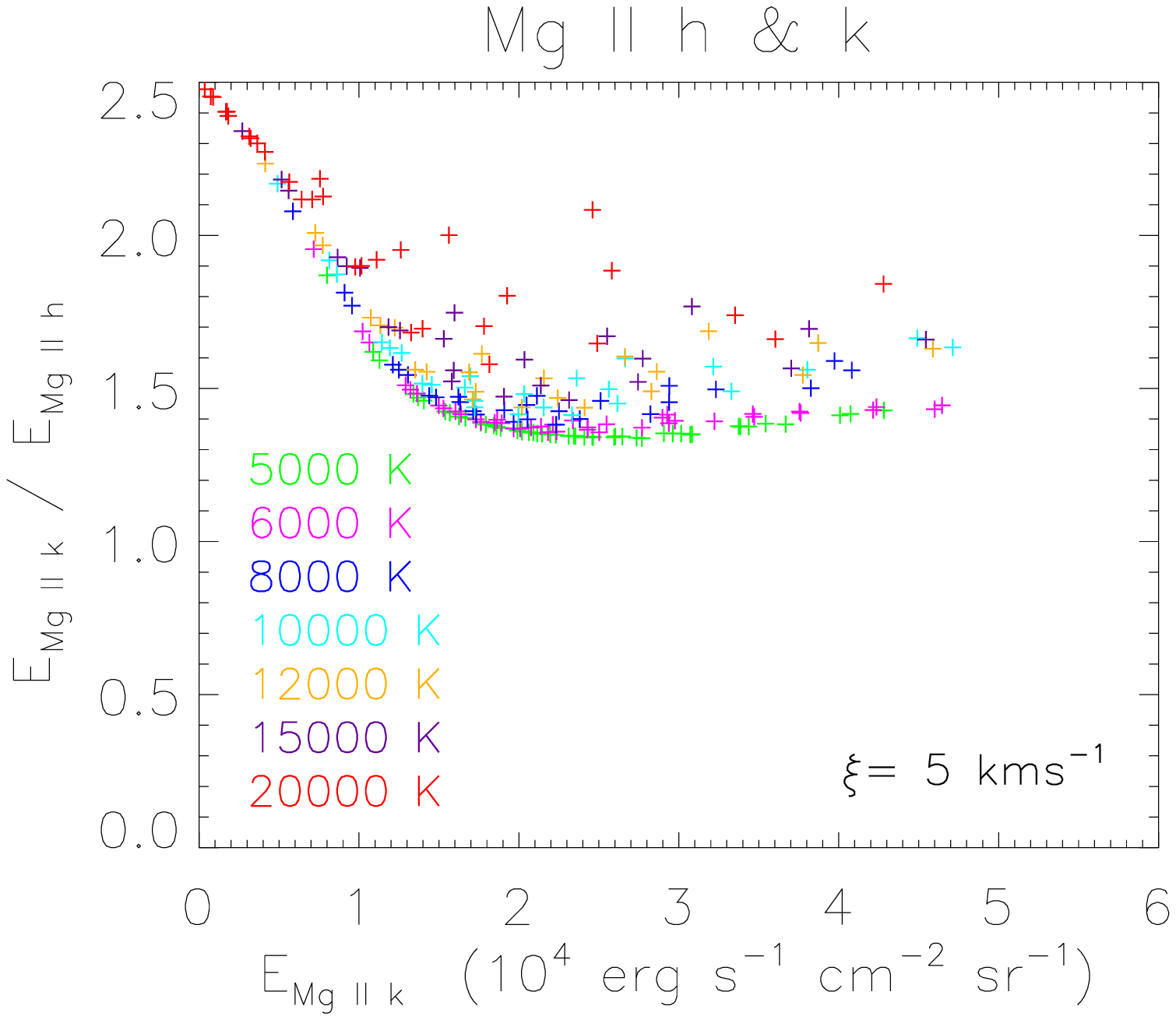}
            }
\vspace{0.01\textwidth}
\centerline{\includegraphics[width=0.27\textwidth,clip=]{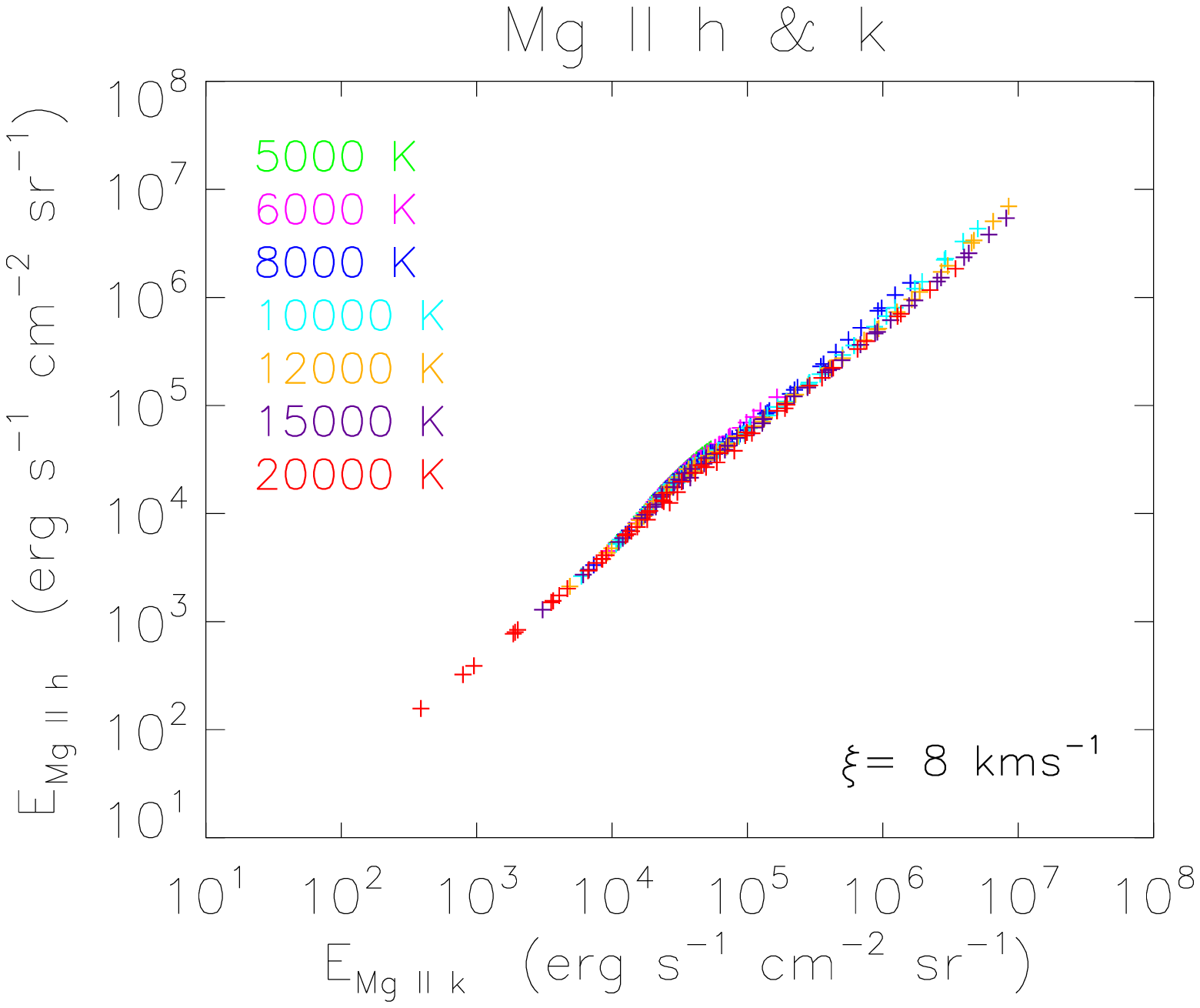}
            \hspace*{-0.03\textwidth}
            \includegraphics[width=0.27\textwidth,clip=]{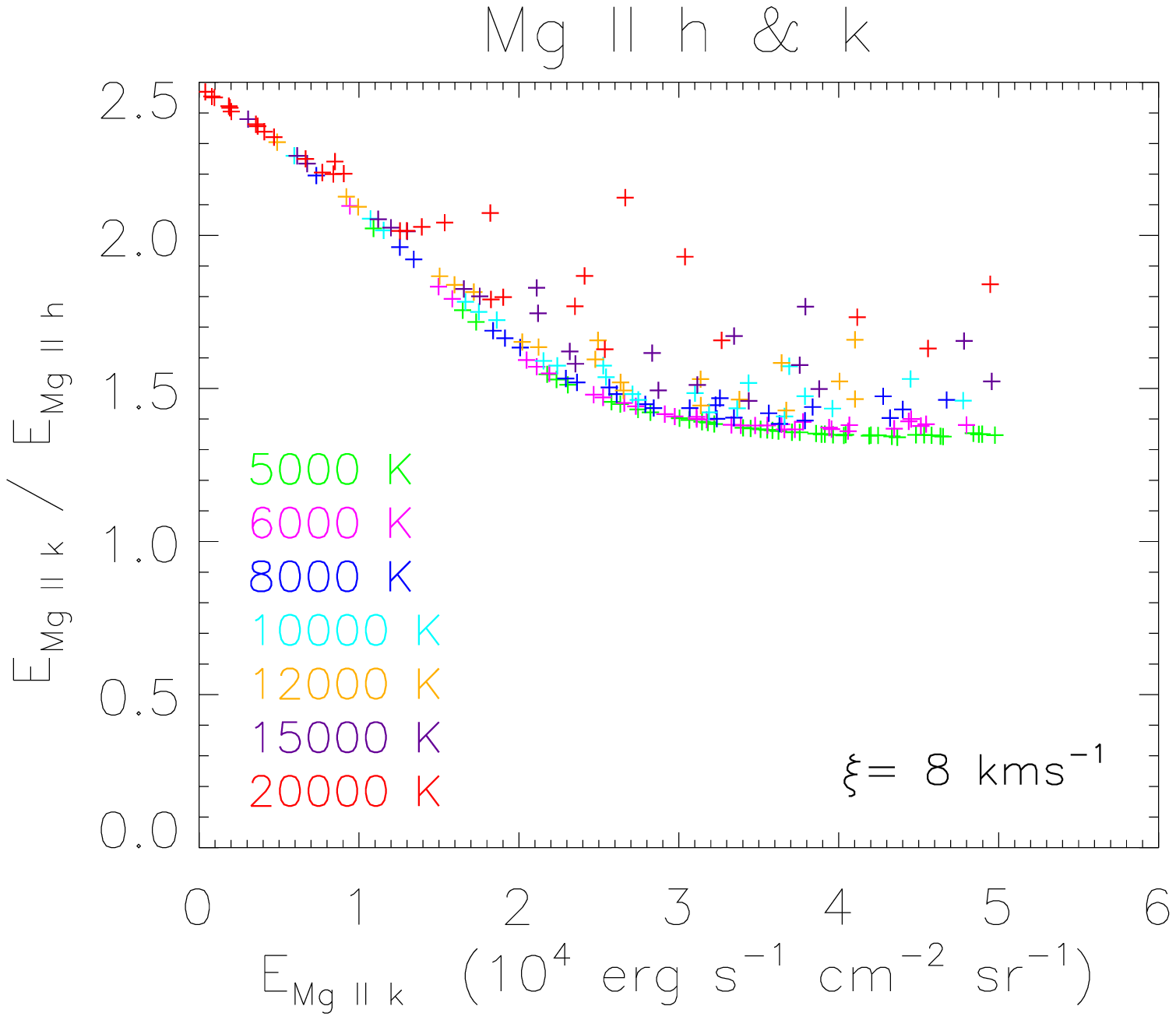}
            }
\caption{Integrated intensity emitted in \mg\ h line versus \mg\ k line for all 343 model points at different temperatures (left panels).
The right panels show the ratio  of integrated intensity of the \mg\ k line to the \mg\ h line versus the integrated intensity of the \mg\ k line for different temperatures 
at three representative microturbulent velocities.}
\label{f-eratio}
\end{figure}

\begin{figure*} 
\centerline{\includegraphics[width=0.35\textwidth,clip=]{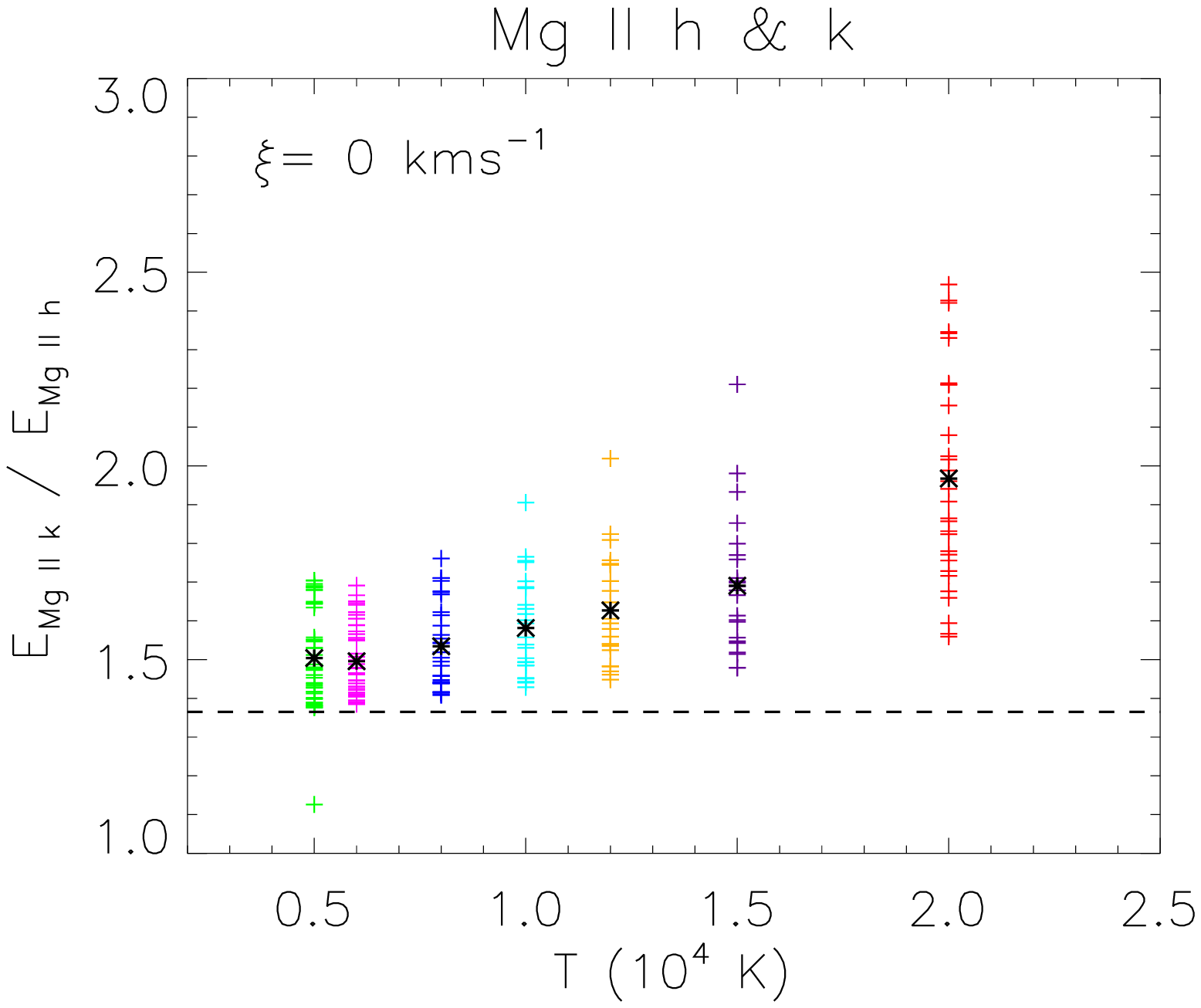}
            \hspace*{-0.02\textwidth}
            \includegraphics[width=0.35\textwidth,clip=]{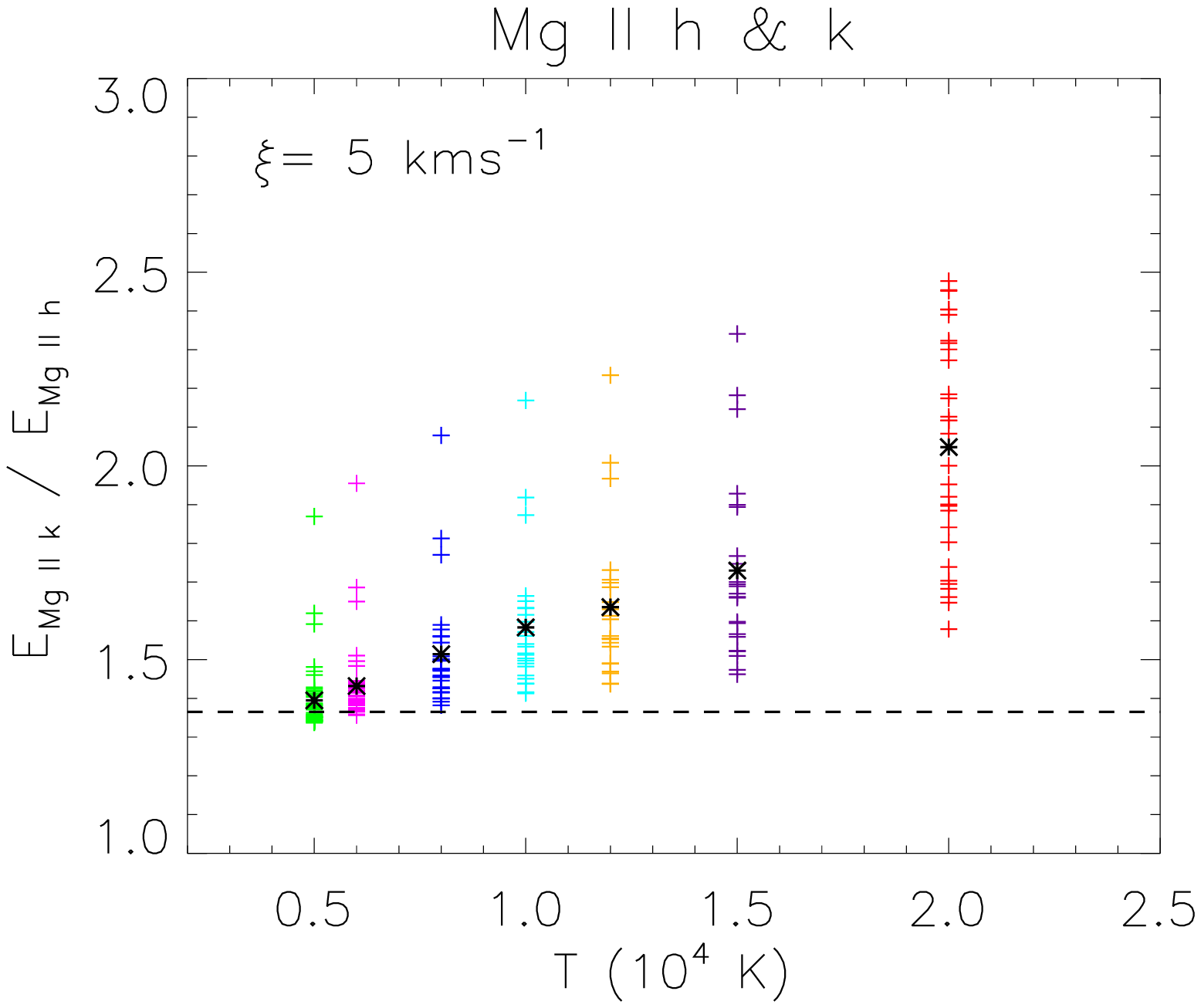}
            \hspace*{-0.02\textwidth}
            \includegraphics[width=0.35\textwidth,clip=]{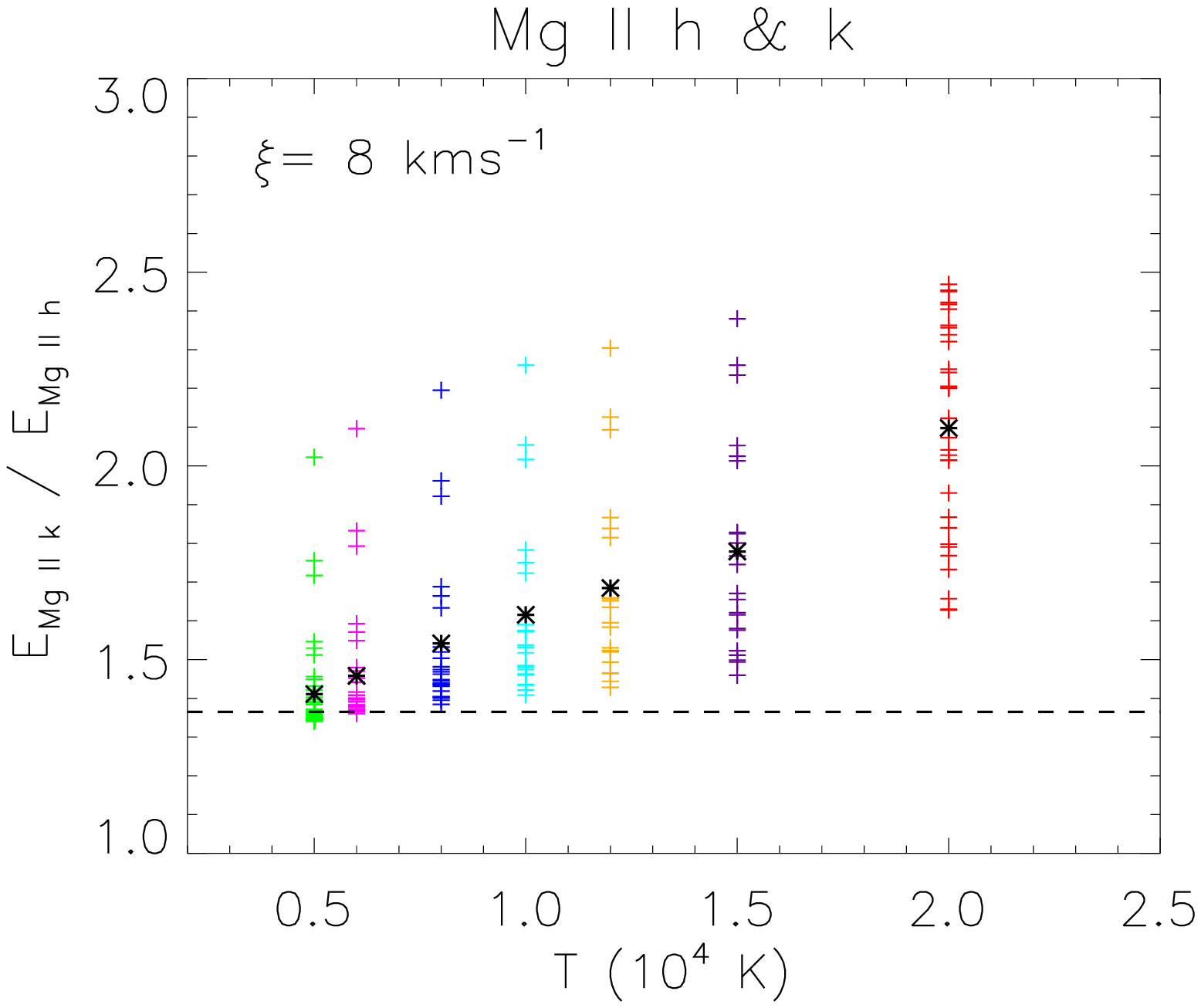}
            }
\caption{Ratio of \mg\ k to \mg\ h integrated intensities as function of temperature at zero microturbulence (left panel), 5~km~s$^{-1}$ (middle panel), 
and 8~km~s$^{-1}$ (right panel). The black asterisk symbols show the average ratio at a given temperature. The horizontal black dashed line marks the observed average 
value of 1.365.}
\label{f-avg}
\end{figure*}

\begin{figure}    
\centerline{\includegraphics[width=0.27\textwidth,clip=]{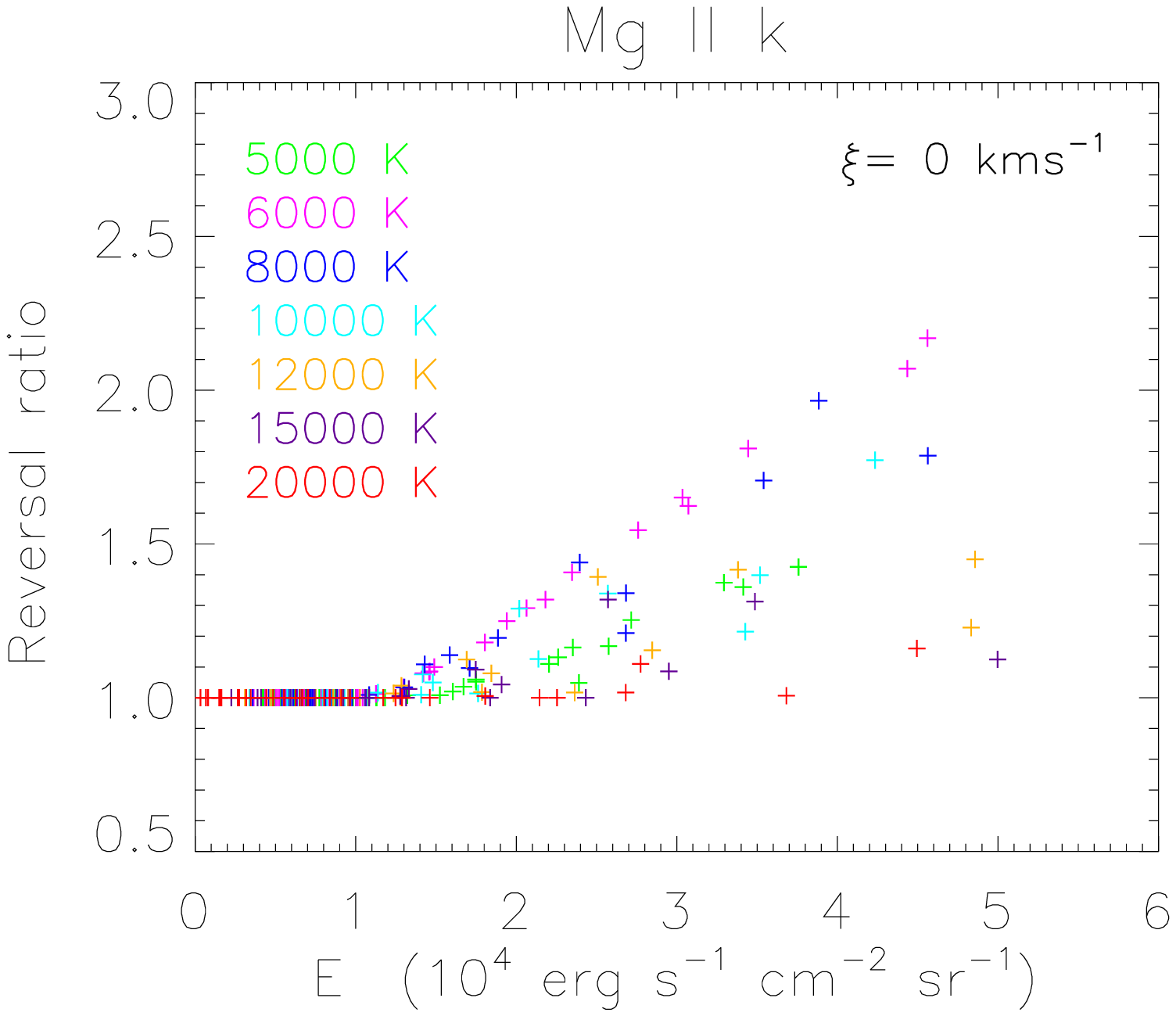}
            \hspace*{-0.03\textwidth}
            \includegraphics[width=0.27\textwidth,clip=]{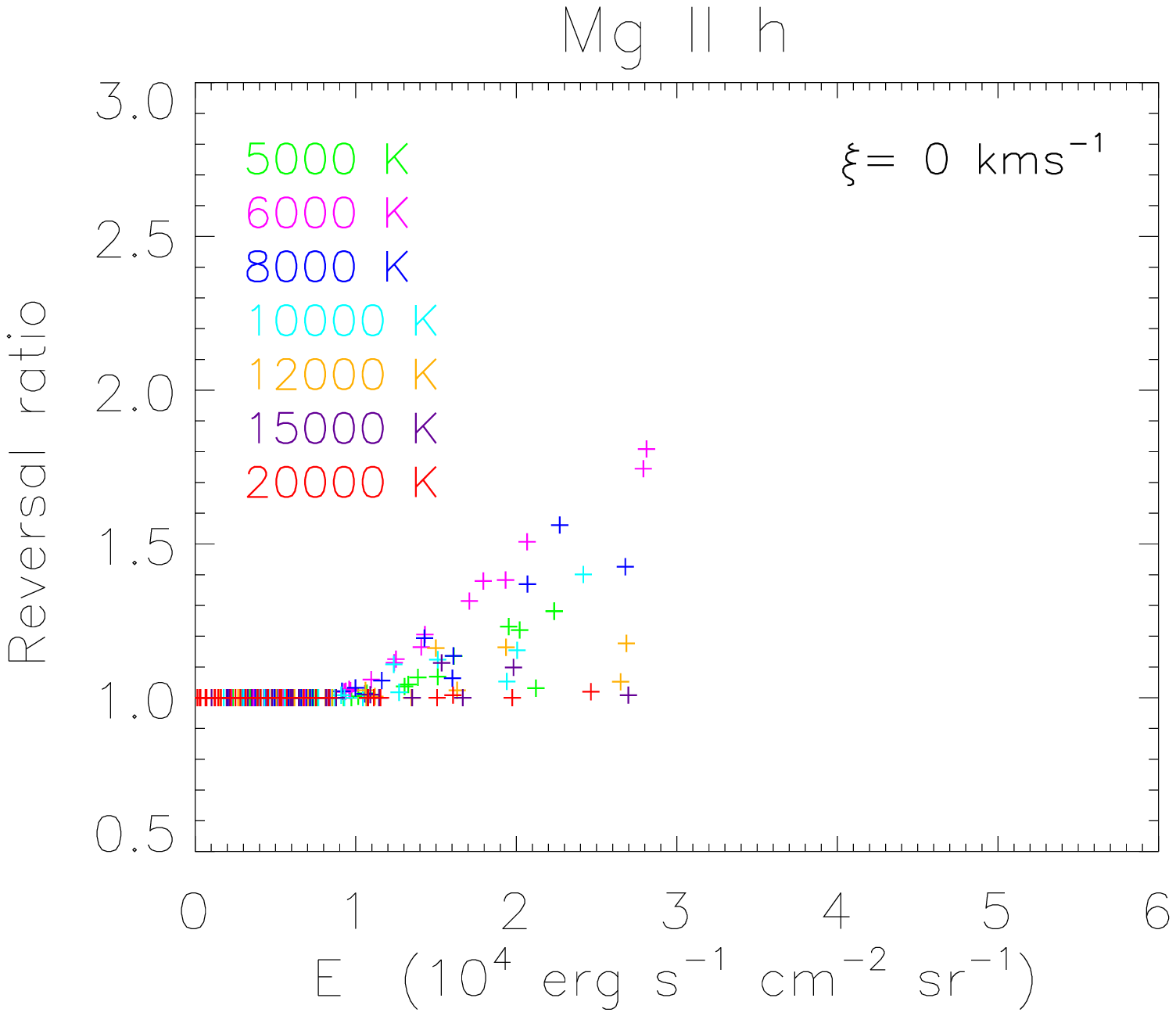}
            }
\vspace{0.01\textwidth}
\centerline{\includegraphics[width=0.27\textwidth,clip=]{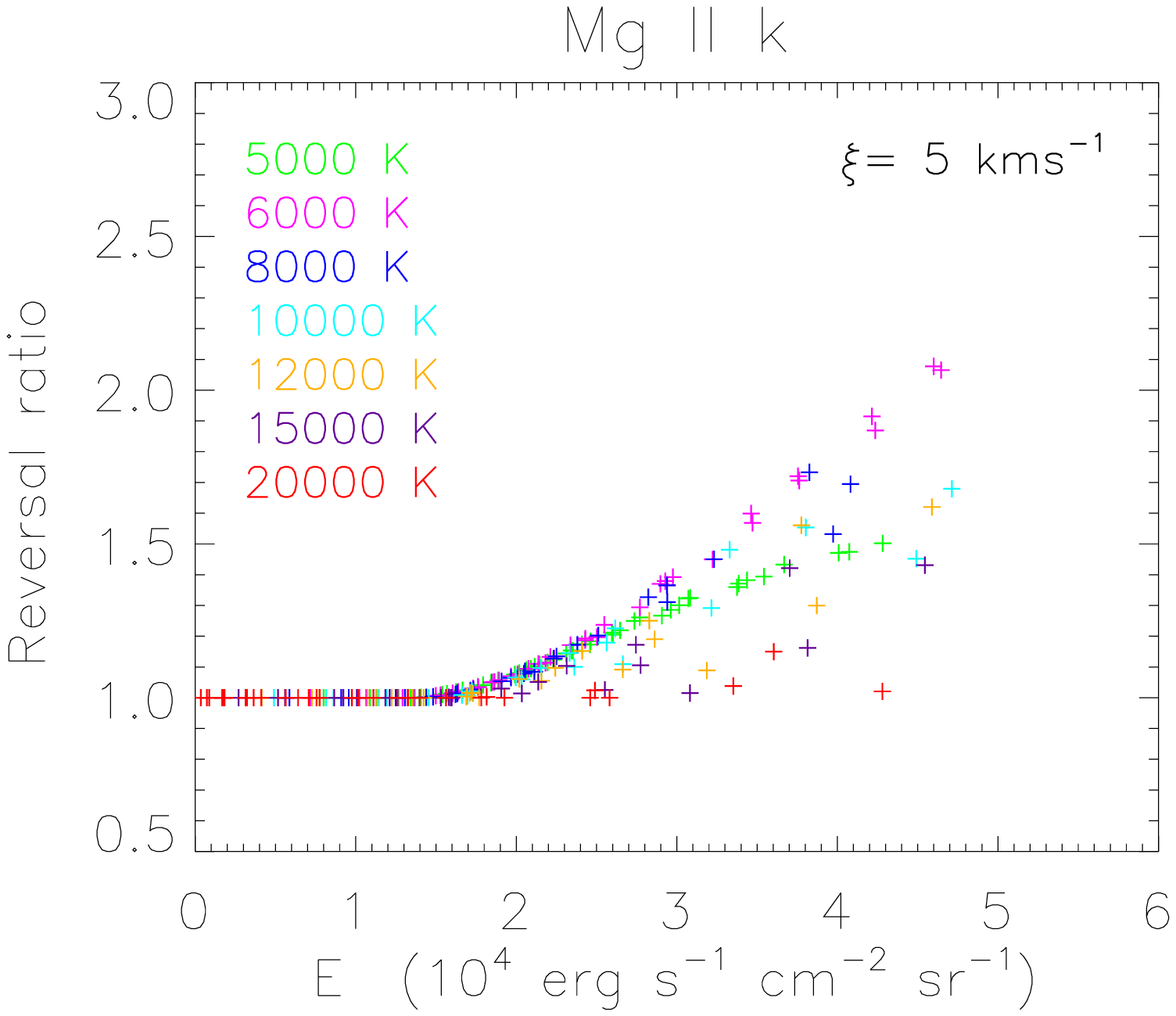}
            \hspace*{-0.03\textwidth}
            \includegraphics[width=0.27\textwidth,clip=]{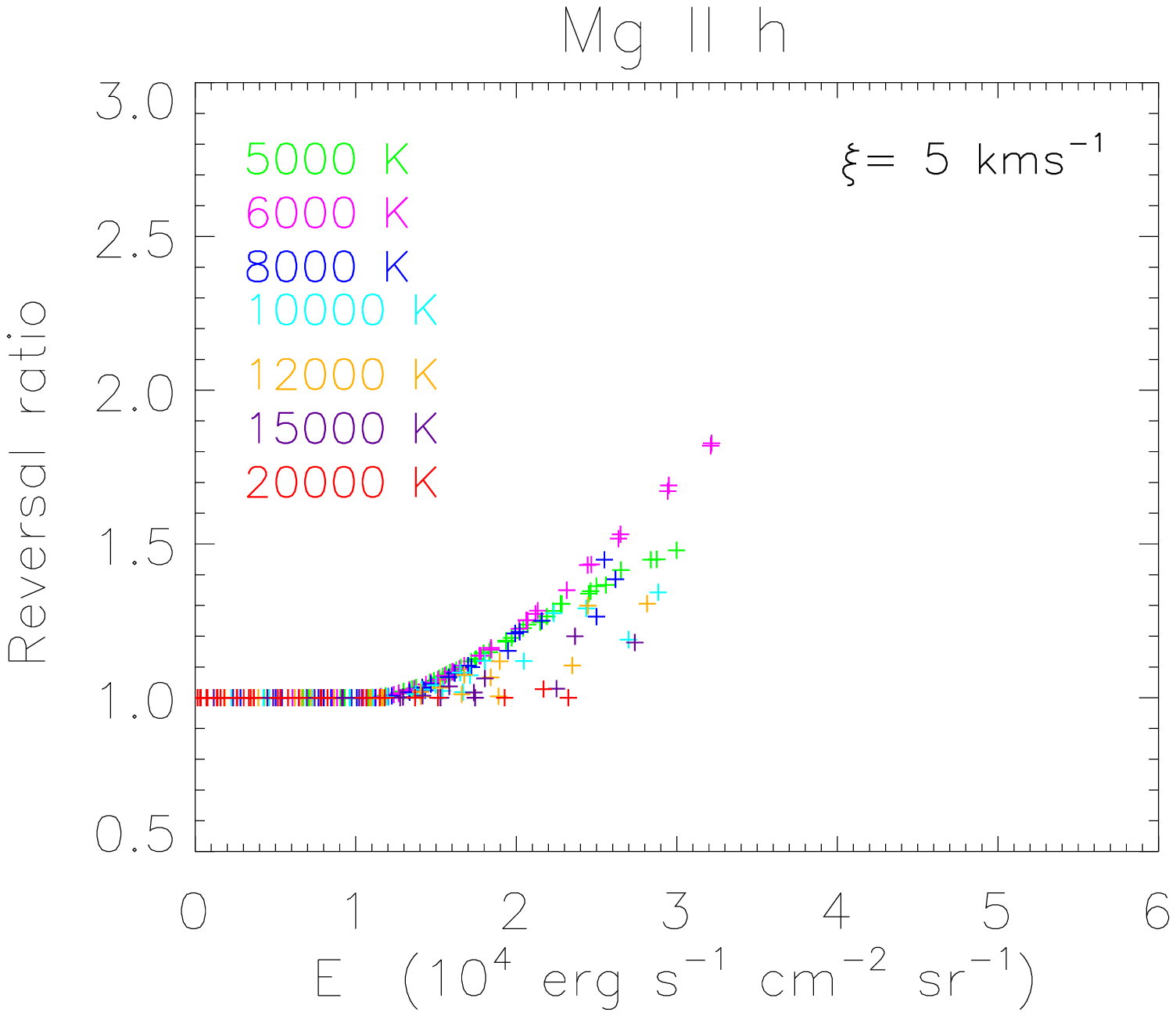}
            }
\vspace{0.01\textwidth}
\centerline{\includegraphics[width=0.27\textwidth,clip=]{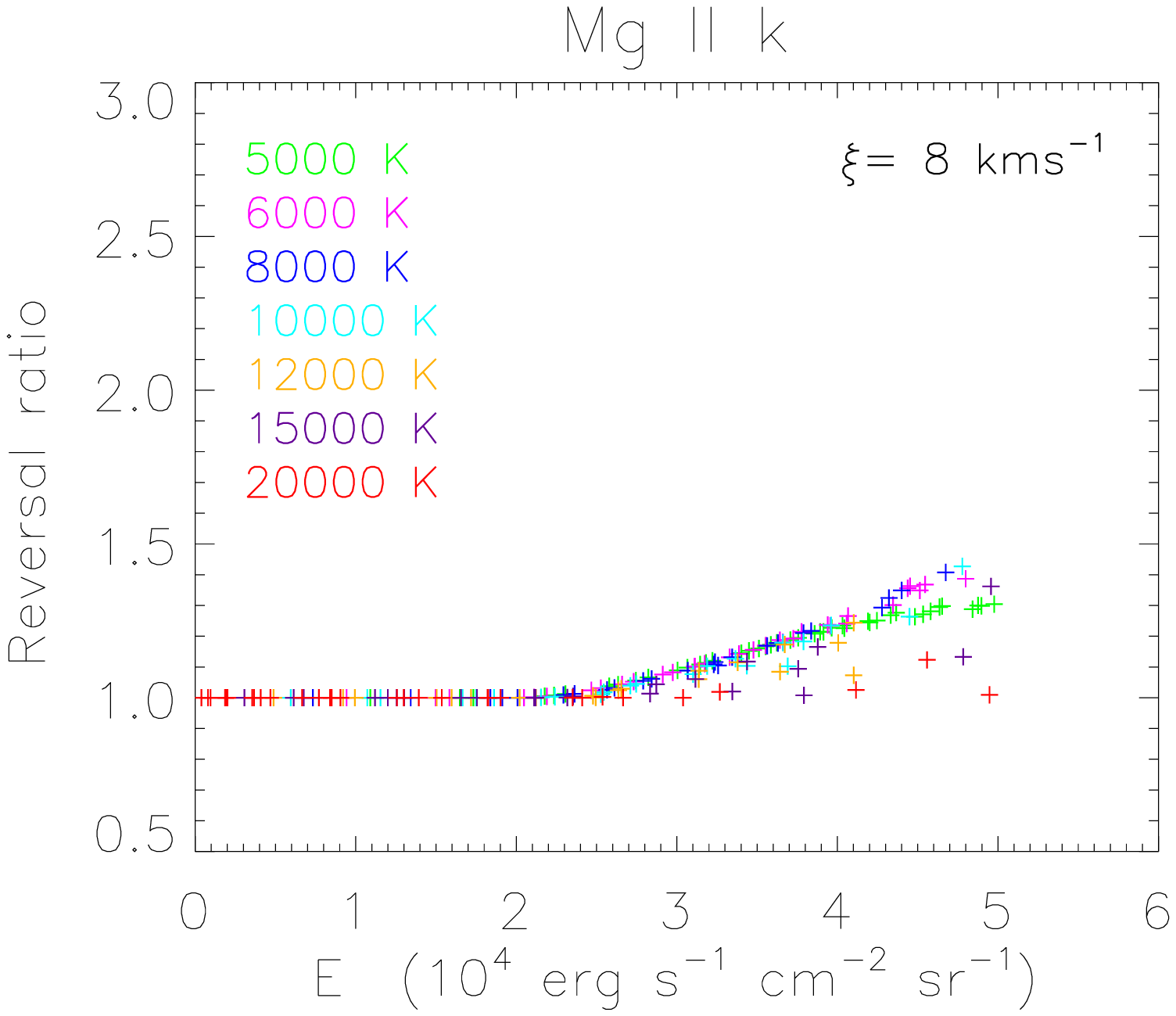}
            \hspace*{-0.03\textwidth}
            \includegraphics[width=0.27\textwidth,clip=]{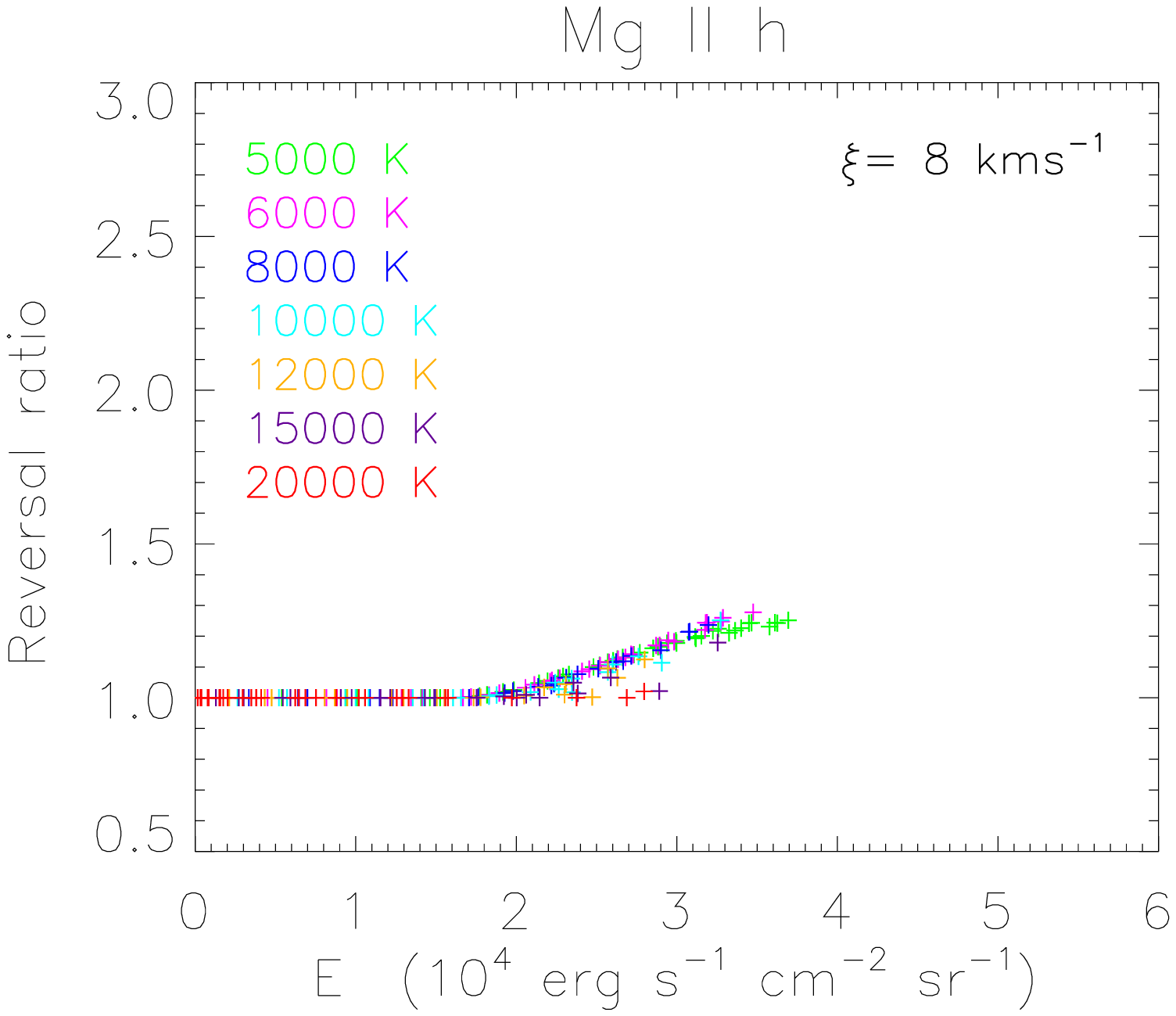}
            }
\caption{Reversal ratio versus integrated intensity of \mg\ lines for different temperatures at three representative microturbulent velocities.}
\label{f-rrem}
\end{figure}

\begin{figure}    
\centerline{\includegraphics[width=0.27\textwidth,clip=]{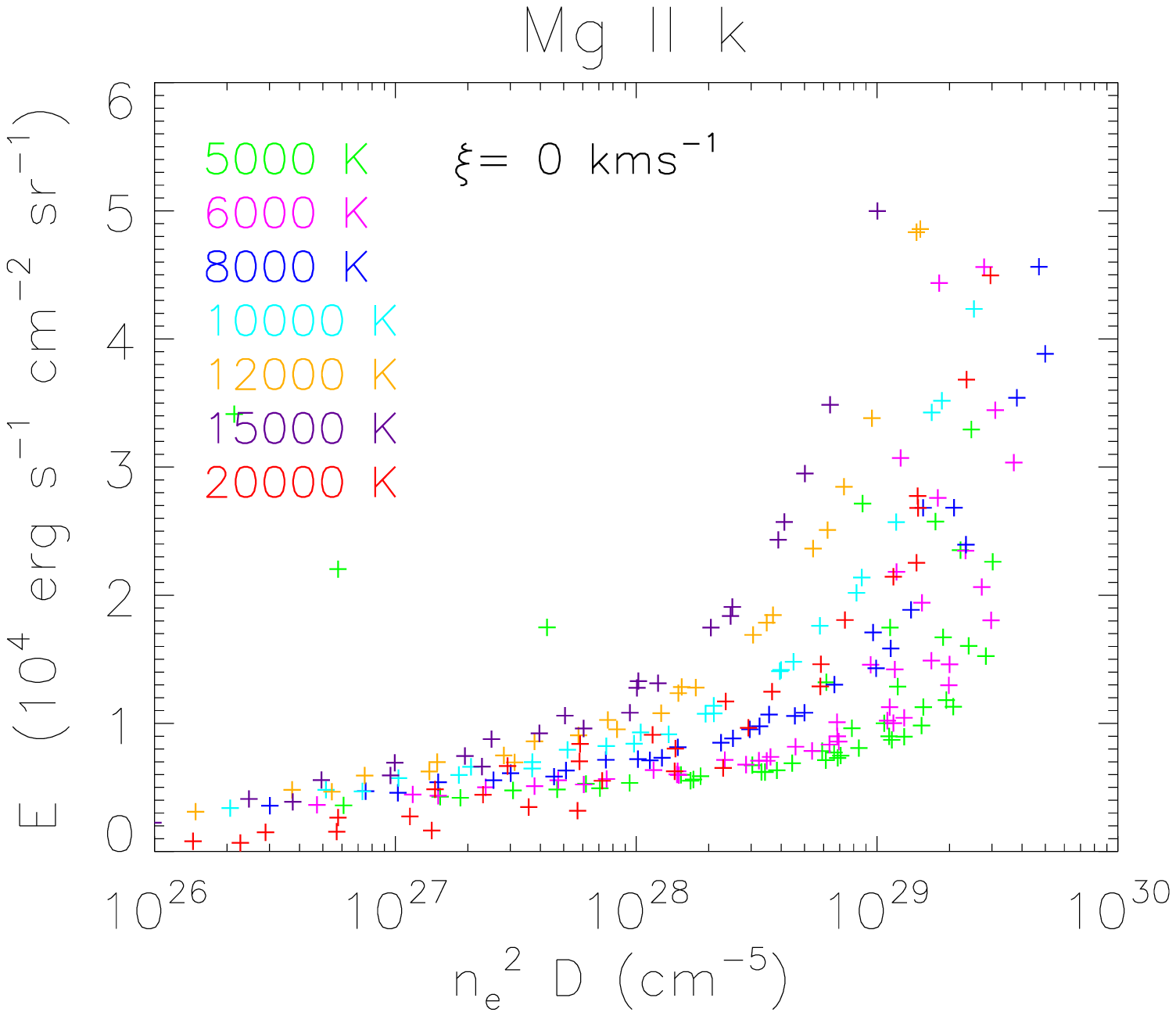}
            \hspace*{-0.03\textwidth}
            \includegraphics[width=0.27\textwidth,clip=]{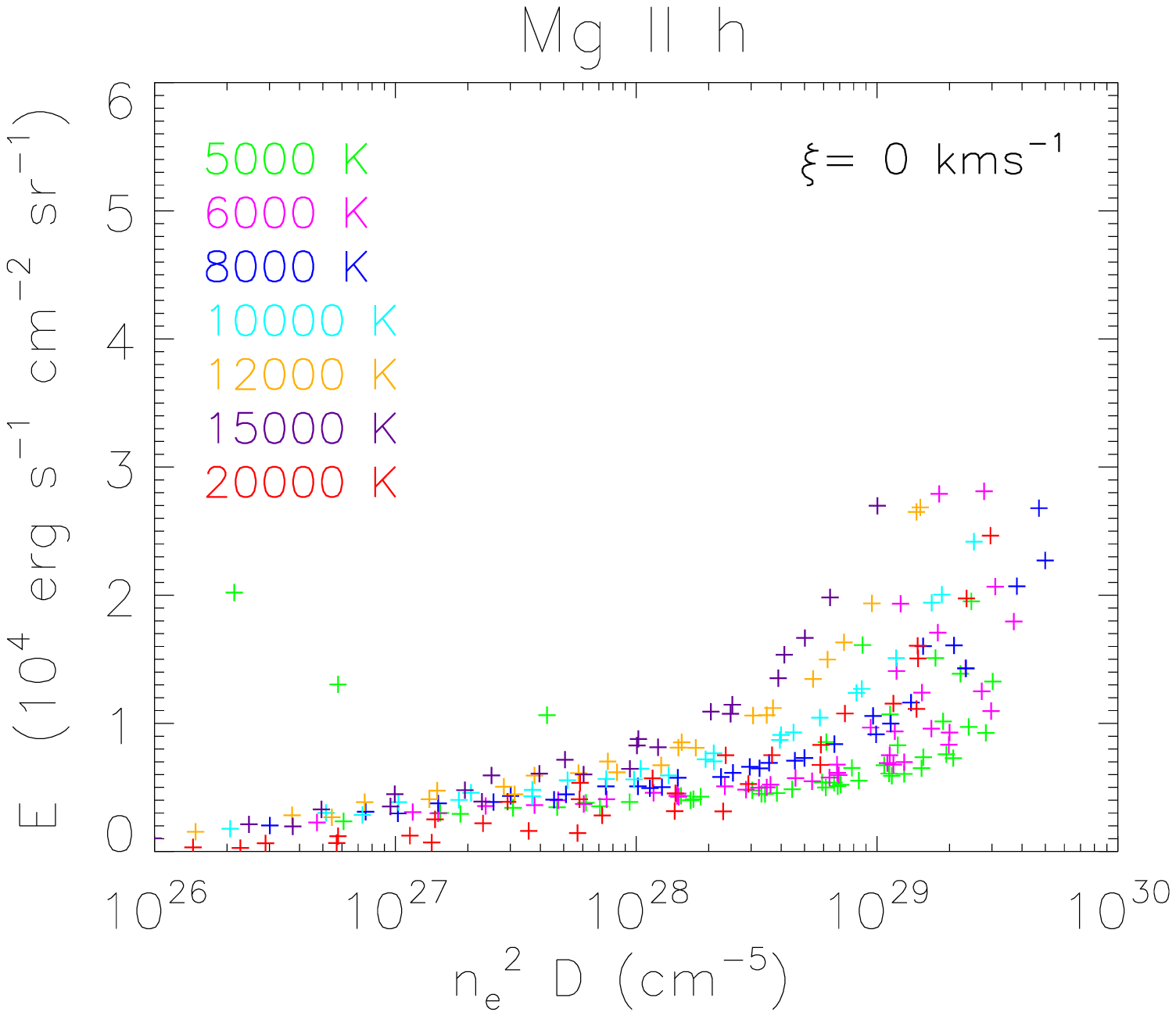}
            }
\vspace{0.01\textwidth}
\centerline{\includegraphics[width=0.27\textwidth,clip=]{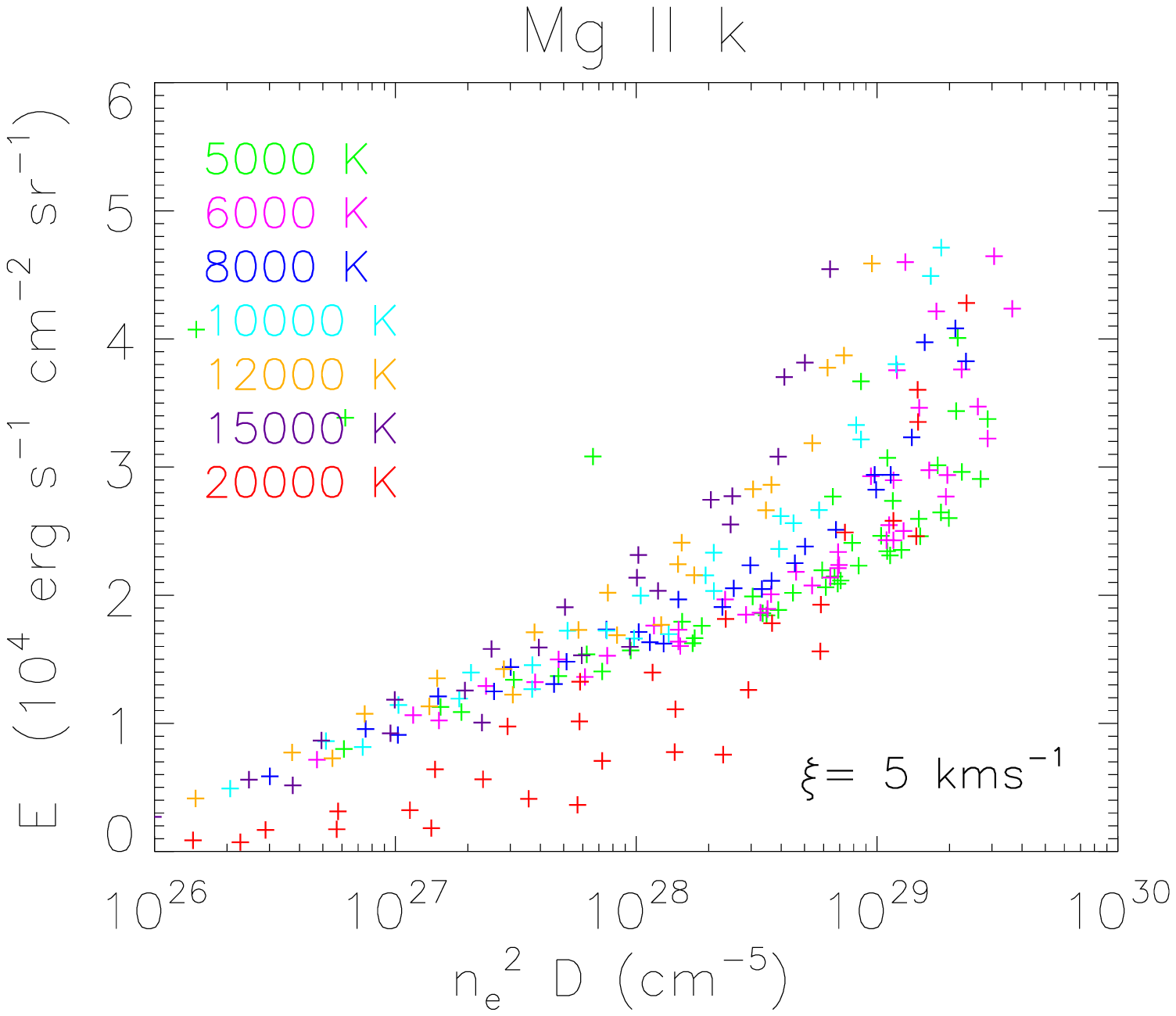}
            \hspace*{-0.03\textwidth}
            \includegraphics[width=0.27\textwidth,clip=]{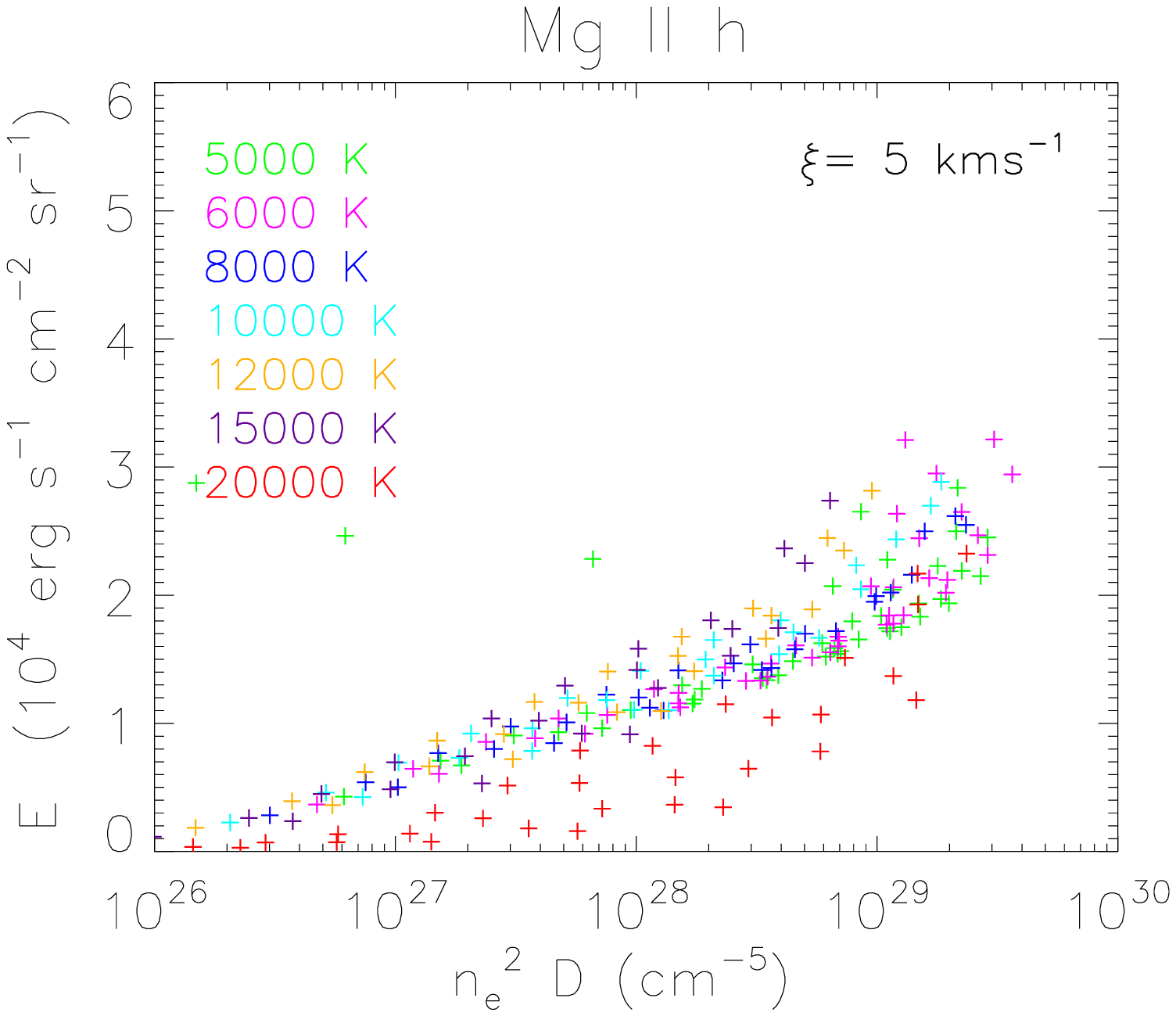}
            }
\vspace{0.01\textwidth}
\centerline{\includegraphics[width=0.27\textwidth,clip=]{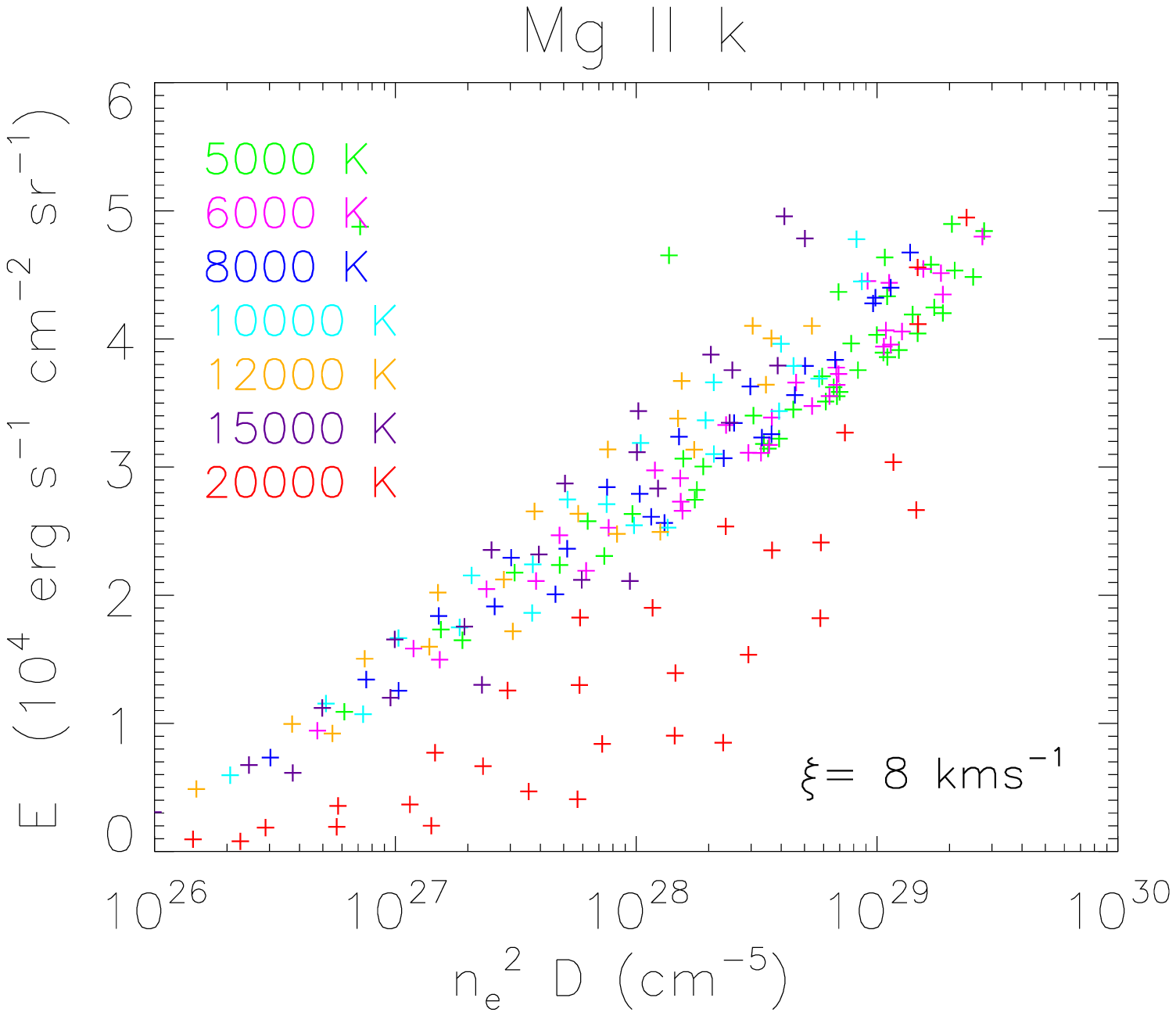}
            \hspace*{-0.03\textwidth}
            \includegraphics[width=0.27\textwidth,clip=]{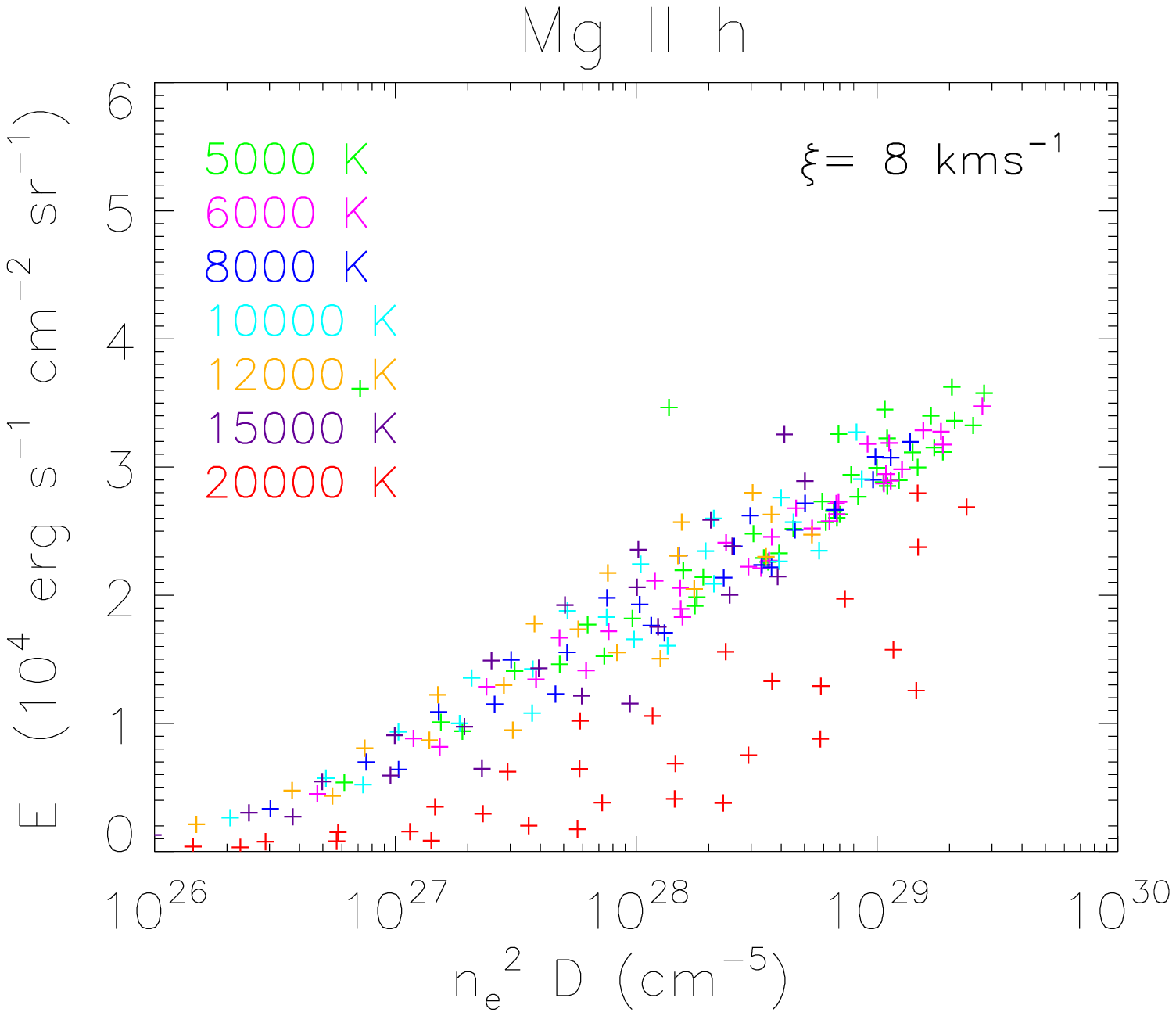}
            }
\caption{Integrated intensity emitted in \mg\ line as function of emission measure for different temperatures at three representative microturbulent velocities.}
\label{f-eem}
\end{figure} 

Figure~\ref{f-taue} presents the relationship between the integrated intensities of the \mg\ lines and the optical thicknesses at the line centers. Plots clearly show that for our 
range of observed integrated intensities (roughly between 5000 and 50\,000~\cgs), the observed prominence is optically thick with optical thicknesses up to 10$^{4}$. Points are more scattered in the case of the \mg\ k line (see the left panels) than the \mg\ h line because of a higher \mg\ k intensity. 
At lower temperatures the optical thickness reaches higher values in all plots. On the contrary, at higher temperatures, optical thickness 
becomes lower. At higher temperatures more \mg\ ions are ionized, which decreases the opacity and thus the optical thickness. However, at the same time, at higher temperatures 
we have more collisional excitations (which also increase with higher pressure) and for a given optical thickness we get higher emission in the \mg\ lines. 
As a result we see scatter in the plot 
because at a given temperature we have seven different pressures. The integrated intensity of both \mg\ lines at higher microturbulence for a given optical thickness
reaches higher values due to turbulent Doppler brightening \citep{hei14}. 

Figure~\ref{f-taur} shows a comparison between the reversal ratio and optical thickness at the line center for both \mg\ lines. The reversal ratio gradually increases with 
optical thickness at a given temperature. Profiles with optical thickness between 1 and 10 are not yet reversed but become flat. All flat and 
unreversed profiles (single) have a reversal ratio equal to one according to our definition. 
Points are more probably scattered due to the effect of pressure and thickness because the reversal ratio 
depends on optical thickness and optical thickness depends on the density of the lower atomic state, which is proportional to gas pressure at a given temperature. 
At higher microturbulence the reversal ratio and optical thickness are lower and points are less scattered. The \mg\ k line generally reaches higher values of reversal 
ratio than the \mg\ h line.

The left panel of Fig.~\ref{f-eratio} shows a comparison between the computed integrated intensities in the \mg\ h and \mg\ k lines for all 343 models at a given microturbulent 
velocity to show the trend. The correlation between both lines is quite good, points practically lie on a straight line. Slight deviations are related to different 
temperatures. The plot is consistent 
with observations (see upper panel of Fig.~\ref{f-spmg}). The right panels of Fig.~\ref{f-eratio} show the ratio of energy emitted in both \mg\ lines as a function of energy 
emitted in the \mg\ k line. The ratio is initially high, around 2.5, and then it decreases with increasing integrated intensity of the \mg\ k line to reach a practically constant value for a given 
temperature. Higher microturbulent velocity gives a lower line ratio. This plot is also consistent with observations 
(see lower panel of Fig.~\ref{f-spmg}) where the average value is 1.365 and scatter is mainly due to different temperatures of observed prominence structures.

Figure~\ref{f-avg} presents the ratio of energy emitted in the \mg\ k line to \mg\ h line at a given temperature for three representative microturbulent 
velocities (here we do not consider the additional effect of random motion of unresolved fine structures). From this plot we can conclude that the ratio betwen the lines (see the dashed horizontal line that shows the averaged observed value) can provide a 
useful diagnostic for the determination of a characteristic temperature of the studied prominence. The best fit is obtained for low temperatures reaching 5\,000 K and for
a microturbulent velocity of 5 km s$^{-1}$. 

Reversal ratio as a function of integrated intensity for both \mg\ lines is presented in Fig.~\ref{f-rrem}. The reversal ratio is equal to one for integrated intensities 
up to about 10$^4$ - 2~$\times$~10$^4$~\cgs. At higher intensities, the reversal ratio increases gradually for a given temperature. 
The reversal ratio is higher at a lower microturbulent velocity because the microturbulence smears the line profiles. 
Scatter mainly depends on temperature and microturbulence. There is no similarity between the models and observations (see the left and middle panels in 
Fig.~\ref{f-rre}), which poses a problem for the modeling; we discuss this issue later.

All plots in Fig.~\ref{f-eem} show that integrated intensity increases with the emission measure ${\rm EM} = n_{\rm e}^2 D$ at a given temperature (here we use
the standard definition of  emission measure). 
For a given emission measure the integrated intensity is higher at higher microturbulence. 
Plots show that \mg\ becomes fully ionized  around 20\,000~K.

\subsection{Analysis of \si\ spectra using CHIANTI} 
                   \label{s-chianti}

Besides the chromospheric UV lines of \mg\ and \ca, the IRIS spectrograph observed also hotter transition region (TR)
spectral lines among which the \si\ line is the most visible at the prominence location. Although the S/N ratio of the spectral intensities 
of this line is much lower than that of the \mg\ and \ca\ lines, the integrated intensity of the \si\ line averaged from all five sections along the slit 
and from all 16 slit positions within 27 selected rasters can be used to analyze this line emission. The idea is to estimate the geometrical 
extension of hotter PCTR regions where the \si\ line is formed. Assuming that the \si\ line is optically 
thin and that it is formed under the coronal approximation, we can use the CHIANTI atomic database \citep{der97} to calculate theoretical values 
of the integrated intensity. For the calculations 
in this paper we used version 8 of CHIANTI \citep{del15}. Total emissivity of the spectral line into unit solid angle is 
defined as a product of the spontaneous transition rate $A_{\mathrm{ul}}$ $n_{\mathrm{u}}$ (Einstein coefficient times the population density
of the upper atomic level) and the photon energy 
\begin{equation}
\varepsilon=\frac{h \nu}{4\pi}\,A_{\mathrm{ul}}\,n_{\mathrm{u}} \, .
\label{e-esich}
\end{equation}
The line intensity is then obtained by integration of the emissivity along the LOS through the whole optically-thin PCTR, which basically reduces
to a simple multiplication by the effective geometrical thickness $D$. The number density of ions in the upper level of transition is calculated using 
the statistical equilibrium. In CHIANTI the ionization and excitation equilibria are solved independently and the relative population of 
a specific ion is the function of temperature only while the relative population of ions in the upper level depends both on the temperature and electron 
density. Using the contribution function $G(n_{\rm e},T)$, the line integrated intensity is
\begin{equation}
E = G(n_{\rm e},T)\,\mathrm{EM'}\,,
\label{e-gf}
\end{equation}
where  $\mathrm{EM'}$ is the emission meassure defined in CHIANTI as 
\begin{equation}
\mathrm{EM'}=n_{\rm e}\,n_{\rm H}\,D\,
\label{e-emh}
,\end{equation}
with $n_{\rm H}$ being the total hydrogen density (neutrals and protons). Adopting a common solar value of 0.83 for the ratio $n_{\rm H}/n_{\rm e}$,  
the emission measure is then calculated as 
$0.83\,n_{\rm e}^2\,D$. 
We have calculated the contribution function for \si\ line at its formation temperature 80\,000~K for three representative
values of the gas pressure, 0.01, 0.05, and 0.1 dyn~cm$^{-2}$. For hydrogen and helium mixture at high temperatures, 
gas pressure is related to electron density through 
the equation of state $p= 1.91~n_{\rm e}\,k_{\rm B}\,T$, where $k_{\rm B}$ is the Boltzmann constant. Finally, the geometrical thickness is derived 
using the observed averaged integrated intensity of the \si\ line. Resulting values of the LOS PCTR thicknesses are shown in Table~\ref{t-dpctr}.
From Fig. \ref{f-eem} we can roughly estimate the prominence gas pressure as 0.1 dyn cm$^{-2}$ in central cool parts where the \mg\ lines are formed.
Assuming an isobaric prominence equilibrium, we get rather thin PCTR of the order of tens of kilometers. However, if the PCTR gas pressure
is smaller, for example, like in the hydrostatic equilibrium of prominence magnetic dips, the PCTR would be more extended. An extended PCTR is
expected in the dipped structures along the magnetic field lines \citep[see, e.g.,][]{gun10}. 
\begin{table} 
\caption{Values of typical gas pressures for quiescent prominences, electron densities, contribution functions, and geometrical thicknesses of hotter envelope of the prominence 
at approximate \si\ line formation temperature 80\,000\,K.}
\label{t-dpctr}
\begin{tabular}{c|ccc}
\hline \hline
$p$~(dyn~cm$^{-2}$) & $0.01$ & $0.05$ & $0.1$ \\
$n_{\rm e}$~(10$^{9}$~cm$^{-3}$) & 0.47 & 2.4  & 4.7  \\
$G(n_{\rm e},T)$~(10$^{-25}~$erg~cm$^3$~s$^{-1}$~sr$^{-1}$) & 8.1 & 4.4 & 4.0 \\ 
$D^{\rm PCTR}$~(km)            & $3300$ & $240$ & $70$ \\ \hline 
\end{tabular}
\end{table}

\section{Discussion and conclusions}
        \label{s-con}

In this study we have analyzed one of the first IRIS observations of a quiescent prominence. The prominence was observed on October 22, 2013 in NUV 
wavelengths at two strong 
\mg\ k and h lines at 2796 and 2803~\AA, as well as at FUV wavelengths where rather weak \ca\ lines at 1334 and 1336~\AA~and a faint and noisy \si\ line at 
1394 \AA~were detected. Level 2 data were calibrated to absolute radiometric units. Altogether we got 2160 observational points in five selected 
sections along the slit with $y$ binning equal to ten pixels due to low \sn\ ratio in \ca, and \si\ lines, and for 16-slit
raster with 27 repetitions. 
In this paper we address several key questions related to prominence structure and dynamics: What is the characteristic temperature and microturbulent
velocity, what is the role of PCTR, 
what are the dynamics of this prominence (i.e. fine-structure dynamics versus global oscillations), and how are the structure and dynamics reflected in the IRIS line
profiles and intensities? Having a large number of the observed points, we focused in this study on statistical correlations between various spectral line parameters,
rather than on detailed analysis of individual line profiles. In order to understand the physical behavior of such correlations, we also computed a large grid
of 1D isothermal-isobaric models using the \mg\ code described in \citet{hei14}. A comparison between observed and computed characteristics of \mg\ lines
points to limitations of these models. For example, the line-core intensity of the \mg\ and \ca\ line will certainly be influenced by the presence of a PCTR around
cool fine structures as indicated by our analysis of the \si\ line. 

Analysis shows that more than two-thirds of \mg, and almost one-half of \ca\ 1334 profiles are reversed, which indicates that the prominence structure is mostly 
optically thick, while \si\ profiles have a single peak indicating small optical thickness of the line. 
Correlations between integrated intensities of both \mg\ lines and both \ca\ lines show that time variations in line intensities are real 
and not caused by the noise. 
Detailed analysis shows that there are no global oscillations in the \mg, and \ca\ lines. From that we conclude that time 
variations in integrated intensities are caused by fine-structure random motion with LOS velocities up to 10~km~s$^{-1}$. 
Correlation between LOS velocities of \mg\ lines is better than between \mg\ and \ca\ lines because \mg\ lines are supposed to be formed in 
a similar region and are less noisy than \ca\ lines.
Correlations between integrated intensities and central intensities of \mg\ and \ca\ lines 
(the \ca\ 1336 line, which is a blended line, and the \si\ line, which is weak and noisy, are not considered)
show that LOS variations of plasma parameters are reflected only in integrated intensities but not in the line center intensities that seem to form 
in the foremost fine structures. 

The ratio of \mg\ k to \mg\ h line intensity is rather uniform, an average value is  equal to 1.365 and is consistent with the values reported by \citet{hei14}.
Compared to the present grid of 1D models this indicates that (i) the \mg\ lines are optically thick, (ii) the kinetic temperature can be as low as 5000 K 
(Fig.~\ref{f-avg}), and (iii) the
microturbulent velocity is around 5 km s$^{-1}$ or lower rather than the highest one. Such a low temperature would be consistent with extended 
higher-pressure structures and would correspond to radiative equilibrium models as presented in \citet{hei14}. In the follow-up study 
we will also include a PCTR in the modeling to check the robustness of this result, although PCTR will affect only the inner line cores.

Scatter plots for the integrated line intensities of different combinations of elements show weak or no correlations 
indicating that the lines are not formed in the same plasma volume. There is certain correlation between \mg\ k and \ca\ 1334 line intensity (Fig.~\ref{f-spmix}) 
and this is consistent with the situation in the quiet chromosphere where both lines are formed under similar conditions \citep{rat15}. In contrast,
the \si\ line is formed at much higher temperature and thus does not correlate with the \mg\ and \ca\ 1334 lines (Fig.~\ref{f-spmix}).
Scatter plots between the reversal ratio and integrated intensity of the \mg\ and \ca\ 1334 lines give clouds of points above unity because, by our definition, 
single and flat profiles have values of one and reversed profiles have values higher than one. Plots show that with increasing energy emitted in a given line the 
reversal ratio first increases and then gradually decreases. The observed 
reversal ratio is definitely larger than the computed one. For the \mg\ k line with energy between 1 - 2 $\times$ 10$^4$ \cgs\ many observed points show reversal 
between 1 and 1.5, while 1D models predict no reversal in this range of intensities for $\xi$= 5 or 8 km~s$^{-1}$ and a small reversal for zero microturbulence. 
If we  added a PCTR to the 1D models \citep[see, e.g.,][]{hei15}, the theoretical reversal ratio
would even decrease. Here we thus face a principal problem of why 1D models cannot reproduce the statistics of reversals in \mg\ lines. 
The situation is improved if we consider 1D models with zero microturbulent velocity (see Fig.~\ref{f-rrem}). One thing is that the microturbulence generally 
smears the line peaks so that for its lower values the peaks should be higher. However, we also found another reason for strengthening the peaks, which is the 
effect of partial redistribution (PRD)
on peak intensity at low values of microturbulent velocity. As mentioned in \citet{hei14}, the \mg\ line peaks are formed roughly under 
complete-redistribution (CRD) conditions
for models with $\xi$=5~km~s$^{-1}$. However, if $\xi$ is small, reaching zero, the peaks are formed already in the PRD regime and thus they reflect
the quasi-coherent scattering of the incident peaked radiation, similar to the case of hydrogen L$\alpha$. A comparison between PRD and CRD clearly shows
this additional effect of the peak enhancement. Our conclusion regarding the \mg\ line reversals is that the profiles synthesized from 1D isothermal-isobaric 
models have smaller reversals than the observed profiles, even when the microturbulent velocity is zero at the extreme. We note that synthetic profiles 
are sharply peaked in many cases, but these peaks are smeared by the convolution with the IRIS instrumental profile, which is necessary to compare with 
observations. As a next step, we therefore need to perform simulations of heterogeneous prominence structures with random velocities \citep{gun08} 
to see whether they can give results more consistent with current observations.

We have analyzed statistically the peak asymmetry for \mg\ and \ca\ 1334 reversed profiles. The histograms show  that the red-peak asymmetry 
slightly dominates, but  for the \mg\ h line the histogram is rather symmetrical. These results indicate that the peak asymmetry might be due to the random motions
of prominence fine structures rather than due to some systematic flows (oscillations or propagating waves). A similar kind of asymmetry was already detected
in prominences for hydrogen Lyman lines, where the asymmetry varied from pixel to pixel. This was successfully modeled by \citet{gun10} and \citet{sch15}
in terms of 2D multi-thread structures
with prescribed LOS random velocities up to 10 km s$^{-1}$. In a following paper we will use such 2D models to predict the \mg\ line profiles. We note that
10  km s$^{-1}$ corresponds to a Doppler shift of 0.1 \AA \, for \mg\ lines, which is comparable to peak separation from the line center. We therefore expect an important
effect of such fine-structure motions on both the line reversal as well as on the peak asymmetry and this cannot be achieved using simple 1D static slab models.
 
Finally, our analysis of \si\ line intensities using CHIANTI near the line formation temperature, for characteristic gas pressures in a quiescent prominence, gives
the LOS effective thicknesses of the PCTR. We plan to extend this kind of analysis also to our 2D fine-structure models with PCTR
depending on the orientation of the prominence magnetic field  \citep[see, e.g.,][]{gun10,gun14mag,gun15}. A possible role of radiative excitation in the \si\ line
is also a subject to be investigated \citep[see, e.g.,][]{gv2016}.


In a following study we will also use the same grid of 1D models to synthesize \ca\ lines (work in progress) and to compare the results with current
observations. However, our ultimate goal is to perform 2D multi-thread radiative transfer simulations for prescribed distribution of the LOS velocities and 
shapes of the PCTR and to synthesize all IRIS lines analyzed in this paper. Similar simulations have been done already for hydrogen Lyman lines
\citep{gun10,ber11,gun12,sch15}. To compare with observations, we will use SOHO/SUMER Lyman spectra and our ground-based observations in the various lines that are
available for this prominence.

\begin{acknowledgements}
SJ, PH, and SG acknowledge support from the Czech Science Foundation (GA\v CR) through the grant No. 16-18495S. SJ acknowledges financial support from the Slovenian 
Research Agency No. P1-0188. PS acknowledges support from the project VEGA 2/0004/16 of the Science Agency. PS and SG acknowledge support from the Joint Mobility 
Project of Academy of Sciences of the Czech Republic and Slovak Academy of Sciences No.~SAV-AV\v{C}R-18-03. PH and SG acknowledge support from grant
16-17586S of the Czech Science Foundation (GA\v CR). SJ, PH, MZ, and SG acknowledge support from project RVO:67985815 of the
Astronomical Institute of the Czech Academy of Sciences. CHIANTI is a collaborative project involving George Mason University (USA), 
the University of Michigan (USA) and the University of Cambridge (UK).  
We thank the anonymous referee for useful comments.

\end{acknowledgements}

\bibliographystyle{aa}
\bibliography{biblio}

\begin{thebibliography}{39}
\expandafter\ifx\csname natexlab\endcsname\relax\def\natexlab#1{#1}\fi

\bibitem[{{Arregui} {et~al.}(2012){Arregui}, {Oliver}, \& {Ballester}}]{arr12}
{Arregui}, I., {Oliver}, R., \& {Ballester}, J.~L. 2012, Living Reviews in
  Solar Physics, 9, 2

\bibitem[{{Berlicki} {et~al.}(2011){Berlicki}, {Gun{\'a}r}, {Heinzel},
  {Schmieder}, \& {Schwartz}}]{ber11}
{Berlicki}, A., {Gun{\'a}r}, S., {Heinzel}, P., {Schmieder}, B., \& {Schwartz},
  P. 2011, \aap, 530, A143

\bibitem[{{De Pontieu} {et~al.}(2014){De Pontieu}, {Title}, {Lemen}, {Kushner},
  {Akin}, {Allard}, {Berger}, {Boerner}, {Cheung}, {Chou}, {Drake}, {Duncan},
  {Freeland}, {Heyman}, {Hoffman}, {Hurlburt}, {Lindgren}, {Mathur}, {Rehse},
  {Sabolish}, {Seguin}, {Schrijver}, {Tarbell}, {W{\"u}lser}, {Wolfson},
  {Yanari}, {Mudge}, {Nguyen-Phuc}, {Timmons}, {van Bezooijen}, {Weingrod},
  {Brookner}, {Butcher}, {Dougherty}, {Eder}, {Knagenhjelm}, {Larsen},
  {Mansir}, {Phan}, {Boyle}, {Cheimets}, {DeLuca}, {Golub}, {Gates}, {Hertz},
  {McKillop}, {Park}, {Perry}, {Podgorski}, {Reeves}, {Saar}, {Testa}, {Tian},
  {Weber}, {Dunn}, {Eccles}, {Jaeggli}, {Kankelborg}, {Mashburn}, {Pust},
  {Springer}, {Carvalho}, {Kleint}, {Marmie}, {Mazmanian}, {Pereira}, {Sawyer},
  {Strong}, {Worden}, {Carlsson}, {Hansteen}, {Leenaarts}, {Wiesmann},
  {Aloise}, {Chu}, {Bush}, {Scherrer}, {Brekke}, {Martinez-Sykora}, {Lites},
  {McIntosh}, {Uitenbroek}, {Okamoto}, {Gummin}, {Auker}, {Jerram}, {Pool}, \&
  {Waltham}}]{pon14}
{De Pontieu}, B., {Title}, A.~M., {Lemen}, J.~R., {et~al.} 2014, \solphys, 289,
  2733

\bibitem[{{Del Zanna} {et~al.}(2015){Del Zanna}, {Dere}, {Young}, {Landi}, \&
  {Mason}}]{del15}
{Del Zanna}, G., {Dere}, K.~P., {Young}, P.~R., {Landi}, E., \& {Mason}, H.~E.
  2015, \aap, 582, A56

\bibitem[{{Dere} {et~al.}(1997){Dere}, {Landi}, {Mason}, {Monsignori Fossi}, \&
  {Young}}]{der97}
{Dere}, K.~P., {Landi}, E., {Mason}, H.~E., {Monsignori Fossi}, B.~C., \&
  {Young}, P.~R. 1997, \aaps, 125, 149

\bibitem[{{Gontikakis} \& {Vial}(2016)}]{gv2016}
{Gontikakis}, C. \& {Vial}, J.-C. 2016, \aap, 590, A86

\bibitem[{{Gun{\'a}r}(2014)}]{gun14}
{Gun{\'a}r}, S. 2014, in IAU Symposium, Vol. 300, IAUS, ed. B.~{Schmieder},
  J.~{Malherbe}, \& S.~{Wu}, 59

\bibitem[{{Gun{\'a}r} {et~al.}(2008){Gun{\'a}r}, {Heinzel}, {Anzer}, \&
  {Schmieder}}]{gun08}
{Gun{\'a}r}, S., {Heinzel}, P., {Anzer}, U., \& {Schmieder}, B. 2008, \aap,
  490, 307

\bibitem[{{Gun{\'a}r} \& {Mackay}(2015)}]{gun15}
{Gun{\'a}r}, S. \& {Mackay}, D.~H. 2015, \apj, 803, 64

\bibitem[{{Gun{\'a}r} {et~al.}(2012){Gun{\'a}r}, {Mein}, {Schmieder},
  {Heinzel}, \& {Mein}}]{gun12}
{Gun{\'a}r}, S., {Mein}, P., {Schmieder}, B., {Heinzel}, P., \& {Mein}, N.
  2012, \aap, 543, A93

\bibitem[{{Gun{\'a}r} {et~al.}(2014){Gun{\'a}r}, {Schwartz}, {Dud{\'{\i}}k},
  {Schmieder}, {Heinzel}, \& {Jur{\v c}{\'a}k}}]{gun14mag}
{Gun{\'a}r}, S., {Schwartz}, P., {Dud{\'{\i}}k}, J., {et~al.} 2014, \aap, 567,
  A123

\bibitem[{{Gun{\'a}r} {et~al.}(2010){Gun{\'a}r}, {Schwartz}, {Schmieder},
  {Heinzel}, \& {Anzer}}]{gun10}
{Gun{\'a}r}, S., {Schwartz}, P., {Schmieder}, B., {Heinzel}, P., \& {Anzer}, U.
  2010, \aap, 514, A43

\bibitem[{{Heinzel}(2007)}]{hei07}
{Heinzel}, P. 2007, in ASPC Ser., Vol. 368, ASPC, ed. {P.~Heinzel,
  I.~Dorotovi{\v c}, \& R.J.~Rutten}, 271

\bibitem[{{Heinzel} {et~al.}(2015){Heinzel}, {Schmieder}, {Mein}, \&
  {Gun{\'a}r}}]{hei15}
{Heinzel}, P., {Schmieder}, B., {Mein}, N., \& {Gun{\'a}r}, S. 2015, \apjl,
  800, L13

\bibitem[{{Heinzel} {et~al.}(2016){Heinzel}, {Susino}, {Jej{\v c}i{\v c}},
  {Bemporad}, \& {Anzer}}]{hei16cme}
{Heinzel}, P., {Susino}, R., {Jej{\v c}i{\v c}}, S., {Bemporad}, A., \&
  {Anzer}, U. 2016, \aap, 589, A128

\bibitem[{{Heinzel} {et~al.}(2014{\natexlab{a}}){Heinzel}, {Vial}, \&
  {Anzer}}]{hei14}
{Heinzel}, P., {Vial}, J.-C., \& {Anzer}, U. 2014{\natexlab{a}}, \aap, 564,
  A132

\bibitem[{{Heinzel} {et~al.}(2014{\natexlab{b}}){Heinzel}, {Zapi{\'o}r},
  {Oliver}, \& {Ballester}}]{hei14b}
{Heinzel}, P., {Zapi{\'o}r}, M., {Oliver}, R., \& {Ballester}, J.~L.
  2014{\natexlab{b}}, \aap, 562, A103

\bibitem[{{Jej{\v c}i{\v c}} {et~al.}(2017){Jej{\v c}i{\v c}}, {Susino},
  {Heinzel}, {Dzif{\v c}{\'a}kov{\'a}}, {Bemporad}, \& {Anzer}}]{jej17}
{Jej{\v c}i{\v c}}, S., {Susino}, R., {Heinzel}, P., {et~al.} 2017, \aap, 607,
  A80

\bibitem[{{Kleint} {et~al.}(2016){Kleint}, {Heinzel}, {Judge}, \&
  {Krucker}}]{kle16}
{Kleint}, L., {Heinzel}, P., {Judge}, P., \& {Krucker}, S. 2016, \apj, 816, 88

\bibitem[{{Kotr{\v c}}(2009)}]{kot09}
{Kotr{\v c}}, P. 2009, Central European Astrophysical Bulletin, 33, 327

\bibitem[{{Ku{\v c}era} {et~al.}(2010){Ku{\v c}era}, {Ambr{\' o}z},
  {G{\"o}m{\"o}ry}, {Koz{\' a}k}, \& {Ryb{\' a}k}}]{kuc10}
{Ku{\v c}era}, A., {Ambr{\' o}z}, J., {G{\"o}m{\"o}ry}, P., {Koz{\' a}k}, M.,
  \& {Ryb{\' a}k}, J. 2010, Contributions of the Astronomical Observatory
  Skalnate Pleso, 40, 135

\bibitem[{{Labrosse} {et~al.}(2010){Labrosse}, {Heinzel}, {Vial}, {Kucera},
  {Parenti}, {Gun{\'a}r}, {Schmieder}, \& {Kilper}}]{lab10}
{Labrosse}, N., {Heinzel}, P., {Vial}, J.-C., {et~al.} 2010, \ssr, 151, 243

\bibitem[{{Levens} {et~al.}(2017){Levens}, {Labrosse}, {Schmieder}, {L{\'o}pez
  Ariste}, \& {Fletcher}}]{lev17}
{Levens}, P.~J., {Labrosse}, N., {Schmieder}, B., {L{\'o}pez Ariste}, A., \&
  {Fletcher}, L. 2017, \aap, 607, A16

\bibitem[{{Levens} {et~al.}(2016){Levens}, {Schmieder}, {Labrosse}, \&
  {L{\'o}pez Ariste}}]{lev16}
{Levens}, P.~J., {Schmieder}, B., {Labrosse}, N., \& {L{\'o}pez Ariste}, A.
  2016, \apj, 818, 31

\bibitem[{{Lin} {et~al.}(2005){Lin}, {Engvold}, {Rouppe van der Voort}, {Wiik},
  \& {Berger}}]{lin05}
{Lin}, Y., {Engvold}, O., {Rouppe van der Voort}, L., {Wiik}, J.~E., \&
  {Berger}, T.~E. 2005, \solphys, 226, 239

\bibitem[{{Liu} {et~al.}(2015){Liu}, {De Pontieu}, {Vial}, {Title}, {Carlsson},
  {Uitenbroek}, {Okamoto}, {Berger}, \& {Antolin}}]{liu15}
{Liu}, W., {De Pontieu}, B., {Vial}, J.-C., {et~al.} 2015, \apj, 803, 85

\bibitem[{{Mackay} {et~al.}(2010){Mackay}, {Karpen}, {Ballester}, {Schmieder},
  \& {Aulanier}}]{mac10}
{Mackay}, D., {Karpen}, J., {Ballester}, J., {Schmieder}, B., \& {Aulanier}, G.
  2010, \ssr, 151, 333

\bibitem[{{Oliver} \& {Ballester}(2002)}]{oli02}
{Oliver}, R. \& {Ballester}, J.~L. 2002, \solphys, 206, 45

\bibitem[{{Parenti}(2014)}]{par14}
{Parenti}, S. 2014, Living Reviews in Solar Physics, 11, 1

\bibitem[{{Rathore} {et~al.}(2015){Rathore}, {Carlsson}, {Leenaarts}, \& {De
  Pontieu}}]{rat15}
{Rathore}, B., {Carlsson}, M., {Leenaarts}, J., \& {De Pontieu}, B. 2015, \apj,
  811, 81

\bibitem[{{Scargle}(1982)}]{sca82}
{Scargle}, J.~D. 1982, \apj, 263, 835

\bibitem[{{Schmieder} {et~al.}(2014{\natexlab{a}}){Schmieder}, {Malherbe}, \&
  {Wu}}]{iau14}
{Schmieder}, B., {Malherbe}, J.-M., \& {Wu}, S.~T., eds. 2014{\natexlab{a}},
  IAU Symposium, Vol. 300, {Nature of Prominences and their role in Space
  Weather}

\bibitem[{{Schmieder} {et~al.}(2014{\natexlab{b}}){Schmieder}, {Tian},
  {Kucera}, {L{\'o}pez Ariste}, {Mein}, {Mein}, {Dalmasse}, \& {Golub}}]{sch14}
{Schmieder}, B., {Tian}, H., {Kucera}, T., {et~al.} 2014{\natexlab{b}}, \aap,
  569, A85

\bibitem[{{Schwartz} {et~al.}(2015){Schwartz}, {Gun{\'a}r}, \& {Curdt}}]{sch15}
{Schwartz}, P., {Gun{\'a}r}, S., \& {Curdt}, W. 2015, \aap, 577, A92

\bibitem[{{Terradas} {et~al.}(2002){Terradas}, {Molowny-Horas}, {Wiehr},
  {Balthasar}, {Oliver}, \& {Ballester}}]{ter02}
{Terradas}, J., {Molowny-Horas}, R., {Wiehr}, E., {et~al.} 2002, \aap, 393, 637

\bibitem[{{Vial} \& {Engvold}(2015)}]{vial15}
{Vial}, J.-C. \& {Engvold}, O., eds. 2015, ASS Library, Vol. 415, {Solar
  Prominences}

\bibitem[{{Wilhelm} {et~al.}(1995){Wilhelm}, {Curdt}, {Marsch}, {Sch{\"u}hle},
  {Lemaire}, {Gabriel}, {Vial}, {Grewing}, {Huber}, {Jordan}, {Poland},
  {Thomas}, {K{\"u}hne}, {Timothy}, {Hassler}, \& {Siegmund}}]{wil95}
{Wilhelm}, K., {Curdt}, W., {Marsch}, E., {et~al.} 1995, \solphys, 162, 189

\bibitem[{{Zapi{\'o}r} {et~al.}(2015){Zapi{\'o}r}, {Kotr{\v c}}, {Rudawy}, \&
  {Oliver}}]{zap15}
{Zapi{\'o}r}, M., {Kotr{\v c}}, P., {Rudawy}, P., \& {Oliver}, R. 2015,
  \solphys, 290, 1647

\bibitem[{{Zapi{\'o}r} {et~al.}(2016){Zapi{\'o}r}, {Oliver}, {Ballester}, \&
  {Heinzel}}]{zap16}
{Zapi{\'o}r}, M., {Oliver}, R., {Ballester}, J.~L., \& {Heinzel}, P. 2016,
  \apj, 827, 131

\end{thebibliography}

%
%

\end{document}